\documentclass{class/jfm}
\usepackage{graphicx}
\usepackage{amsmath}
\usepackage{epstopdf,epsfig}
\usepackage[normalem]{ulem}

\usepackage{graphicx} 
\usepackage{dcolumn} 
\usepackage{bm} 
\usepackage{hyperref} 
\usepackage{ragged2e} 
\usepackage{setspace} 
\usepackage{fancyhdr} 
\usepackage{upgreek} 
\usepackage{float} 
\usepackage{tikz} 
\usepackage[skip=0pt]{subcaption}
\usetikzlibrary{arrows,shapes,trees,calc,positioning}
\usepackage{gensymb} 
\usepackage{mathrsfs} 
\usepackage{color} 
\usepackage{stackengine} 
\usepackage{soul} 
\usepackage{mathtools}
\usepackage{bigints}
\usepackage{relsize}
\usepackage{braket}
\usepackage{cancel}
\usepackage{relsize}
\usepackage{xcolor}
\hypersetup{colorlinks,urlcolor=blue,linkcolor=blue,citecolor=blue} 

\fancypagestyle{tstyle}{\chead{}} 

\newcommand{\tc}{\textcolor}

\definecolor{orange}{RGB}{255,93,11}
\definecolor{bg}{RGB}{0,0,255}
\definecolor{dg}{RGB}{0,150,0}
\definecolor{lg}{RGB}{46,204,113}
\definecolor{dr}{RGB}{192,57,43}
\usepackage{algorithm}
\usepackage[noend]{algpseudocode}
\newcommand{\B}[1]{\mathbf #1}

\newcommand{\cA}{\mathcal{A}}
\newcommand{\cB}{\mathcal{B}}
\newcommand{\cC}{\mathcal{C}}
\newcommand{\cD}{\mathcal{D}}
\newcommand{\cH}{\mathcal{H}}

\newcommand{\cO}{\mathcal{O}}
\newcommand{\cR}{\mathcal{R}}
\newcommand{\cN}{\mathcal{N}}
\newcommand{\cW}{\mathcal{W}}
\newcommand{\cI}{\mathcal{I}}
\newcommand{\bx}{\mathbf{x}}
\newcommand{\bd}{\mathbf{d}}
\newcommand{\bu}{\mathbf{u}}
\newcommand{\bphi}{\mbox{\boldsymbol{$\phi$}}}
\newcommand{\bpsi}{\mbox{\boldsymbol{$\psi$}}}
\newcommand{\bPsi}{\mbox{\boldsymbol{$\Psi$}}}
\newcommand{\bPhi}{\mbox{\boldsymbol{$\Phi$}}}

\newcommand{\mre}{\mathrm{e}}
\newcommand{\mrd}{\mathrm{d}}
\newcommand{\mri}{\mathrm{i}}
\newcommand{\DefinedAs}[0]{\mathrel{\mathop:}=}

\newsavebox\mybox
\newcommand{\matbegin}{
        \left[
}
\newcommand{\matend}{
        \right]
}

\newcommand{\tbo}[2]{
  \matbegin \begin{array}{c}
       #1 \\ #2
       \end{array} \matend }
\newcommand{\tbt}[4]{
  \matbegin \begin{array}{cc}
       #1 & #2 \\ #3 & #4
       \end{array} \matend }

\shorttitle{Oblique transition in hypersonic double-wedge flow}
\shortauthor{Dwivedi, Sidharth, Jovanovi\'c}

\title{\Large \tc{black}{Oblique transition in hypersonic double-wedge flow}}

\author{Anubhav Dwivedi\aff{1}
  \corresp{\email{anubhavd91@gmail.com}}, G. S. Sidharth\aff{2}, Mihailo R. Jovanovi\'c\aff{1}}
\affiliation{\aff1{Ming-Hsieh Department of Electrical and Computer Engineering, 
\\
University of Southern California, Los Angeles, CA, USA} 
\aff2{X-Computational Physics, Los Alamos National Laboratory, NM, USA}
\\}

\begin{document}
\maketitle

	\begin{abstract}
	We utilize \tc{black}{resolvent} and weakly nonlinear analyses in combination with direct numerical simulations (DNS) to identify mechanisms for oblique transition in a Mach $5$ hypersonic flow over an adiabatic slender double-wedge. Even though the laminar separated flow is globally stable, \tc{black}{resolvent} analysis demonstrates significant amplification of unsteady external disturbances \tc{black}{to the linearized flow equations}. These disturbances are introduced upstream of the separation zone and they lead to the appearance of oblique waves further downstream. We demonstrate that large amplification of oblique waves arises from \tc{black}{the growth of fluctuation shear stress due to streamline curvature of the laminar base flow in the separated shear layer.} This is in contrast to the attached boundary layers, where no such mechanism exists. We also use a weakly nonlinear analysis to show that the resolvent operator associated with linearization around the laminar base flow governs the evolution of steady reattachment streaks that arise from quadratic interactions of unsteady oblique waves. These quadratic interactions generate vortical excitations in the reattaching shear layer which lead to the formation of streaks in the recirculation zone and their subsequent amplification, breakdown, and transition to turbulence downstream. Our analysis of the energy budget shows that deceleration of the base flow near reattachment is primarily responsible for amplification of steady streaks. Finally, we employ DNS to examine latter stages of transition to turbulence and demonstrate the predictive power of \tc{black}{a weakly nonlinear input-output framework} in uncovering triggering mechanisms for oblique transition in separated high-speed boundary layer flows. 
	\end{abstract}
	
	\vspace*{-4ex}
\section{Introduction}

	Slender double-wedges are commonly encountered in intakes, control surfaces, and junctions in high-speed supersonic and hypersonic vehicles~\citep{dolling2001fifty}. In this geometry, laminar boundary layer can separate at the corner because of the pressure rise that arises from deflection of the inviscid free stream. The resulting flow is characterized by separation-reattachment shocks as well as a recirculation zone and it provides a canonical setup for studying shock-wave-boundary-layer interaction (SWBLI)~\citep{simeonides1995experimental}. In spite of spanwise homogeneity of laminar base flows over compression corners, both experiments~\citep{chuvakhov2017effect,roghelia2017experimental,dwivedi2020three} and numerical simulations~\citep{navarro2005numerical,dwivedi2017optimal,caoJFM2021} identify three-dimensional (3D) features in time-averaged separated flow. In particular, streamwise streaks associated with persistent local peaks of heat flux or wall temperature, that appear near reattachment, can trigger transition to turbulence downstream~\citep{simeonides1995experimental,roghelia2017experimental}. 

The development of 3D flow structures in hypersonic flows was recently studied by examining the growth of small perturbations in the presence of a recirculation zone~\citep{anubhav-phd20}. For example, 2D SWBLI 
\tc{black}{can become} unstable inside the separation bubble when the strength of interaction increases beyond a critical value~\citep{gs2017global}. The spanwise modulation that arises from global instability introduces streaks over compression corners~\citep{sidharth2018onset} \tc{black}{as well as} oblique shocks impinging on a flat plate~\citep{hildebrand2017simulation} \tc{black}{and it can trigger transition to turbulence~\citep{caohaokliwenoliheu_JFM2022}}. \tc{black}{Similar 3D flow features have also been observed in hypersonic regimes where non-continuum effects are important~\citep{sawant2022linear}. However, recent numerical simulations and global stability analysis demonstrate that hypersonic compression corner flows can be stabilized by increasing radius of the leading edge (i.e., its bluntness)~\citep{cao_hao_klioutchnikov_olivier_heufer_wen_2021} or by increasing \mbox{the wall temperature~\citep{hao_cao_wen_olivier_2021}.}}

Even in the absence of global instability, high-speed separated flows are highly sensitive to upstream vortical disturbances~\citep{dwisidniccanjovJFM19} and small fluctuations around the laminar 2D base flow can experience significant non-modal amplification that leads to the appearance of steady reattachment streaks~\citep{dwivedi2020transient}. Furthermore, recent experiments on the cone-flare configurations~\citep{benitez2020instability,butler_laurence_2021}, which represent axisymmetric counterparts of slender double-wedges, identify unsteady fluctuations in the separation zone. These fluctuations are significantly amplified in the recirculation zone and they play an important role in transition to turbulence~\citep{butler_laurence_2021}.

In this paper, we examine amplification of unsteady fluctuations around the laminar 2D base flow in the separation/reattachment zone and investigate subsequent transition to turbulence.
Free-stream disturbances~\citep{choudhari1996boundary,berlin1999nonlinear,maslov2001leading} that arise from wind tunnel noise in ground experiments~\citep{schneider2015developing} or from atmospheric disturbances in free flights~\citep{bushnell1990notes,skinner2020situ} can lead to the appearance of unsteady fluctuations in boundary layer flows. \tc{black}{It is well-documented that unsteady oblique waves provide a potent mechanism for initiating transition in low-speed incompressible~\citep{berlin1999numerical,rigas_sipp_colonius_2021} and compressible~\citep{chang_malik_1994,mayer2011direct} boundary layers.} Even though the importance of oblique fluctuations in initiating transition in attached high-speed boundary layers has received significant attention~\citep{ma2005receptivity,sivasubramanian_fasel_2015,hader_fasel_2019}, their role in separated high-speed flows has not been studied. Recent experiments~\citep{benitez2020instability} suggest that their amplification within the recirculation zone can trigger unsteadiness in transitional SBWLI flows. We utilize \tc{black}{global resolvent and weakly nonlinear analyses} to quantify amplification of unsteady upstream disturbances in a Mach $5$ flow over a slender double-wedge and characterize their role in initiating transition to turbulence in high-speed separated boundary layers.

\tc{black}{Resolvent} analysis provides a framework for evaluating responses (outputs) of stable dynamical systems to \tc{black}{time-periodic} external disturbances (inputs)~\citep{trefethen1993hydrodynamic,schhen01,schmid2007nonmodal}. For time-independent globally stable base flows, the steady-state response of the linearized Navier-Stokes (NS) equations to a harmonic input with frequency $\omega$ is also harmonic with the same frequency and the frequency response operator maps the input forcing to the resulting steady-state output~\citep{jovARFM21}. The singular value decomposition (SVD) of the frequency response characterizes amplification across frequency $\omega$ and decomposes inputs and outputs into modes whose significance is ordered by the magnitude of the corresponding singular values~\citep{schmid2007nonmodal}. In addition to providing insights into dynamics of canonical incompressible flows~\citep{jovanovic2005componentwise,mckeon2010critical,brandt2011effect,sipp2013characterization,ranzarhacjovPRF19a,ranzarhacjovPRF19b}, input-output analysis has also been utilized to discover mechanisms for noise generation in turbulent jets~\tc{black}{\citep{garnaud_lesshafft_schmid_huerre_2013,jeun2016input,schmidt2018spectral}, separation control on airfoils~\citep{yeh_taira_2019},} and the appearance of reattachment streaks in \mbox{hypersonic flows~\citep{dwisidniccanjovJFM19}.}
	
\tc{black}{In a Mach $5$ double-wedge flow subject to unsteady disturbances, we employ resolvent analysis to demonstrate that oblique waves represent the most energetic response of the compressible linearized NS equations. We utilize the compressible energy norm~\citep{chu1965energy,hanifi1996transient} to quantify energy amplification and show that unsteady upstream disturbances that are localized before flow separation induce oblique waves downstream of the double-wedge corner. Our analysis of the transport equation for the streamwise specific kinetic energy of oblique waves reveals that concave flow curvature of the separated/reattaching laminar 2D base flow is the primary source of amplification in the presence of SWBLI. We also utilize a weakly nonlinear analysis to demonstrate that quadratic interactions of oblique waves generate vortical excitations that induce reattachment streaks in the recirculation bubble. We show that the resolvent operator associated with linearization around the laminar 2D base flow governs the evolution of steady reattachment streaks and use SVD to demonstrate that the streaks are well approximated by the second output resolvent mode. Our analysis of the energy budget shows that the base flow deceleration near reattachment is primarily responsible for amplification of reattachment streaks. Finally, we conduct DNS to confirm the predictive power of our approach and provide insight into latter stages of transition to turbulence.}

\tc{black}{Recently,~\citet{rigas_sipp_colonius_2021} utilized a variational framework to extend input-output analysis in the frequency domain to the nonlinear NS equations. For fundamental forcing, the disturbance that triggers transition and yields the largest skin-friction coefficient in incompressible boundary layer is given by a pair of oblique waves with temporal frequency and spanwise wavenumber which are very close to the ones identified by the resolvent analysis of the linearized NS equations~\citep{rigas_sipp_colonius_2021}. While the Orr-mechanism~\citep{schhen01} and the Tollmien-Schlichting linear instability~\citep{sipp2013characterization} initiate the early stages of oblique transition in the attached low-speed boundary layers~\citep{rigas_sipp_colonius_2021}, even linear amplification mechanisms are poorly understood in separated compressible flows. Recent numerical simulations with inlet stochastic excitations in axisymmetric cylinder flare geometry showed that the local ``first mode'' instability~\citep{mac84b} can cause growth of oblique waves in the upstream boundary layer (i.e., before separation) and initiate transition in high-speed compressible flows with SWBLI~\citep{lugrin_beneddine_leclercq_garnier_bur_2021}. However, in the absence of local and global boundary layer instabilities, the role of flow separation in amplification of unsteady fluctuations and the ensuing transition has not been previously investigated.}

\begin{figure}
     \centering
       \begin{tabular}{c}
       \includegraphics[width=0.7\linewidth, trim= 4 4 5 6, clip]{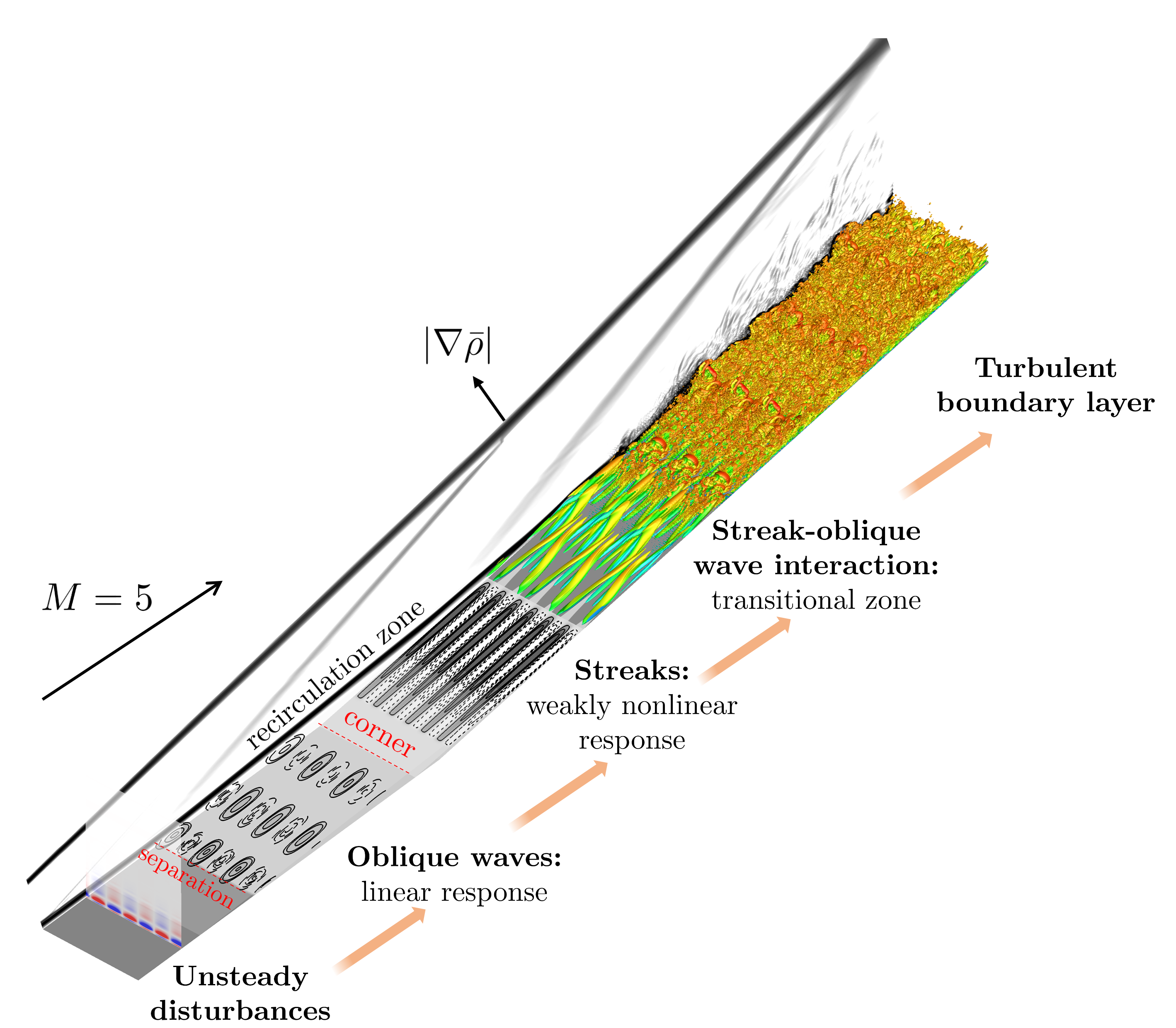}
     \end{tabular}
     \caption{\tc{black}{Preview of key results: spatially localized unsteady upstream forcing triggers oblique waves in the separated shear layer and their quadratic interactions lead to the appearance of steady reattachment streaks further downstream. DNS validates our theoretical predictions and demonstrates the efficacy of unsteady oblique disturbances in triggering transition in globally stable separated high-speed boundary layer flows.}} 
     \label{fig:summary}
 \end{figure}

\tc{black}{Figure~\ref{fig:summary} provides a summary of our key findings. We utilize resolvent analysis of the laminar 2D base flow {with spatially localized forcing introduced in the streamwise plane immediately upstream of separation} to identify the spatial structure of unsteady external disturbances that yield the most energetic response of the compressible linearized NS equations. The resulting  forcing is used to trigger non-modal amplification of oblique waves in the separated shear layer and generate steady reattachment streaks, which are routinely observed in SWBLI experiments, further downstream through weakly nonlinear interactions. Interaction of streaks with oblique waves is observed after reattachment and DNS is used to demonstrate that unsteady upstream oblique disturbances can indeed trigger transition to turbulence in separated high-speed compressible flows.} 

Our presentation is organized as follows. In \S~\ref{sec:setup}, we \tc{black}{introduce} the slender double-wedge geometry along with a finite-volume compressible flow solver that we use in our computations. In \S~\ref{sec:io}, we \tc{black}{describe resolvent and weakly nonlinear analyses} that we use to evaluate frequency responses of the double-wedge flow in the presence of 3D disturbances. We \tc{black}{also utilize resolvent analysis to} demonstrate large amplification of unsteady oblique \tc{black}{disturbances to the linearized flow equations} and identify the underlying physical mechanism. In \S~\ref{ref:strk}, we employ a weakly nonlinear analysis to demonstrate that quadratic interactions of oblique waves induce steady reattachment streaks and discuss physical mechanism responsible for their amplification in recirculation zone. In \S~\ref{sec:secinstab}, we employ DNS to validate utility of our theoretical predictions and examine latter stages of transition induced by unsteady upstream disturbances. In \S~\ref{sec:transition}, we analyze statistical features of the resulting transitional and turbulent boundary layers. We provide summary of our contributions and offer concluding remarks in \S~\ref{sec:discuss}.

	\vspace*{-4ex}
\section{Hypersonic flow over an adiabatic slender double-wedge}
\label{sec:setup}

\begin{figure}
    \centering
    \includegraphics[width=0.7\linewidth]{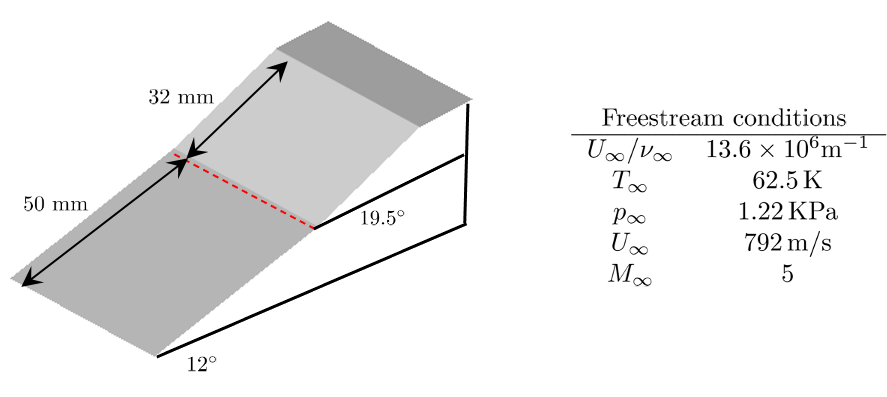}
    \caption{Slender double-wedge geometry and the associated free-stream conditions.}
    \label{fig:flowsetup}
\end{figure}

Hypersonic flow over a slender double-wedge with free-stream conditions shown in figure~\ref{fig:flowsetup} corresponds to the experiments of~\citet{yang2012investigation}. Since the enthalpy and the temperature in the flow field are low, we utilize the ideal gas law abstraction and employ the finite-volume compressible flow solver US3D~\citep{candler2015development} to compute the solution of the compressible NS equations in conservative form,
	\begin{align}
    \label{eq:nse}
    \tc{black}{\frac{\partial \bPsi}{\partial t} 
    \; = \; 
    \mathcal{F} (\bPsi).}
\end{align}
\tc{black}{Here, $\mathcal{F} (\bPsi) \DefinedAs - \nabla \cdot {\B F} (\bPsi)$ is the dynamical generator of the compressible NS equations, ${\B F} (\bPsi)$ is the flux vector, $\nabla$ is the gradient, and $\bPsi = (\rho, \rho \B u, E_t)$} is the vector of conserved variables that represent density, momentum, and total energy per unit volume of the gas. \tc{black}{In equation~\eqref{eq:nse} and throughout the paper, spatial coordinates are non-dimensionalized by the boundary layer thickness at separation $\delta_{99} = 9.8 \times 10^{-4} \, \mathrm{m}$, velocity by the free-stream velocity $U_\infty$, pressure by $p_{\infty}$, temperature by $T_{\infty}$, and time  by $\delta_{99}/U_\infty$. The Reynolds number based on separation boundary layer thickness is $13.3 \times 10 ^{3}$ and the Mach number is $5.0$.} 

We discretize the inviscid fluxes using the second-order-accurate modified Steger-Warming fluxes with the MUSCL limiters~\citep{candler2015development}. In previous studies, the numerical method for the computation of the base state was validated using hypersonic double-wedge and double-cone setups~\citep{nompelis2003effect,nompelis2009numerical}. The laminar 2D base flow \tc{black}{$\bar{\bPsi}$} is computed as the steady-state solution of equation~\eqref{eq:nse}, 
	\begin{align}
    \label{eq:U0}
    \tc{black}{
    \mathcal{F} (\bar{\bPsi})
    \; = \;
    0,}
\end{align}
by implicit time-marching with 249 cells in the normal and 535 cells in the streamwise direction. As illustrated in~\citet{sidharth2018onset}, this resolution is sufficient to capture separated flow and resolve the evolution of small perturbations.

\begin{figure}
    \centering
    \includegraphics[width=0.8\linewidth, trim= 4 4 4 4, clip]{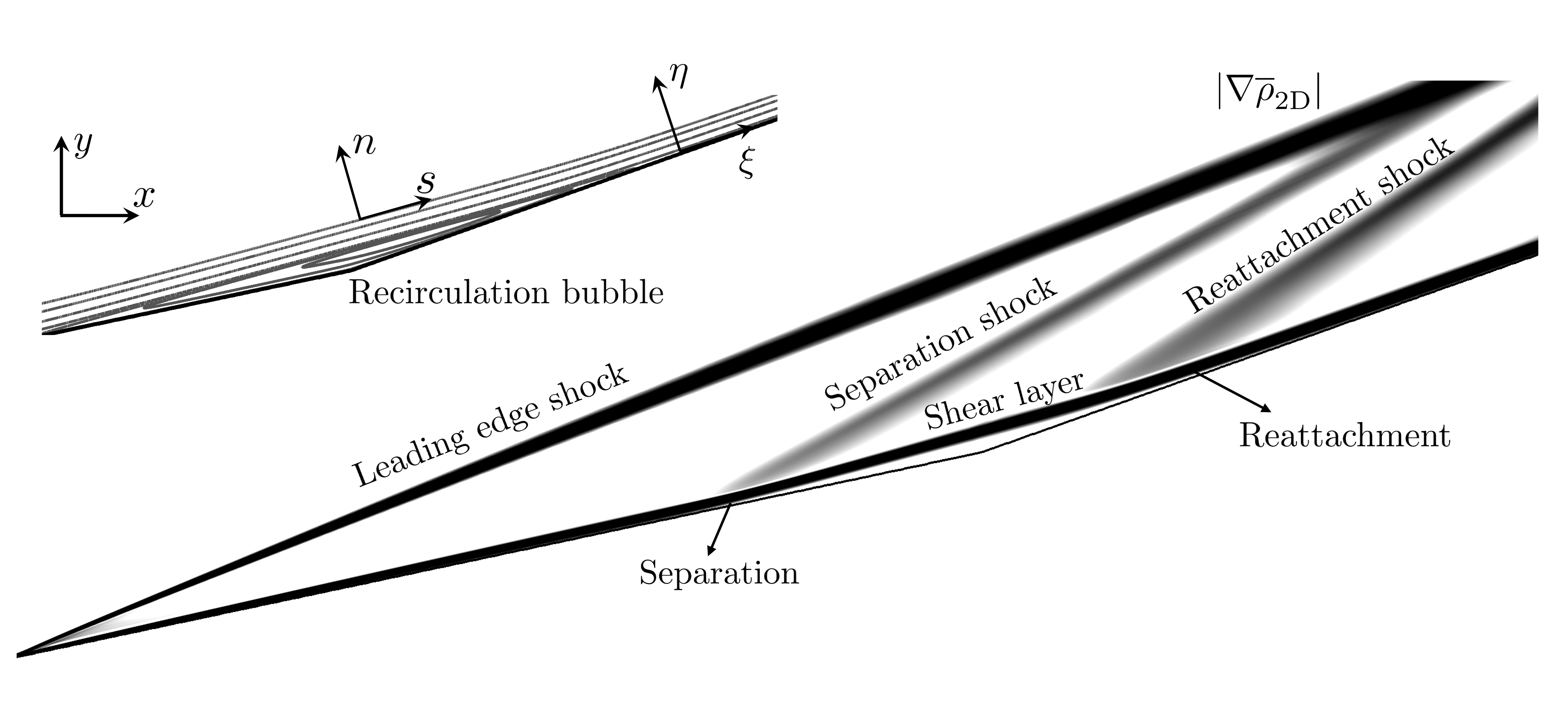}
    \caption{Contours of density gradient magnitude. The inset illustrates zoomed in view of the separation bubble and the schematic of various coordinate systems associated with the double-wedge geometry.}
    \label{fig:flowsetup1}
\end{figure}

	Figure~\ref{fig:flowsetup1} shows the contours of density gradient magnitude on the compression corner. The separation and the reattachment locations in the laminar 2D base flow \tc{black}{$\bar{\bPsi}$} are marked by S and R, respectively. \tc{black}{The 2D flow separates upstream of the corner on the first wedge, it reattaches on the second wedge, and the separated and reattaching shear layers are respectively associated with the formation of the separation and reattachment shocks.} Figure~\ref{fig:flowsetup1} also shows an inset of the separation zone along with the wall-aligned coordinate system (where $\xi$ and $\eta$ denote directions that are parallel and perpendicular to the wall) and a coordinate system that is locally aligned with the streamlines of the laminar 2D flow. Both coordinate systems are used in our study of the evolution of flow fluctuations.

\citet{sidharth2018onset} demonstrated global linear stability of the laminar 2D base flow \tc{black}{$\bar{\bPsi}$.} Recent studies of similar SWBLI configurations, such as compression ramps, revealed extreme sensitivity to upstream disturbances even in the absence of global instability~\citep{dwisidniccanjovJFM19}. Leading-edge roughness and free-stream disturbances provide persistent sources of external excitation and they are inevitable in realistic flows. To evaluate the role of such uncertainty in triggering early stages of transition to turbulence, we utilize input-output analysis to quantify amplification of unsteady disturbances in a hypersonic flow over slender double-wedge.

	\vspace*{-4ex}
\section{Input-output analysis of a high-speed double-wedge flow}
\label{sec:io}

In this section, we employ input-output analysis to quantify amplification of \tc{black}{small} unsteady external disturbances in globally stable 2D SWBLI over a slender double-wedge and uncover physical mechanisms that trigger early stages of transition to turbulence.   

	\vspace*{-2ex}
\subsection{Externally forced compressible NS equations}
	\label{sec:nse-d}

\tc{black}{To account for the rate of change of perturbation density, momentum, and total energy, we model unsteady external disturbances $\mathbf{{d}}(\B x, t)$ to the compressible NS equations  in the conservative form~\eqref{eq:nse} as volumetric sources of excitation, 
	\begin{equation}
	\frac{\partial \bPsi (\B x, t)}{\partial t} 
	\; = \; 
	\mathcal{F} (\bPsi (\B x, t)) 
	\; + \;
	\mathbf{{d}}(\B x, t),
	\label{eq:nse1-tmp}
	\end{equation}
where $\B x \DefinedAs (x,y,z)$ is the vector of streamwise, normal, and spanwise spatial coordinates. By decomposing the flow field $\bPsi$ into the sum of the base $\bar{\bPsi}$ and fluctuating $\bpsi$ parts, 
	\begin{align}
	\bPsi (\B x, t) 
	\; = \;
	\bar{\bPsi} (\B x) 
	\; + \;
	\bpsi (\B x, t), 
	\label{eq.Psibar+psi}
	\end{align}
we obtain the equation that governs the dynamics of flow fluctuations around $\bar{\bPsi} (\B x)$, 
	\begin{equation}
	\frac{\partial \bpsi (\B x, t)}{\partial t} 
	\; = \; 
	\mathcal{F} (\bar{\bPsi} (\B x) \, + \, \bpsi (\B x, t)) 
	\; + \;
	\mathbf{{d}}(\B x, t).
	\label{eq:nse1}
	\end{equation}
For disturbances with small amplitude $\epsilon$,
	\begin{subequations}
		\label{eq:distpert}
\begin{equation}
	\mathbf{{d}}(\B x, t)
	\; = \;
	\epsilon \mathbf{{d}}^{(1)}(\B x, t),
\label{eq:distpert1}
\end{equation}
a weakly nonlinear analysis can be utilized to examine externally forced compressible NS equations~\eqref{eq:nse1} and determine the corrections to the steady laminar 2D base flow $\bar{\bPsi} (\B x)$. Up to a second order in $\epsilon$, the vector of flow fluctuations $\bpsi$ can be represented as, 
\begin{equation}
	\bpsi (\B x, t)
	\; = \; 
	\epsilon \bpsi^{(1)}(\B x, t) 
	\; + \; 
	\epsilon^2 \bpsi^{(2)}(\B x, t) 
	\; + \; 
	{\cal O} (\epsilon^3),
\label{eq:psi-eps}
\end{equation}
	\end{subequations}
where $\bpsi^{(1)}(\B x, t)$ satisfies the linearized flow equations
	\begin{subequations} 
\begin{align}
	\left[\frac{\partial}{\partial t} 
	\; - \; 
	\mathcal{A} \! \left( \bar{\bPsi} \right)\right] \bpsi^{(1)}
	\; = \; 
	\mathbf{{d}}^{(1)}, 
	\label{eq:pert1}
\end{align}
and $\bpsi^{(2)}(\B x, t)$ satisfies
	\begin{align}
	\left[\frac{\partial}{\partial t} 
	\; - \; 
	\mathcal{A} \! \left( \bar{\bPsi} \right)\right] \bpsi^{(2)}
	\; = \; 
	\mathcal{N}^{(2)} \! \left(  \bpsi^{(1)} \right). 
	\label{eq:pert2}
\end{align}
	\end{subequations}
Equations~\eqref{eq:pert1} and~\eqref{eq:pert2} are respectively obtained by neglecting ${\cal O} (\epsilon^2)$ and ${\cal O} (\epsilon^3)$ terms upon substitution of~\eqref{eq:distpert1} and~\eqref{eq:psi-eps} to the compressible NS equations~\eqref{eq:nse1}. The compressible NS operator resulting from linearization around the base flow $\bar{\bPsi}$ is determined by $\mathcal{A} (\bar{\bPsi})$ (see~\citet[equation~(23)]{candler2015book} and \citet[equations~(A1)-(A2)]{sidharth2018onset}) and $\mathcal{N}^{(2)}  (  \bpsi^{(1)} )$ accounts for quadratic interactions at ${\cal O} (\epsilon^2)$; see appendix~\ref{app.nonlin} for details.}

\tc{black}{Several recent studies demonstrated the utility of the compressible energy norm~\citep{chu1965energy,hanifi1996transient},} 
	\begin{subequations}
	\label{eq:Echu}
\begin{align}
	\tc{black}{
	E 
	\; \DefinedAs \;
	\dfrac{1}{2} 
	\, 
	\int_{\Omega} 
	\left(
	\dfrac{\bar{p}}{\bar{\rho}^{2}} \, \rho^{\prime 2}
	\, + \, 
	\bar{\rho} \left|\mathbf{u}^{\prime}\right|^{2} 
	\, + \, 
	\dfrac{C_{v} \bar{\rho}}{\bar{T}} \, T^{\prime 2} 
	\right)
	\mathrm{d} \Omega,}
	\label{eq:norm}
\end{align} 
\tc{black}{in quantifying growth of fluctuations in high-speed boundary layer flows~\citep{franko2013breakdown,sidharth2018onset,quintanilha_paredes_hanifi_theofilis_2022}. This quantity is determined by the weighted $L_2$ norm of the vector of flow fluctuations $\bphi \DefinedAs (\phi_{1}, \bphi_{2}, \phi_{3}) = (\rho^\prime, \bu^\prime, T^\prime)$ in primitive variables,}
	\begin{align}
	\tc{black}{
	E 
	\; = \;
	\| \bphi \|_{E}^{2}
	\; = \;
	\langle \bphi , \bphi \rangle_E
	\; = \;
	\langle \bphi , \cW \bphi \rangle_2,}
	\label{eq:norm}
\end{align} 
\tc{black}{where $\langle \, \cdot \, , \, \cdot \, \rangle_2$ is the standard $L_2$ inner product over the domain $\Omega$, $C_{v}$ is the specific heat at constant volume in~$\Omega$, and} 
	\begin{align}
	\tc{black}{\cW 
	\; \DefinedAs \;
	\dfrac{1}{2} 
	\left[
	\begin{array}{ccc}
	\bar{p} / \bar{\rho}^{2} & 0 & 0 
	\\[0.1cm]
	0 & \bar{\rho} & 0
	\\[0.1cm]
	0 & 0 & C_{v} \bar{\rho}/\bar{T}
	\end{array}
	\right],}
	\end{align}
	\end{subequations}
\tc{black}{is the multiplication operator determined by the pressure $\bar{p}$, density $\bar{\rho}$, and temperature $\bar{T}$ of the base flow $\bar{\bPsi}$. For small amplitude disturbances, we can represent $\bphi$ as}
	\begin{subequations}
	\begin{equation}
	\tc{black}{\bphi (\B x, t)
	\; = \; 
	\epsilon \bphi^{(1)}(\B x, t) 
	\; + \; 
	\epsilon^2 \bphi^{(2)}(\B x, t) 
	\; + \; 
	{\cal O} (\epsilon^3),}
	\label{eq:phi-eps}
	\end{equation}
\tc{black}{where,  
\begin{equation}
	\begin{array}{rcl}
 	\bphi^{(1)}(\B x, t)
 	& \!\! = \!\! & 
 	\cC \, \bpsi^{(1)}(\B x, t),
	\\
	\bphi^{(2)}(\B x, t)
 	& \!\! = \!\! & 
 	\cC \, \bpsi^{(2)}(\B x, t) 
 	\; + \; 
 	\cD
 	\left[
	\begin{array}{c}
	-{\phi^{(1)}_{1}} {\bphi^{(1)}_{2}}  \\
	-2 C_{v} {\phi^{(1)}_{1}} {\phi^{(1)}_{3}} - \bar{\Phi}_{1} |{\bphi^{(1)}_{2}}|^{2}
	\end{array}
	\right].
	\end{array}
	\end{equation}}
	\end{subequations}
\tc{black}{As shown in appendix~\ref{app.conv2Nconv}, $\cC$ and $\cD$ are multiplication operators parameterized by the laminar 2D base flow $\bar{\bPsi}$; see equation~\eqref{eq.CD} for their definition.}

\tc{black}{In the remainder of this section, we identify oblique waves as the most energetic responses of the linearized flow equations~\eqref{eq:pert1} in the presence of unsteady harmonic disturbances $\mathbf{{d}}^{(1)}$. In \S~\ref{ref:strk}, we utilize equation~\eqref{eq:pert2} to demonstrate that steady streaks can arise from quadratic interactions of unsteady oblique waves.}  

	\vspace*{-2ex}
\subsection{Amplification of exogenous disturbances \tc{black}{to the linearized flow equations}}
	\label{sec:svd}

\tc{black}{The linearized equations~\eqref{eq:pert1} describe the evolution of fluctuation vector $\bpsi^{(1)}$ in the presence of external disturbances $\mathbf{d}^{(1)}$ and they can be written using the state-space formulation~\citep{schhen01},
\begin{align}
\begin{split}
\label{eq:SScomp}
	\frac{\partial \bpsi^{(1)}}{\partial t} 
	& \; = \; 
	\cA \bpsi^{(1)} \; + \; \cB \mathbf{d}^{(1)}, \\
	\bphi^{(1)} 
	& \; = \; 
	\cC \bpsi^{(1)}.
\end{split}
\end{align}
Here, $\mathbf{d}^{(1)}$ is a spatially distributed and temporally varying disturbance (input), $\bpsi^{(1)}$ is the state of the linearized system (which is determined by the vector of flow fluctuations in conserved variables), $\bphi^{(1)}$ is the quantity of interest (output) whose weighted $L_2$ norm determines energy of flow fluctuations~\eqref{eq:Echu}, and $\cA$ is the generator of the linearized compressible NS dynamics. The input operator $\cB$ in~\eqref{eq:SScomp} allows to specify spatial support of body forcing inputs and the output operator $\cC$ relates the state of the linearized system $\bpsi^{(1)}$ to the vector of flow fluctuations in primitive variables $\bphi^{(1)}$.} 

\tc{black}{For the parameters shown in figure~\ref{fig:flowsetup}, the linearized system is globally stable} and for a time-periodic input with frequency $\omega$, $\bd^{(1)} (\bx, t) = \hat{\B d}^{(1)} (\bx, \omega) \mathrm{e}^{\mri \omega t}$, the steady-state output of~\eqref{eq:SScomp} is determined by $\bphi^{(1)} (\bx,t) = \hat{\bphi}^{(1)} (\bx, \omega) \mre^{\mri \omega t}$ where
	\begin{equation}
	\hat{\bphi}^{(1)} (\bx, \omega) 
	\; = \; 
	\left[ \cH (\omega) \hat{\bd}^{(1)} ( \, \cdot \, , \omega) \right]
	(\bx), 
	\end{equation}
$\cH (\omega)$ is the frequency response operator, 
\begin{align}
    \label{eq:output}
    \cH (\omega) 
    & 
    \; = \; \cC(\mri \omega \cI \, - \, \cA)^{-1} \cB,
\end{align}
and $\cR (\omega) = (\mri \omega \cI - \cA)^{-1}$ is the resolvent operator associated with the linearized model~\eqref{eq:SScomp}. At any $\omega$, the singular value decomposition of $\cH (\omega)$ can be used to quantify amplification of time-periodic inputs~\citep{schmid2007nonmodal,jovARFM21},
	\begin{align}
	\hat{\bphi}^{(1)} (\bx, \omega) 
	\; = \; 
	\left[ \cH (\omega) \hat{\bd}^{(1)} ( \, \cdot \, , \omega) \right]
	(\bx)
	\; = \;
	\displaystyle{\sum_i}
	\,
	\sigma_i (\omega)
	\bphi_{i} (\bx, \omega)
	\langle
	\mathbf{d}_{i} (\, \cdot \, , \omega), \hat{\B d} (\, \cdot \, , \omega)
	\rangle_E,
	 \label{eq:svd}
	 \end{align}
where $\sigma_i (\omega)$ denotes the $i$th singular value of $\cH (\omega)$, $\langle \, \cdot \, , \, \cdot \, \rangle_E$ is the inner product \tc{black}{in~\eqref{eq:Echu} that induces the compressible energy norm,} and $\mathbf{d}_{i} (\bx, \omega)$ and $\bphi_{i} (\bx, \omega)$ are the left and right singular functions of $\cH (\omega) $ which provide orthonormal bases of the corresponding input and output spaces (with respect to $\langle \, \cdot \, , \, \cdot \, \rangle_E$). 

The frequency response operator $\cH (\omega)$ maps the $i$th input mode $\mathbf{d}_{i} (\bx, \omega)$ into the response whose spatial profile is specified by the $i$th output mode $\bphi_{i} (\bx, \omega)$ and the amplification is determined by the corresponding singular value $\sigma_{i} (\omega)$. \mbox{In other words, for} 
	\begin{equation}
	\hat{\bd}^{(1)} (\bx, \omega) 
	\; = \; 
	\bd_i (\bx,\omega)
	~\; \Rightarrow ~\;
	\hat{\bphi}^{(1)} (\bx, \omega) 
	\; = \; 
	\left[ \cH (\omega) \bd_i ( \, \cdot \, , \omega) \right] (\bx) 
	\; = \; 
	\sigma_i (\omega) \bphi_{i} (\bx, \omega),
	\end{equation}
and $\| \hat{\bphi}^{(1)} ( \, \cdot \, , \omega) \|_{E} = \sigma_i (\omega)$. Note that, at any $\omega$, 
	\begin{equation}
	G (\omega) 
	\; \DefinedAs \; 
	\sigma_{1} (\omega) 
	\; = \; 
	\dfrac{\| \cH (\omega) \mathbf{d}_{1} (\, \cdot \, , \omega)\|_{E}}{\| \mathbf{d}_{1} ( \, \cdot \, ,\omega)\|_{E}}
	\; = \; 
	\dfrac{\| \sigma_{1} (\omega) \bphi_{1} ( \, \cdot \, , \omega) \|_{E}}
	{\| \mathbf{d}_{1} ( \, \cdot \, ,\omega)\|_{E}},
	\label{eq:G}
	\end{equation}
determines the largest induced gain with respect to a compressible energy norm, where ($\mathbf{d}_1 (\bx, \omega),\bphi_1 (\bx, \omega)$) identify the spatial structure of the dominant input-output pair. \tc{black}{We use a second-order central finite-volume discretization~\citep{sidharth2018onset} to obtain a finite dimensional approximation of the linearized model~\eqref{eq:SScomp} and employ matrix-free Arnoldi iterations~\citep{jeun2016input,anubhav-phd20} to compute the singular values $\sigma_i (\omega)$ of $\cH (\omega)$.}

	\vspace*{-2ex}
\subsection{Frequency response analysis}
\label{sec:iowedge}

We utilize \tc{black}{the resolvent analysis} to study amplification of harmonic disturbances with frequency $\omega$ \tc{black}{to the linearized flow equations. In double-wedge geometry, the laminar 2D base flow $\bar{\bPsi}$ is a function of streamwise normal coordinates, $\bar{\bPsi} (x,y)$, and} owing to homogeneity in the spanwise direction, the 3D fluctuations in~\eqref{eq:pert1} take the form,
\begin{equation}
    \bpsi^{(1)}  (x,y,z,t) 
    \; = \; 
    \hat{\bpsi}^{(1)} (x,y; \beta,\omega) \mre^{\mri(\beta z + \omega t)}, 
\end{equation}
where $\beta=2\pi/\lambda_z$ is the spanwise wavenumber. Thus, in addition to $\omega$, the frequency response operator is also parameterized by $\beta$,
\begin{equation}
	\tc{black}{\cH_\beta (\omega)
    	\; = \; 
	\cC 
	(\mri \omega \cI \, - \, \cA_\beta)^{-1} \cB,}
\end{equation} 	
\tc{black}{where $\cA_\beta$ denotes the Fourier symbol of the operator $\cA$ in~\eqref{eq:SScomp} obtained by replacing the spanwise differential operator $\partial_z$ with $\mri \beta$. At any pair $(\omega, \beta)$, $\cH_\beta (\omega)$ maps the input function $\hat{\bd}^{(1)}$ of $x$ and $y$ into the output function $\hat{\bphi}^{(1)}$ of $x$ and $y$,
	\begin{align}
	\hat{\bphi}^{(1)} (x, y; \beta, \omega) 
	\; = \; 
	\left[ \cH_\beta (\omega) \hat{\bd}^{(1)} ( \, \cdot \, , \, \cdot \, ; \beta, \omega) \right]
	(x,y),
	 \label{eq:IObeta}
	 \end{align}
and SVD of $\cH_\beta (\omega)$ can be used to study amplification across spatio-temporal frequencies.}

\tc{black}{We first set ${\cB} = {\cI}$, i.e., we introduce body forcing inputs to excite flow at every spatial location in the computational domain $\Omega$ and we choose the output operator ${\cC}$ to examine the impact of forcing on the compressible energy norm of $\bphi^{(1)}$ in the entire $\Omega$. The resolvent analysis is done using a resolution that yields grid-independent outputs with $545$ cells in the streamwise, $249$ cells in the normal direction, and numerical sponge boundary conditions near the leading edge ($x=1$) and the outflow \mbox{boundaries are utilized.}}

\begin{figure}
    \centering
    {
    \begin{tabular}{cc}
    \begin{tabular}{c}
    {
    \subcaptionbox{}
    {\includegraphics[height= 6 cm, trim= 4 4 6 4, clip]{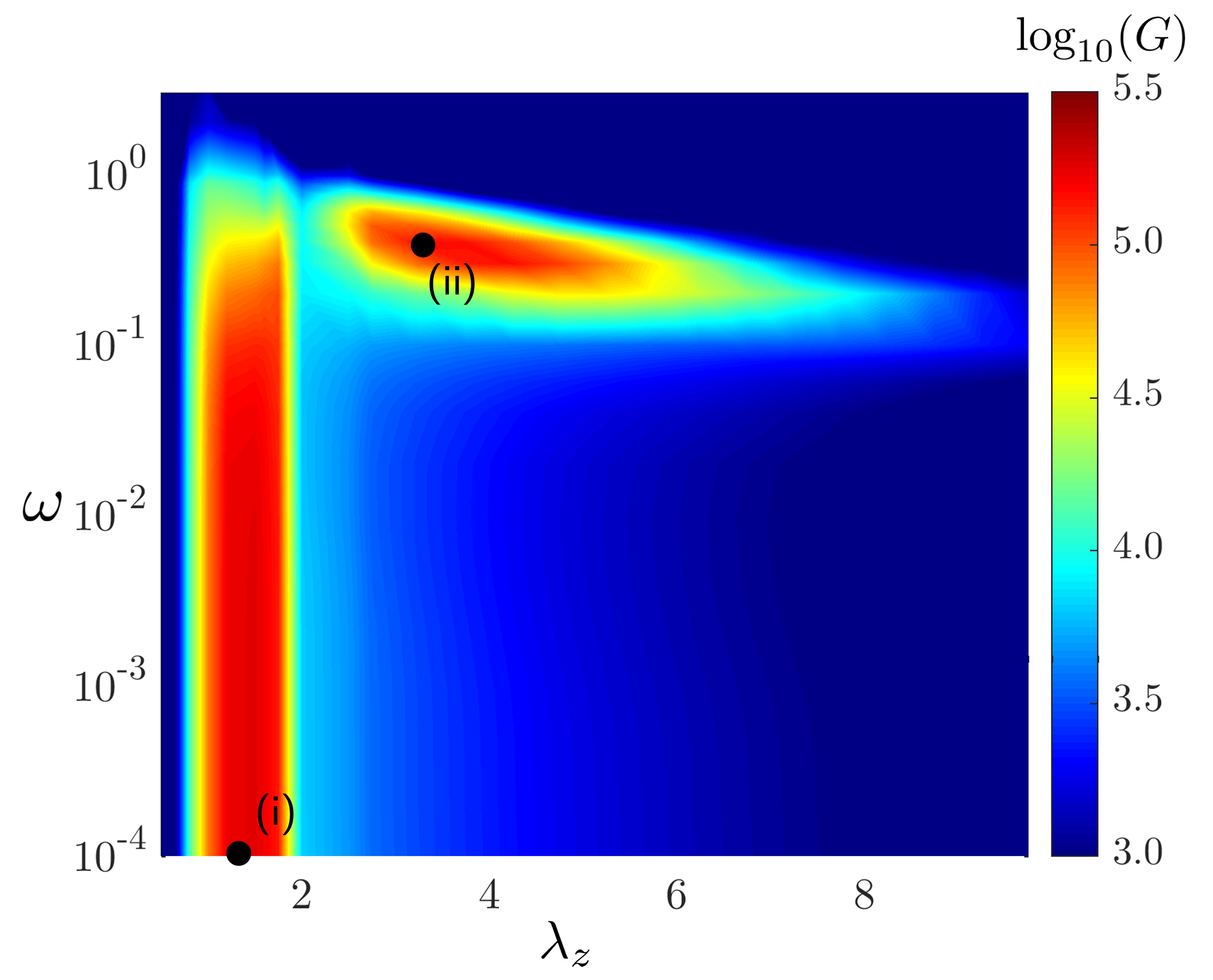}
           \label{fig.res1}}}
    \end{tabular}
    &    
    \begin{tabular}{c}
    {
   \hspace*{-0.4cm}
    \subcaptionbox{}
    {\includegraphics[height= 6 cm, trim= 6 4 4 4, clip]{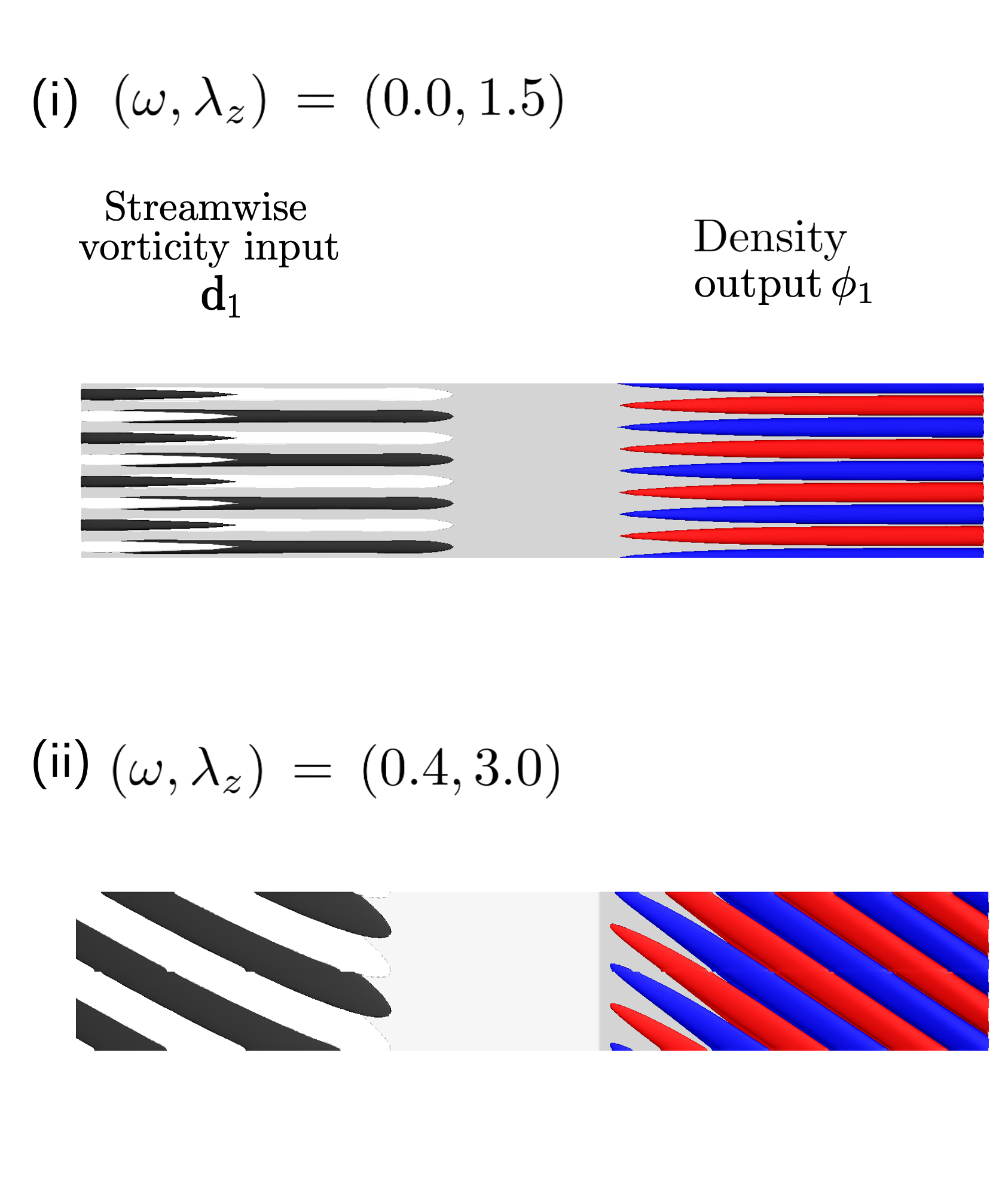}
           \label{fig.res2}}}
    \end{tabular}
    \end{tabular}
    }    
     \caption{(a) Input-output gain $G (\omega,\lambda_z)$ associated with the resolvent operator across temporal frequency $\omega$ and spanwise wavelength $\lambda_z$. (b) Isosurfaces of streamwise vorticity and density fluctuations corresponding to the input-output modes $\mathbf{d}_{1}$ and $\bphi_1$.}
    \label{fig:BCidentity}
    \end{figure}

Figure~\ref{fig:BCidentity} shows the dependence of the input-output gain $G(\omega,\lambda_z)$ on the frequency $\omega$ and the wavelength $\lambda_z$. There are two major amplification regions with the respective peaks at $(\omega=0,\lambda_{z}=1.5)$ and $(\omega=0.4,\lambda_{z}=3)$. The first peak in $G$ identifies the largest amplification and the corresponding output is determined by reattachment streaks that result from steady vortical disturbances upstream of the recirculation zone. We observe selective amplification of disturbances with $\lambda_z \approx 1.5$ and low-pass filtering features over $\omega$. The gain $G$ experiences rapid decay beyond the roll-off frequency $\omega \approx 0.4$ and it attains its largest value at $\omega=0$ for $\lambda_z$ that scales with the reattaching shear layer thickness~\citep{dwisidniccanjovJFM19}. In contrast to~\citet{dwisidniccanjovJFM19}, which focused on disturbances with $\omega = 0$, we examine unsteady disturbances that trigger oblique waves in the reattaching shear layer, as identified by the second peak in $G$. This amplification region takes place in a narrow band of temporal frequencies $\omega$ over a fairly broad range of spanwise wavelengths $\lambda_{z}$. 

\tc{black}{As demonstrated in figure~\ref{fig:BCidentity},} even when we allow disturbances to enter through the entire computational domain the largest amplification is caused by inputs that are localized upstream of the corner and the resulting response is localized downstream of the corner. The upstream disturbances are the most effective way to excite the flow because of large convection velocity of the laminar 2D base flow~\citep{chomaz2005global,schmid2007nonmodal} and the dominant output emerges in the separated and the reattached regions. 

\begin{figure}
    \centering
    {
    \begin{tabular}{cc}
    \begin{tabular}{c}
    {
    \subcaptionbox{}
    {\includegraphics[height= 5.0 cm, trim= 4 4 4 4, clip]{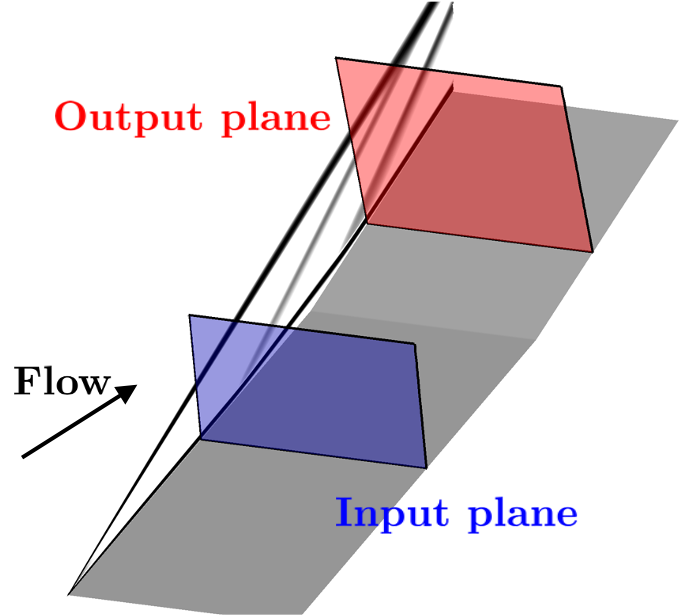}
           \label{fig.res1}}}
    \end{tabular}
    &    
    \begin{tabular}{c}
    {
   \hspace*{-0.3cm}
    \subcaptionbox{}
    {\includegraphics[height= 5.0 cm, trim= 3 3 3 3, clip]{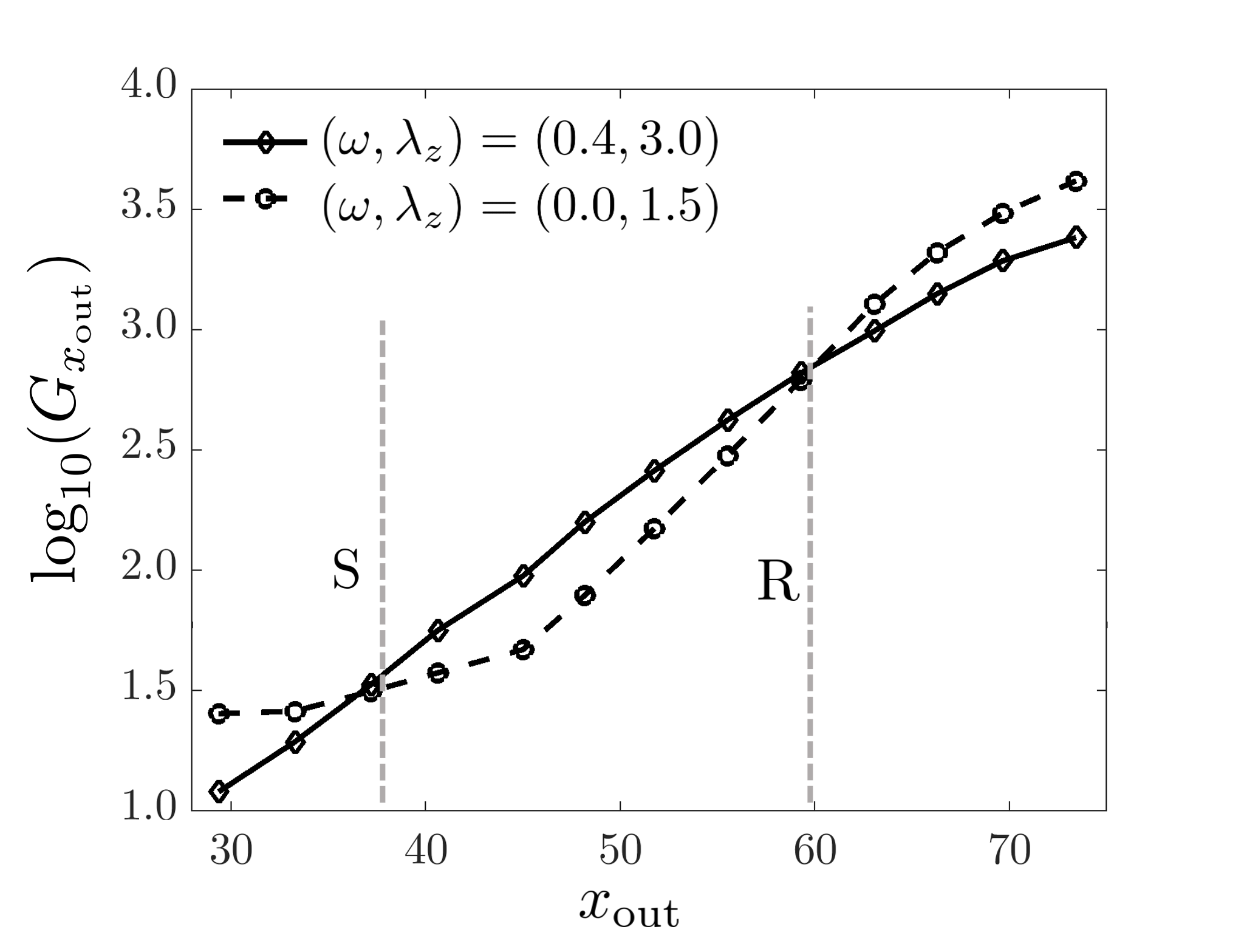}
           \label{fig.res2}}}
    \end{tabular}
    \end{tabular}
    }    
    \caption{Spatial input-output analysis: (a) input is introduced at a streamwise location $x_\mathrm{in} = 25$ before separation and output is evaluated at $x_{\mathrm{out}}$; (b) dependence of the input-output gain $G_{x_{\mathrm{out}}}$ on the streamwise location $x_{\mathrm{out}}$ for streaks and oblique waves. Unsteady oblique waves with $(\omega=0.4,\lambda_{z}=3)$ are strongly amplified throughout the separation zone.}
    \label{fig:IOspatial}
    \end{figure}

\tc{black}{Experimental studies of oblique transition in channel and boundary layer flows~\citep{elofsson1998experimental,berlin1999nonlinear} often utilize streamwise localized disturbances and a common criticism of the resolvent analysis is that the identified global input modes represent excitation sources that are not easy to realize experimentally. In contrast, traditional approach to the analysis of boundary layers utilizes spatially localized fluctuation sources and evaluates the streamwise growth of fluctuations as they convect downstream~\citep{herbert1997parabolized}. However, in the presence of flow separation a parabolized approximation of the NS equations cannot be made. To evaluate amplification in different spatial regions, we restrict inputs and outputs to belong to a plane but still account for the global nature of the separated flow through the resolvent operator $(\mathrm{i}\omega \cI - \cA_{\beta})^{-1}$. As illustrated in figure~\ref{fig:IOspatial}(a), this is accomplished via a proper selection of the operators $\cB$ and $\cC$ in equation~\eqref{eq:SScomp} by fixing the input location before flow separation, at $x_\mathrm{in} = 25$, and by evaluating the output at different locations downstream of the separation, $x_{\mathrm{out}}$. In this setup, $G_{x_{\mathrm{out}}}$ quantifies the largest amplification at $x_{\mathrm{out}}$ of disturbances that are introduced at $x_\mathrm{in} = 25$.}

Figure~\ref{fig:IOspatial}(b) shows \tc{black}{the dependence of $G_{x_{\mathrm{out}}}$ on $x_{\mathrm{out}}$ for upstream} disturbances with $(\omega=0,\lambda_{z}=1.5)$ and $(\omega=0.4,\lambda_{z}=3)$. The gain associated with the steady fluctuations begins to grow in the latter half of the recirculation zone, especially near the reattachment location. In contrast, unsteady perturbations with $\omega = 0.4$ experience significant amplification throughout the separation zone. This observation suggests that the separation zone plays a critical role in the amplification of unsteady fluctuations.

	\vspace*{-2ex}
\subsection{Amplification of oblique waves: physical mechanism}
	\label{sec:oblphys}

\tc{black}{We now analyze physical mechanisms responsible for amplification of flow fluctuations within the separation zone in the presence of upstream unsteady disturbances. In particular, we examine the global response of the linearized equations to the input with ($\omega=0.4,\lambda_{z}=3.0$) introduced prior to separation (i.e., at $x_\mathrm{in}=25$) that triggers the largest amplification in the entire domain $\Omega$. The spatial structure of flow fluctuations is studied in the $(s,n,z)$ coordinate system which is locally aligned with the streamlines of the laminar 2D base flow $\bar{\bPsi}$; see figure~\ref{fig:flowsetup1} for an illustration. In this frame of reference, $\bar{\bu} = (\bar{u}_s,0,0)$ with $\bar{u}_{s} \geq 0$, and, as discussed in~\citet{finnigan1983streamline,patel1997longitudinal,dwisidniccanjovJFM19}, this coordinate system is convenient for the analysis of separated boundary layers especially within the separation zone.}

	The streamwise specific kinetic energy $\mathcal{E}_{s} \DefinedAs u_s^{\prime} u_s^{\prime}$ obeys the transport equation,
\begin{align}
	\frac{\partial \mathcal{E}_{s}}{\partial t} 
	\; + \; 
	\bar{u}_{s}\frac{\partial \mathcal{E}_{s}}{\partial s}
	\; = \; 
	\mathcal{P} 
	\; + \; 
	\mathcal{S} 
	\; + \; 
	\mathcal{V} 
	\; + \; 
	\tc{black}{\mathcal{K}} 
	\; + \; 
	\mathcal{F},
\label{eq:ketrans_symb}
\end{align}
where $\mathcal{P}$, $\mathcal{S}$, $\mathcal{V}$, and \tc{black}{$\mathcal{K}$} are the production, source, viscous, and curvature terms (see appendix~\ref{app.terms}), and $\mathcal{F}$ is the work done by external disturbances~\citep[appendix C]{dwisidniccanjovJFM19}. The production term $\mathcal{P}$ quantifies interactions of fluctuation stresses with the base flow gradients, the source term $\mathcal{S}$ corresponds to the perturbation component of the inviscid material derivative, the viscous term $\mathcal{V}$ determines dissipation of kinetic energy by viscous stresses, and \tc{black}{$\mathcal{K}$} accounts for the curvature that arises from a coordinate transformation. \tc{black}{Our computations indicate that while} the dissipative viscous term $\mathcal{V}$ is negative throughout the domain the production term $\mathcal{P}$ is sign-indefinite, $\mathcal{S}$ and \tc{black}{$\mathcal{K}$} are negligible, and $\mathcal{F}$ is zero downstream of the forcing plane.

Insight into physical mechanisms can be gained by the analysis of dominant production terms in~(\ref{eq:ketrans_symb}) \tc{black}{associated with the global linearized response to upstream oblique disturbances}. Averaging over the time period $T = {2\pi}/{\omega}$ and the spanwise wavelength $\lambda_{z} = 2 \pi / \beta$, $\braket{\cdot} \DefinedAs ( T \lambda_{z} )^{-1}\int^T_{0} \int^{\lambda_{z}}_{0} (\cdot) \, \mrd z \,\mrd t$, and neglecting the terms that do not contribute significantly to the production of the averaged streamwise specific kinetic energy $E_{s} \DefinedAs \langle \mathcal{E}_s \rangle$ yields the following approximation to the transport equation~(\ref{eq:ketrans_symb}),
		\begin{equation}
	\label{eq:Es-transport}
	\bar{u}_{s} \frac{\partial E_{s}}{\partial s}
	\; + \;
	2 \, ( \partial_s \bar{u}_{s} )
	E_{s} 
	\; \approx \; 
	- 
	2 \, ( \partial_n \bar{u}_{s} )
	{R}_{sn},
	\end{equation}
where ${R}_{sn} \DefinedAs \braket{u_s^{\prime} u_n^{\prime}}$ denotes the averaged shear stress of the streamwise velocity fluctuations. The second term on the left-hand-side represents the production of fluctuations' energy that arises from the streamwise gradient of the base flow $\partial_{s}\bar{u}_{s}$ and the term on the right-hand-side determines the production term that originates from interactions of the base flow shear $\partial_n \bar{u}_{s}$ with the fluctuation shear stress ${R}_{sn}$. 

\tc{black}{To understand the mechanism that facilitate the growth of $E_{s}$, we now investigate the streamwise transport of ${R}_{sn}$. In contrast to the transport equation for $E_{s}$, both} the production $\mathcal{P}$ and curvature \tc{black}{$\mathcal{K}$} terms contribute \tc{black}{significantly} to the streamwise transport of ${R}_{sn}$ for fluctuations with $\omega=0.4$ and $\lambda_{z}=3.0$. \tc{black}{As demonstrated in appendix~\ref{app.shear},} omitting negligible terms leads to the following approximate transport equation for ${R}_{sn}$,
\begin{subequations}
\label{eq:Es-Rsn-transport}	
\begin{equation}
	\bar{u}_{s} \frac{\partial {R}_{s n}}{\partial s} 
	\; + \; 
	( \partial_s \bar{u}_{s} \, + \, K_{s} ) {R}_{s n}
	\; \approx \; 
	2 K_{c}  E_{s},
	\label{eq:Rsn-transport1}
\end{equation}
where $K_{c}$ and $K_{s}$ denote contributions that arise from the curvature normal to the streamlines and from deceleration along the streamline direction, respectively,
	\begin{equation}
	\tc{black}{K_c 
	\; = \; 
	-(\bar{\Omega} + \partial_{n}{\bar{u}_{s}}), 
	~~
	K_{s} 
	\; = \; 
	-\partial_{s} \bar{u}_{s}.}
	\label{eq:KsKc}
	\end{equation}
In the $(s,n,z)$ coordinate system,  $\bar{\Omega} = \partial_{x}\bar{v} - \partial_{y}\bar{u}$ denotes the \tc{black}{spanwise} vorticity of the base flow in the Cartesian coordinates~\citep{finnigan1983streamline} \tc{black}{and using the definition of $K_{s}$,} equation~\eqref{eq:Rsn-transport1} simplifies to
\begin{equation}
\label{eq:Rsn-transport}
	\bar{u}_{s} \frac{\partial {R}_{s n}}{\partial s} 
	\; \approx \; 
	2 K_{c}  E_{s}.
\end{equation} 
	\end{subequations}

\begin{figure}
    \centering
    {
    \begin{tabular}{cc}
    \begin{tabular}{c}
    {
    \subcaptionbox{}
    {\includegraphics[height= 4.2 cm, trim= 4 4 4 4, clip]{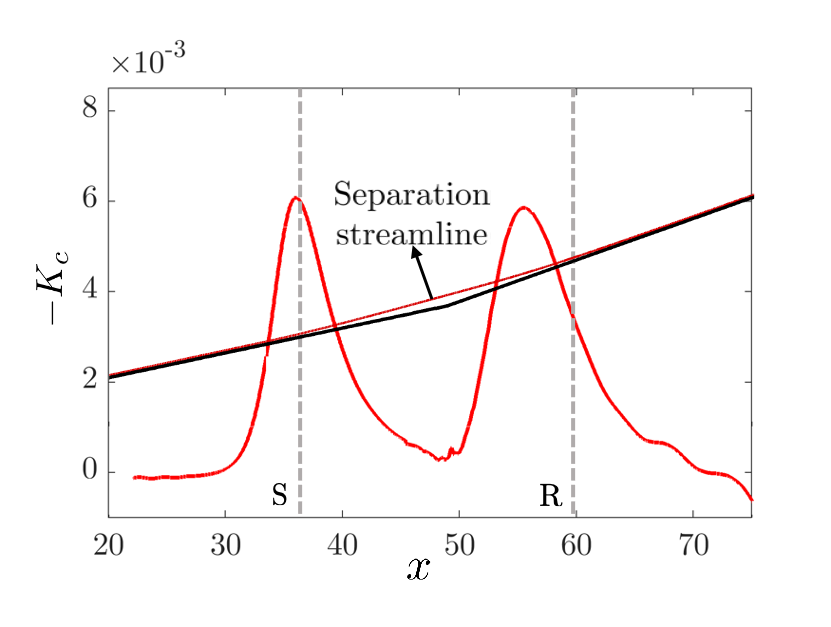}
           \label{fig.bd-periodic}}}
    \end{tabular}
    &    
    \begin{tabular}{c}
    {\hspace*{-0.65cm}  \subcaptionbox{} 
   {\scalebox{.75}{
%
%
%
%
%
\input{figures/Tikz_common_styles}
\noindent
\begin{tikzpicture}[scale=1, auto, >=stealth']
      
     \node[block, minimum height = 1cm, top color=blue!20, bottom color=blue!20] (sys1) {$\begin{array}{c} \mbox{\tc{blue}{\bf transport equation for $E_s$}} \\[0.15cm]    
     \bar{u}_{s} \dfrac{\partial E_{s} }{\partial s} 
	\, + \, 
	2 \, ( \partial_s \bar{u}_{s} )
	E_{s} 
	\, = \,  
	- 
	2 \, ( \partial_n \bar{u}_{s} )
	{R}_{sn} \end{array}$};
     
     \node[block, minimum height = 1cm, top color=red!20, bottom color=red!20] (sys2) at ($(sys1.south) - (0cm,3cm)$) {$\begin{array}{c} \mbox{\tc{red}{\bf transport equation for ${R}_{s n}$}} \\[0.15cm] \bar{u}_{s} \dfrac{\partial {R}_{s n}}{\partial s} \, = \, 2 K_{c} E_{s} \end{array}$};
     
     \node[] (input-node) at ($(sys1.west) - (1cm,0)$) {}; 
     
     \node[] (output-node) at ($(sys1.east) + (2.3cm,0)$) {};
     
     \node[] (mid-node1) at ($(sys1.east) + (1cm,0cm)$) {}; 
     
     \node[] (mid-node2) at ($(input-node) - (0cm,3.7cm)$) {};
     
      \node[block, minimum height = 1.cm, top color=red!20, bottom color=red!20] (sys3) at ($(mid-node1) - (0cm,1.85cm)$) {$\begin{array}{c} \mbox{\tc{red}{\bf base flow}} \\[0.cm] \mbox{\tc{red}{\bf curvature}} \\[0.15cm] 2 K_c \end{array}$};
      
      \node[block, minimum height = 1.cm, top color=red!20, bottom color=red!20] (sys4) at ($(input-node.east) - (0cm,1.85cm)$) {$\begin{array}{c} \mbox{\tc{red}{\bf base flow}} \\[0.cm] \mbox{\tc{red}{\bf shear}} \\[0.15cm] - 2 \, ( \partial_n \bar{u}_{s} ) \end{array}$};

     
%
     
                    	
    \draw [line] (sys1.east) -- ($(sys1.east) + (1cm,0cm)$);
    
    
    \draw [connector] (input-node.east) --  (sys1.west);
    
    \draw [connector] ($(sys1.east) + (1cm,0cm)$) -- node [midway, above] {$E_s$} (output-node);
    
    
    \draw [connector] ($(sys1.east) + (1cm,0cm)$) -- (sys3.north);
    
    \draw [connector] (sys3.south) |- (sys2.east);
    
    \draw [connector] (sys2.west) -| (sys4.south);
    
    \draw [line] (sys2.west) -- node [midway, above] {${R}_{sn}$} (mid-node2);
	
   \draw [line] (sys4.north) -|  (input-node.east);
   	
                                       
\end{tikzpicture}
}
      \label{fig.bd-nuT}}}
    \end{tabular}
    \end{tabular}
    }    
    \caption{(a) Curvature ($-K_{c}$) of the laminar base flow along the separation streamline; (b) illustration of a physical mechanism that facilitates growth of the averaged streamwise specific kinetic energy $ E_{s}$ of oblique fluctuations in the separated shear layer.}
    \label{fig:obl_mech}
\end{figure}

	\tc{black}{In summary,} equations~\eqref{eq:Es-transport} and \eqref{eq:Rsn-transport} determine a coupled system of linear equations that governs the streamwise transport of ${E_{s}}$ and ${R}_{s n}$ in the separation zone for oblique fluctuations with ($\omega=0.4,\lambda_{z}=3.0$),
	\begin{equation}
	\label{eq:Es-Rns}
	\tc{black}{
	\tbt{\bar{u}_s}{0}{0}{\bar{u}_s}
	\frac{\partial}{\partial s}
	\tbo{E_{s}}{{R}_{s n}}
	\, \approx \,
	\tbt{-2 \partial_s \bar{u}_{s}}{-2 \partial_n \bar{u}_{s}}{2 K_c}{0}
	\tbo{E_{s}}{{R}_{s n}}.}
	\end{equation} 
	\tc{black}{Oblique waves experience largest amplification in the separated shear layer above the recirculation bubble, {i.e., in the region where the presence of flow separation leads to concave flow curvature $K_c<0$.}} \tc{black}{Figure~\ref{fig:obl_mech}(a)} shows this negative curvature along the separation streamline and figure~\ref{fig:obl_mech}(b) illustrates the physical mechanism which is absent in attached boundary layers because of negligibly small positive streamwise curvature.

\begin{figure}
    \centering
    {
    \begin{tabular}{cc}
    \begin{tabular}{c}
    {
   \hspace*{-0.5cm}
    \subcaptionbox{}
    {\includegraphics[height= 5.0 cm, trim= 4 4 4 4, clip]{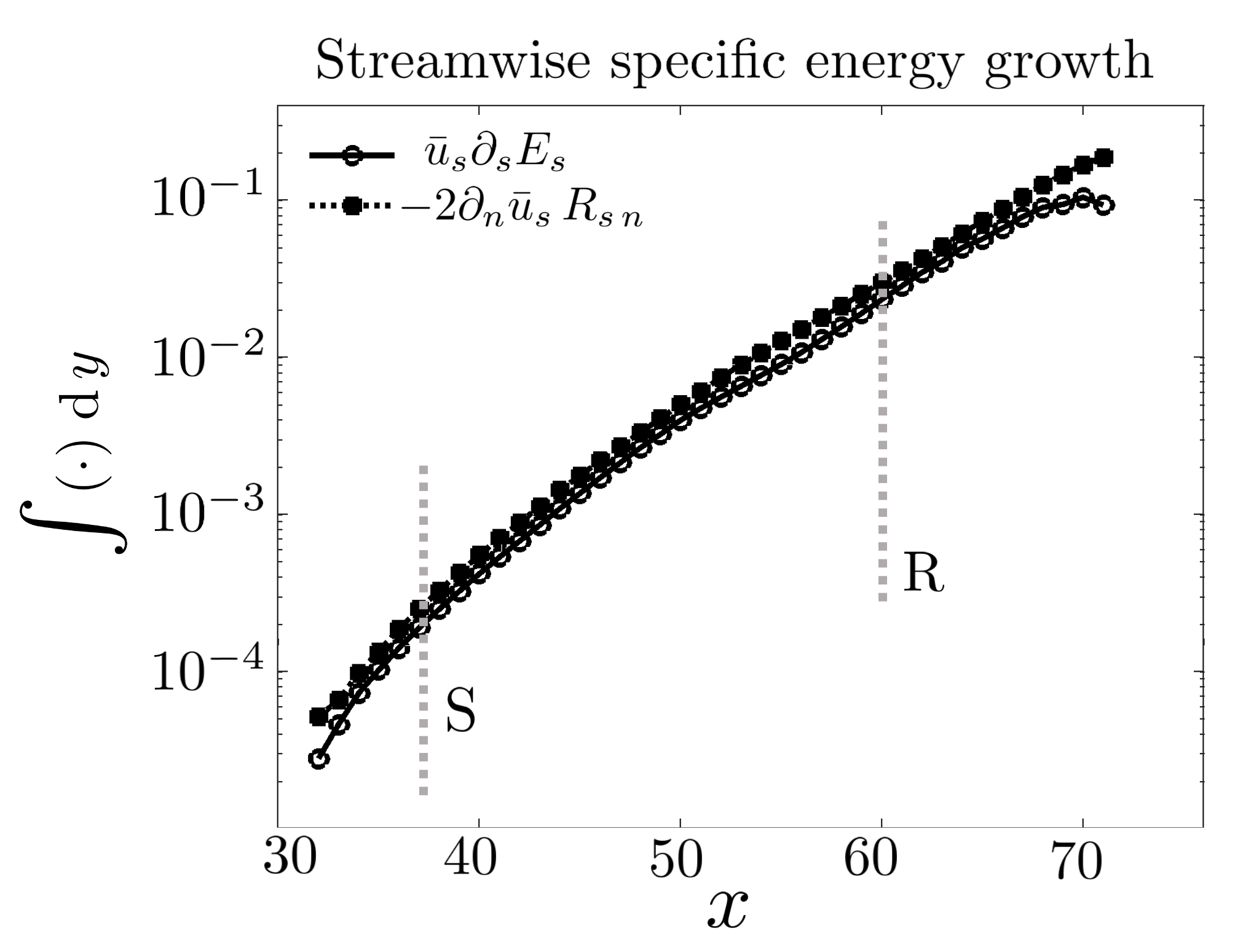}
           \label{fig.res1}}}
    \end{tabular}
    &    
    \begin{tabular}{c}
    {
   \hspace*{-0.45cm}
    \subcaptionbox{}
    {\includegraphics[height= 5.0 cm, trim= 4 4 4 4, clip]{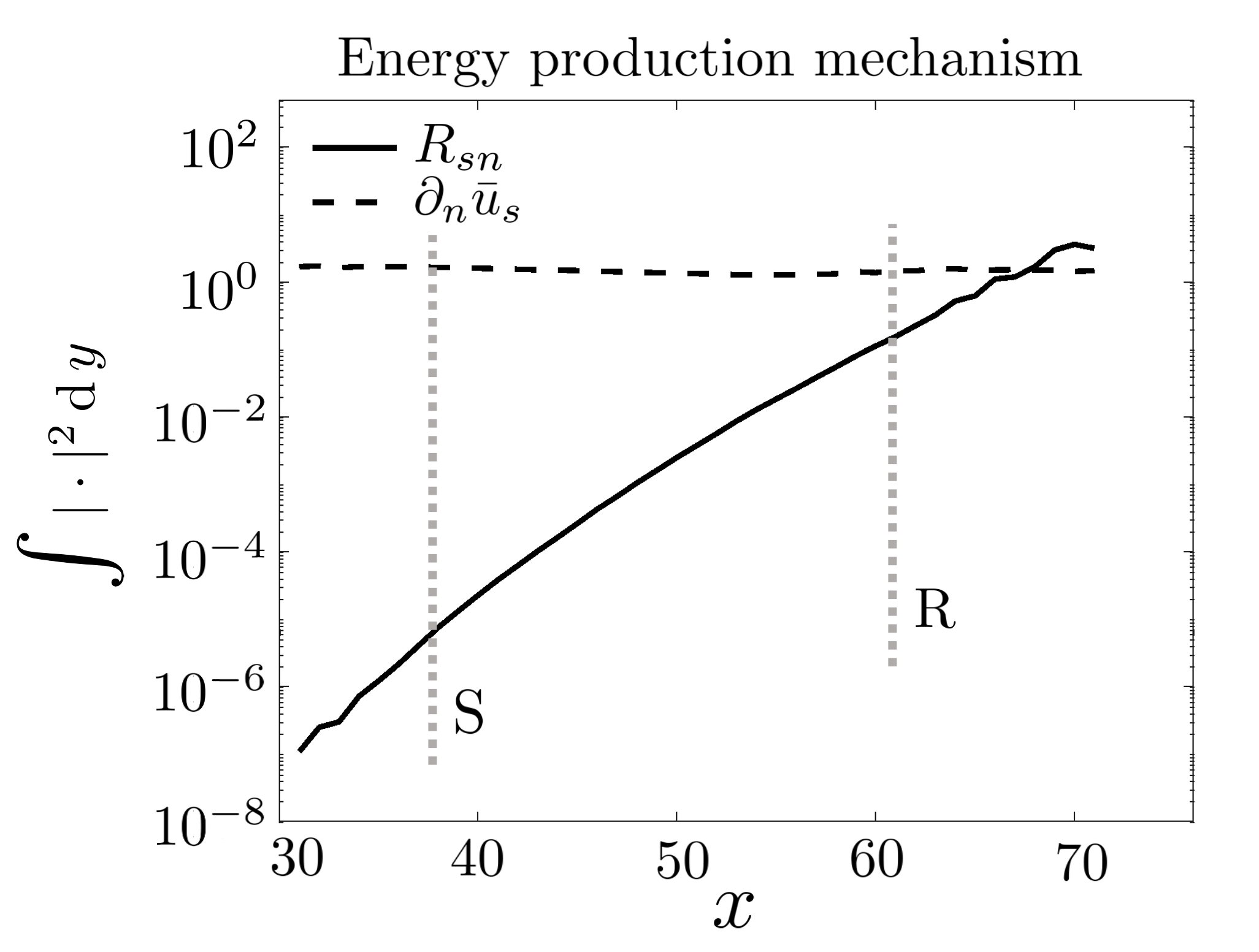}
           \label{fig.res2}}}
    \end{tabular}
    \end{tabular}
    }    
 \caption{Streamwise variation of (a) $\bar{u}_s \partial_{s} E_{s}$ along with the dominant production term in equation~\eqref{eq:Es-transport}; (b) average fluctuation shear stress ${R}_{sn}$ and base flow shear $\partial_{n} \bar{u}_{s}$. }
      \label{fig:obl_budget}
    \end{figure}

\begin{figure}
    \centering
    {
    \begin{tabular}{cc}
    \begin{tabular}{c}
    {
   \hspace*{-0.65cm}
    \subcaptionbox{}
    {\includegraphics[height= 5 cm, trim= 4 4 4 4, clip]{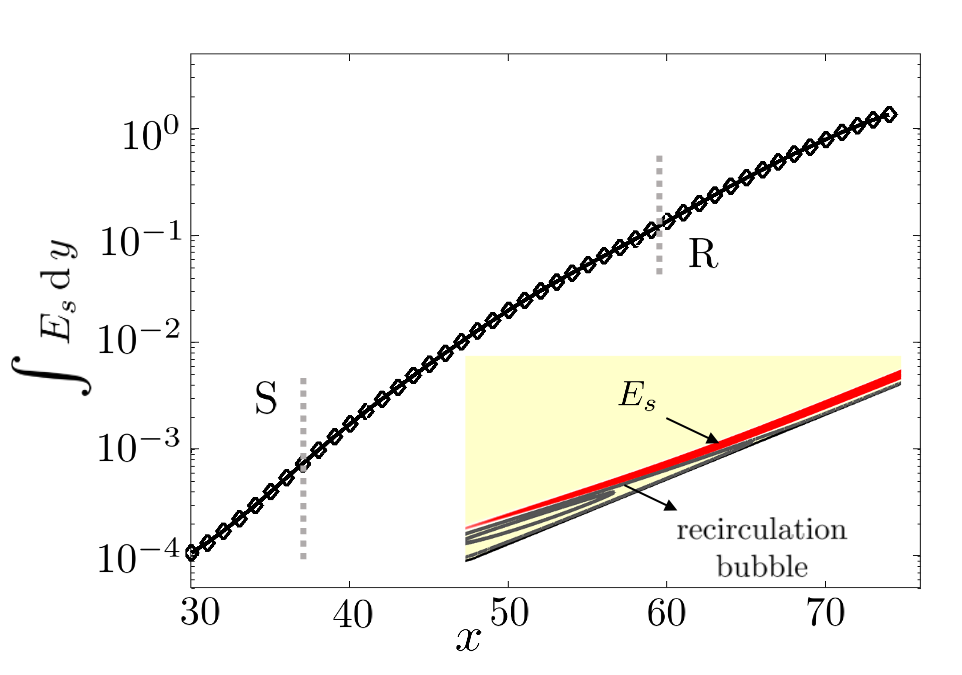}
           \label{fig.res1}}}
    \end{tabular}
    &    
    \begin{tabular}{c}
    {
   \hspace*{-0.65cm}
    \subcaptionbox{}
    {\includegraphics[height= 5 cm, trim= 4 4 4 4, clip]{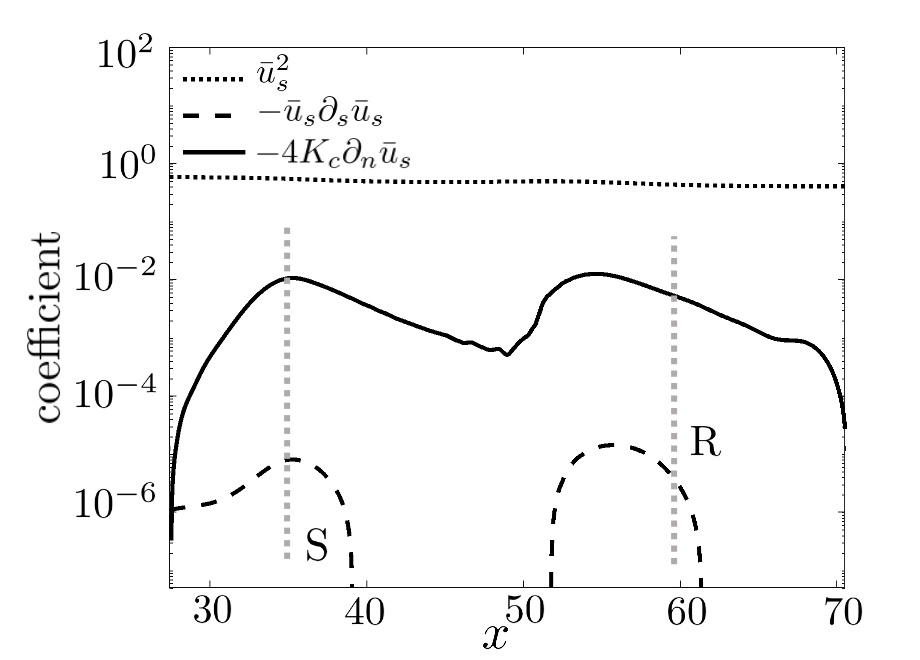}
           \label{fig.res2}}}
    \end{tabular}
    \end{tabular}
    }    
 \caption{(a) Streamwise evolution of the wall-normal integral of $E_{s} \DefinedAs \langle u^\prime_s u^\prime_s \rangle$ for the primary output resolvent mode of oblique fluctuations with ($\omega=0.4,\lambda_{z}=3.0$) along with contours of $E_{s}$ in separated shear layer near reattachment (inset); \tc{black}{(b) the coefficients in equation~\eqref{eq:Es2}} along separation shear layer.}   
    \label{fig:obl_spring}
    \end{figure}

Concave base flow curvature (i.e., $K_{c} < 0$) in the shear layer provides destabilizing effect in system~\eqref{eq:Es-Rns} \tc{black}{which can be understood by simplifying equation~\eqref{eq:Es-Rns} for oblique waves. In figure~\ref{fig:obl_budget}(a), we compare $\bar{u}_{s}  \partial_s E_{s} \DefinedAs \bar{u}_{s} \partial E_{s} / \partial s$ with the dominant production term in~\eqref{eq:Es-transport} to illustrate that $( \partial_n \bar{u}_{s} ){R}_{sn}$ dictates the streamwise growth of $E_{s}$. Furthermore, since the base shear $\partial_n \bar{u}_{s}$ remains almost constant throughout the shear layer (cf.\ figure~\ref{fig:obl_budget}(b)), its streamwise derivative can be neglected, thereby leading~to,}
	\begin{equation}
	\tc{black}{\bar{u}^2_{s} \, \frac{\partial^2 {E}_{s}}{\partial s^2}
	\; + \;
	\bar{u}_{s} (\partial_s \bar{u}_s) \, \frac{\partial {E}_{s}}{\partial s}
	\; + \;
	4 K_c 
	( \partial_n \bar{u}_s )
	{E}_{s}
	\; \approx \; 
	0.}
	\label{eq:Es2}
\end{equation}
\tc{black}{This second-order differential equation for $E_s$ is obtained by taking the derivative of the equation for $E_s$ in~\eqref{eq:Es-Rns}, keeping the dominant terms, and substituting the equation for $R_{sn}$ from~\eqref{eq:Es-Rns} into the resulting expression. Figure~\ref{fig:obl_spring}(a) shows the streamwise evolution of $E_s$ and figure~\ref{fig:obl_spring}(b) compares the coefficients in equation~\eqref{eq:Es2}. Since the effect of $\partial_{s} \bar{u}_{s}$ is negligible, equation~\eqref{eq:Es2} can further be simplified to obtain,}
\begin{equation}
	\tc{black}{
	\bar{u}^2_{s} \, \frac{\partial^2 {E}_{s}}{\partial s^2}
	\; + \;
	4 K_c 
	( \partial_n \bar{u}_s )
	{E}_{s}
	\; \approx \; 
	0.}
	\label{eq:Es2b}
\end{equation}

\tc{black}{As shown in figure~\ref{fig:obl_spring}(b), the concave base flow curvature (i.e., $K_c < 0$) provides the destabilizing influence throughout the separated shear layer and a simple mechanical analogy can be used to explain amplification of oblique waves. In the regions where $K_c < 0$ the ``spring constant'' $4 K_c ( \partial_n \bar{u}_s)$ in equation~\eqref{eq:Es2b} is negative and this system behaves as an inverted pendulum, which enables the spatial growth of $E_s$.}

In summary, \tc{black}{we have utilized the resolvent analysis to identify} the spatial structure of oblique fluctuations that amplify rapidly in the separation zone. Furthermore, by conducting transport analysis of the most energetic  fluctuations, \tc{black}{we have demonstrated that the resulting amplification arises from concave curvature of the laminar 2D base~flow.} 

	\vspace*{-2ex}
\section{Nonlinear interactions of oblique waves}
\label{ref:strk}

In \S~\ref{sec:io}, we used \tc{black}{resolvent analysis} to identify oblique waves as the most energetic responses of the linearized flow equations in the presence of unsteady disturbances. Recent numerical simulations~\citep{lugrin_beneddine_leclercq_garnier_bur_2021} show that, even in the presence of unsteady disturbances, the dominant response near the reattachment appears in the form of streamwise streaks. To investigate the origin of steady streaks in the presence of unsteady external disturbances we utilize a weakly nonlinear formulation based on a perturbation expansion in the amplitude of the oblique disturbances. While previous numerical studies of transition induced by oblique waves in low-speed channel~\citep{schmid1992new} and boundary layer~\citep{chang_malik_1994,berlin1999nonlinear,mayer2011direct} flows show that nonlinear interactions of oblique waves generate streaks, we focus on the origin and spatial growth of these streaks in separated high-speed compressible boundary layer flows.  	
	
	\vspace*{-2ex}
\subsection{\tc{black}{Streaks generated by oblique waves: a weakly nonlinear analysis}}
\label{ref:wnla}

In the presence of a small external disturbance,
\begin{align}
	\B d(x,y,z,t) 
	\;=\; 
	\epsilon \, (\tc{black}{\hat{\bd}^{(1)}_{+}}(x,y) \mre^{\mri \omega t} 
	\, + \, 
	\tc{black}{ \hat{\bd}^{(1)}_{-}}(x,y)\mre^{-\mri \omega t})\,\mre^{\mri \beta z},
	\label{eq.dow}
\end{align}
a weakly nonlinear analysis \tc{black}{of \S~\ref{sec:nse-d} can be utilized to represent the flow state components in compressible NS equations~\eqref{eq:nse1} as
	\begin{equation}
	\begin{array}{rrcl}
	\mbox{${\cal O} (1)$:} 
	& 
	\bar{\bPsi} ({\B x}, t)
	& \!\! = \!\! &
	\bar{\bPsi} (x,y), 
	\\[0.1cm]
	\mbox{${\cal O} (\epsilon)$:} 
	& 
	\bpsi^{(1)}({\B x}, t)
	& \!\! = \!\! &
	\left(
	\hat{\bpsi}^{(1)}_{+}(x,y)\mre^{\mri \omega t} 
	\, + \,
	\hat{\bpsi}^{(1)}_{-}(x,y)\mre^{-\mri \omega t}
	\right)
	\mre^{\mri \beta z},  
	\\[0.2cm]
	\mbox{${\cal O} (\epsilon^2)$:} 
	& 
	\bpsi^{(2)}({\B x}, t)
	& \!\! = \!\! &
	\left(
	\hat{\bpsi}^{(2)}_{0} (x,y) 
	\, + \,
	\hat{\bpsi}^{(2)}_{+} (x,y) \mre^{2 \mri \omega t}
	\, + \,
	\hat{\bpsi}^{(2)}_{-} (x,y) \mre^{-2 \mri \omega t}
	\right)
	\mre^{2 \mri \beta z}.
	\end{array}
	\end{equation}
Here, $\bar{\bPsi} (x,y)$ is the 2D laminar base flow, $\hat{\bd}^{(1)}_{\pm}$ and $\hat{\bpsi}^{(1)}_{\pm}$ are the principal oblique input and state modes resulting from the linearized analysis of \S~\ref{sec:io}, whereas $\hat{\bpsi}^{(2)}_{0}$ and $\hat{\bpsi}^{(2)}_{\pm}$ are the steady and harmonic components of the state at $\mathcal{O}(\epsilon^2)$. At $\mathcal{O}(\epsilon^2)$, the fluctuation's dynamics are governed by equation~\eqref{eq:pert2}, where the steady component $\hat{\bpsi}^{(2)}_{0}$ satisfies}
	\begin{align}
	\tc{black}{
	\left[ \cA_{2 \beta} \, \hat{\bpsi}^{(2)}_{0} (\, \cdot \, , \, \cdot \, ) \right]
	(x,y)
	\; = \; 
	- \hat{\bd}^{(2)}_0 (x,y).}
	\label{eq:weakl2}
\end{align}
\tc{black}{Here, $\cA_{2 \beta}$ is the Fourier symbol of the dynamical generator in the linearized state-space model~\eqref{eq:SScomp} and $\hat{\bd}^{(2)}_0 \DefinedAs \cN^{(2)}_0 ( \hat{\bpsi}^{(1)}_{\pm} )$ is the forcing term that arises from quadratic interactions of $\cO (\epsilon)$ oblique waves with the spanwise wavenumber $\beta$; see appendix~\ref{app.nonlin} for details. Thus, the resolvent operator associated with~\eqref{eq:SScomp} evaluated at ($\omega = 0$, $2 \beta$) maps the nonlinear modulation $\hat{\bd}^{(2)}_0$ of $\mathcal{O}(\epsilon)$ oblique waves to $\mathcal{O}(\epsilon^{2})$ steady streamwise streaks,}
	\begin{align}
	\tc{black}{\hat{\bpsi}^{(2)}_{0}
	(x,y)
	\; = \; 
	\left[ \cR_{2 \beta} (0) \hat{\bd}^{(2)}_{0} (\, \cdot \, , \, \cdot \, ) \right] (x,y)
	\; = \; 
	- 
	\left[ \cA_{2 \beta}^{-1} \, \hat{\bd}^{(2)}_{0} (\, \cdot \, , \, \cdot \, ) \right] (x,y).}
	\label{eq:q^(2)}
\end{align}
 
\tc{black}{To investigate the emergence of streaks from unsteady disturbances, we introduce forcing inputs with ($\omega=\pm 0.4, \lambda_{z}=3$) and examine a weakly nonlinear evolution of the resulting oblique waves. These  forcing inputs are introduced at the upstream plane $x_{\mathrm{in}} = 25$ and their spatial structure is identified using the resolvent analysis of \S~\ref{sec:io} to generate the most energetic response at the reattachment \mbox{(i.e., at $x_\mathrm{out}=60$).}} 
 
\begin{figure}
    \centering
    {
    \begin{tabular}{cc}
    \begin{tabular}{c}
    {
   \hspace*{-0.5cm}
    \subcaptionbox{}
    {\includegraphics[height= 4.4 cm, trim= 5 4 4 4, clip]{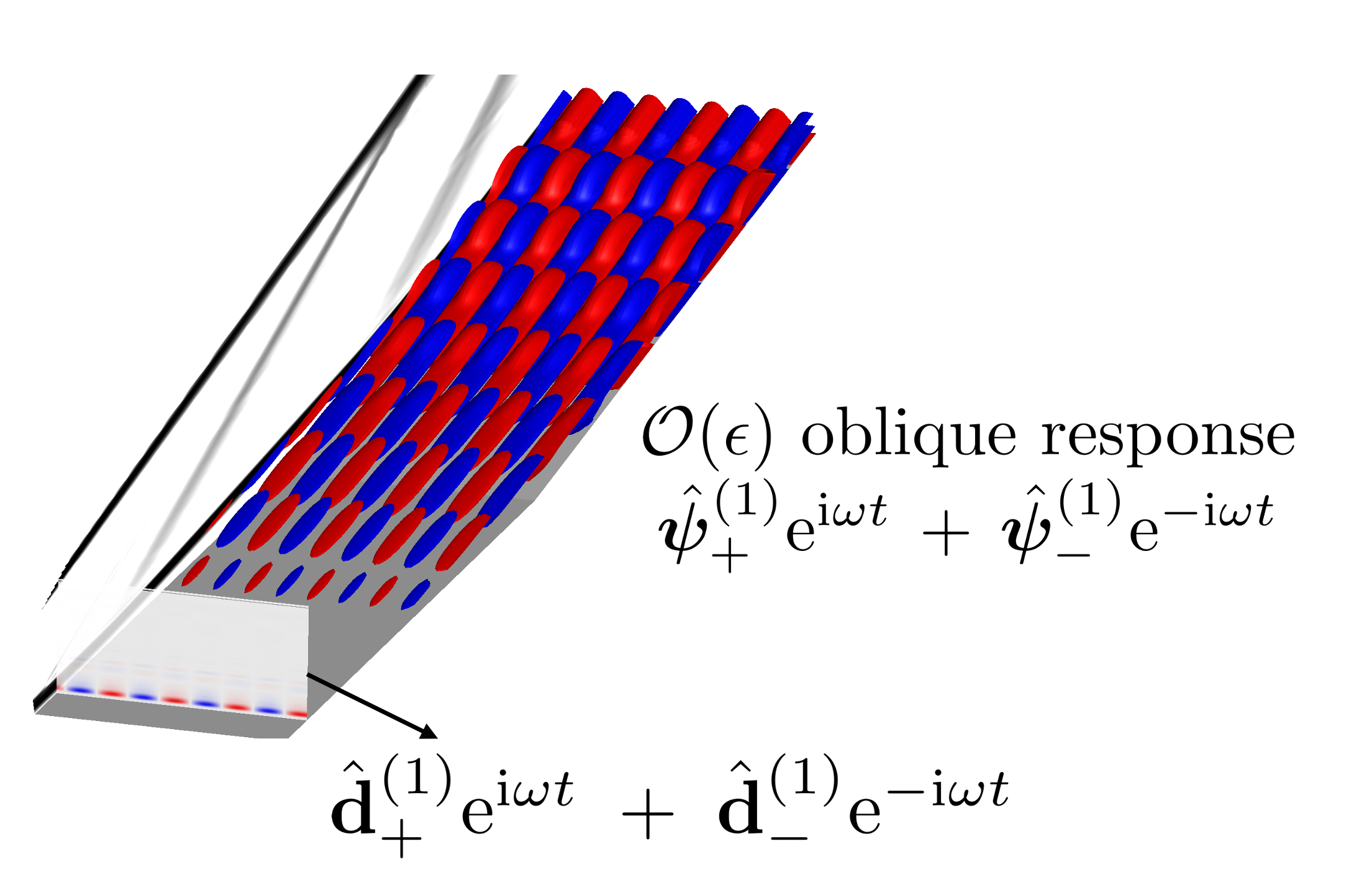}
           \label{fig.setup1}}}
    \end{tabular}
    &    
    \begin{tabular}{c}
    {
   \hspace*{-0.5cm}
    \subcaptionbox{}
    {\includegraphics[height= 4.4 cm, trim= 4 4 4 4, clip]{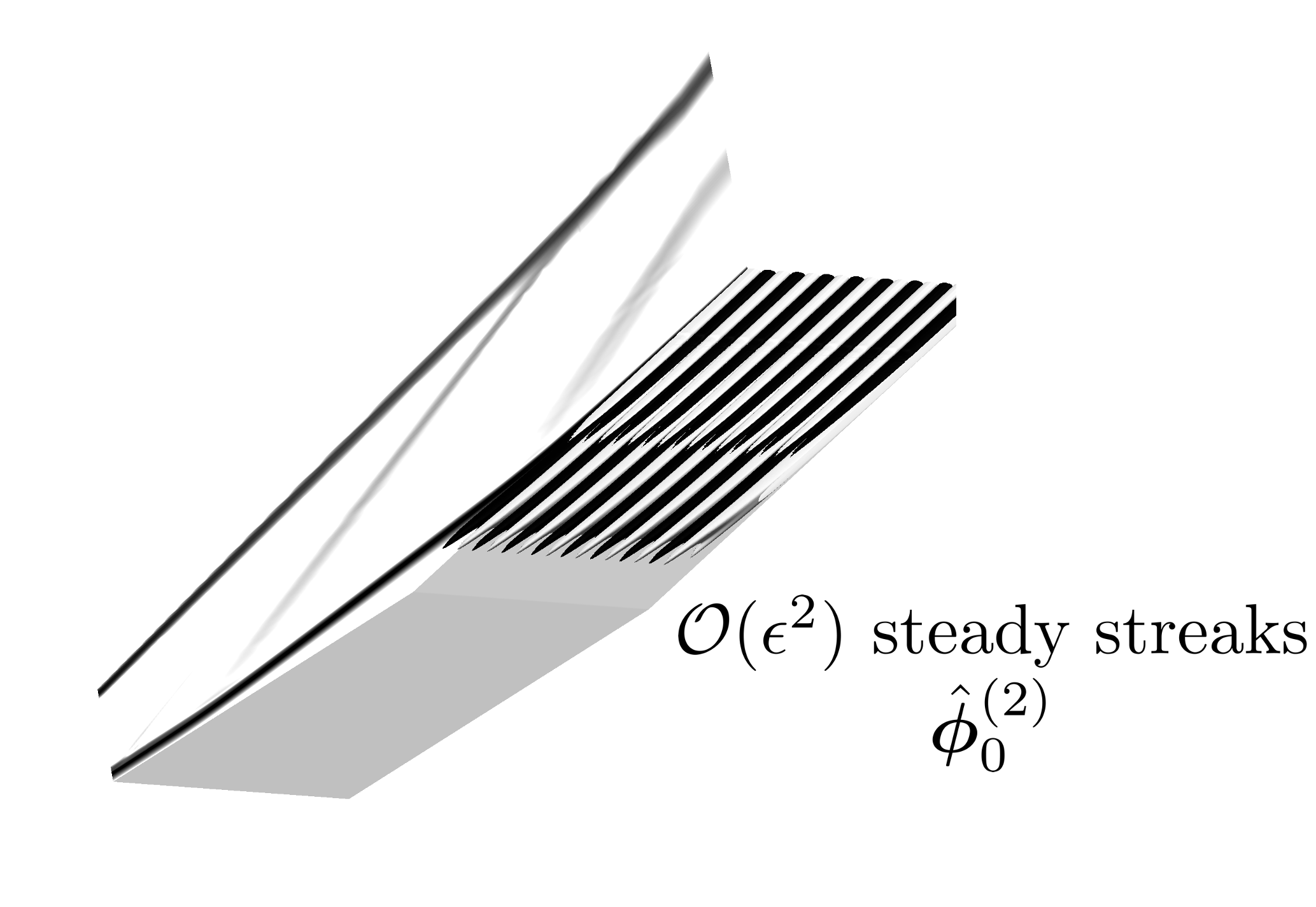}
           \label{fig.setup2}}}
    \end{tabular}
    \end{tabular}
    }    
    \caption{(a) Setup for weakly nonlinear analysis: a pair of dominant input modes with ($\omega=\pm 0.4, \lambda_{z}=3$) resulting from resolvent analysis is introduced at $x_{\mathrm{in}} = 25$ and the corresponding streamwise velocity fluctuations arise as the output of the linearized dynamics. (b) $\mathcal{O} (\epsilon^2)$ steady streamwise streaks with $\lambda_z = 1.5$ are triggered by weakly nonlinear interactions of $\mathcal{O} (\epsilon)$ oblique waves with $\lambda_{z}=3$.}
    \label{fig:lamz_comp}
\end{figure} 
 
\tc{black}{Figure~\ref{fig:lamz_comp}(a) illustrates the setup in which disturbances corresponding to a pair of oblique input modes $\hat{\bd}^{(1)}_{\pm}$ with ($\omega=\pm 0.4,\ \lambda_{z}=3 $) are introduced at $x_\mathrm{in}= 25$. The resulting response of the linearized dynamics consists of oblique waves with opposite phase velocities, leading to a checkerboard wave pattern in the spanwise direction. Figure~\ref{fig:lamz_comp}(b) shows the streamwise velocity component of the steady response $\hat{\bphi}^{(2)}_{0} (x,y)$ at $\cO (\epsilon^2)$ that arises from weakly nonlinear interactions of $\cO (\epsilon)$ oblique waves. The steady response is given by streamwise streaks with half the spanwise wavelength $\lambda^\mathrm{streaks}_{z} = \lambda^\mathrm{oblique}_{z}/2 = 1.5$ of the oblique input.} 
		
\begin{figure}
    \centering
    {
    \begin{tabular}{ccc}
    \begin{tabular}{c}
    {
   \hspace*{-0.5cm}
    \subcaptionbox{}
    {\includegraphics[height= 3.0 cm, trim= 4 4 4 4, clip]{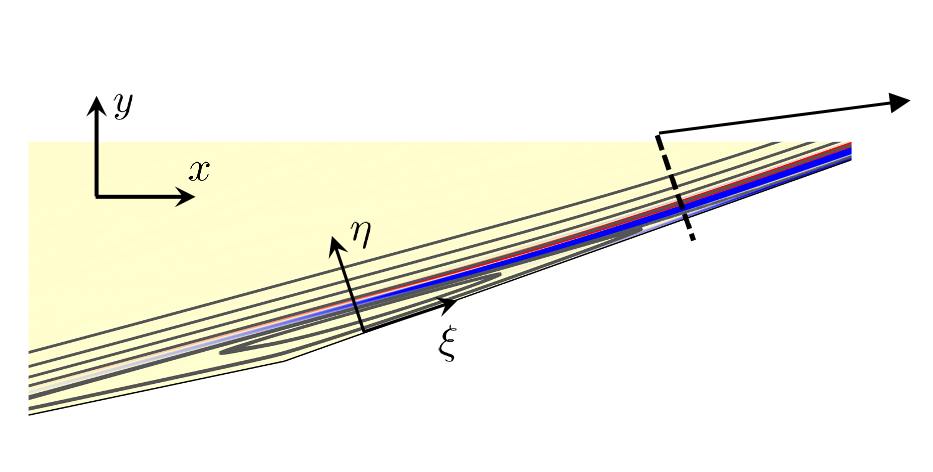}
           \label{fig.force1}}}
    \end{tabular}
    &    
    \begin{tabular}{c}
    {
   \hspace*{-0.75cm}
    \subcaptionbox{}
    {\includegraphics[height= 3.0 cm, trim= 8 4 4 4, clip]{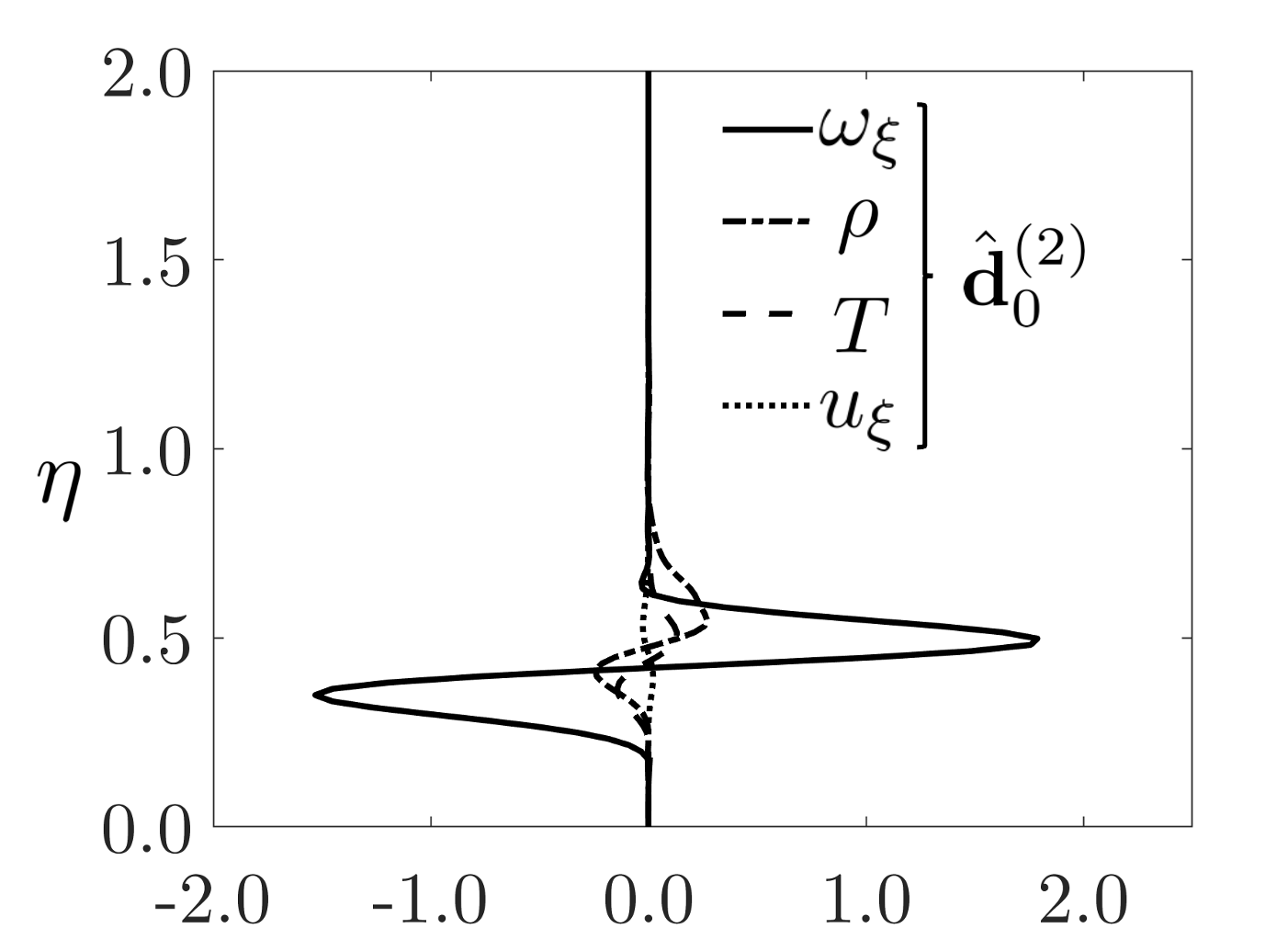}
           \label{fig.force2}}}
    \end{tabular}
    &   
   \begin{tabular}{c}
    {
   \hspace*{-0.5cm}
    \subcaptionbox{}
    {\includegraphics[height= 3.0 cm, trim= 2 2 4 4, clip]{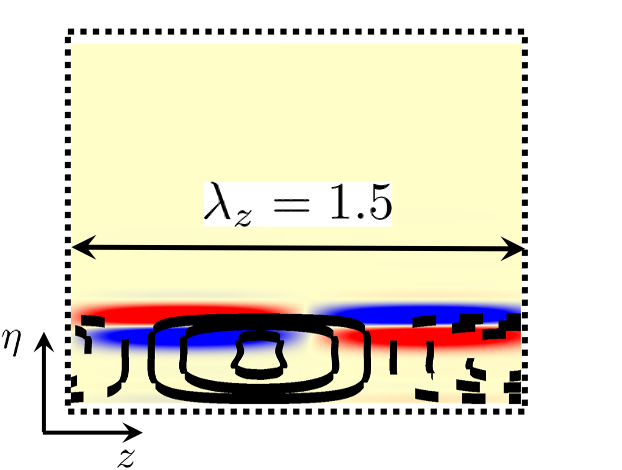}
           \label{fig.force3}}}
    \end{tabular}
    \end{tabular}
    }    
   \caption{(a) Real part of the streamwise vorticity forcing component that arises from weakly nonlinear interactions of oblique waves $\hat{\bd}^{(2)}_0$ in the ($x,y$) plane along with the base flow streamlines. (b) Wall-normal profiles of the forcing terms to the streamwise vorticity, density, temperature, and streamwise velocity equations that originate from $\hat{\bd}^{(2)}_0$ before reattachment, at $x = 58$. (c) Spatial structure of the forcing $\hat{\bd}^{(2)}_0$ near reattachment in the ($z,\eta$) plane (color plots) along with the resulting streak response $\hat{\bphi}^{(2)}_0$ (contour lines) at $x = 58$.}
    \label{fig:vorticity_up_forcing}
    \end{figure}

A weakly nonlinear analysis allows us to demonstrate that steady streaks at $\mathcal{O}(\epsilon^2)$ arise from quadratic interactions of $\mathcal{O}(\epsilon)$ oblique waves. Figure~\ref{fig:vorticity_up_forcing}(a) utilizes a wall-aligned ($\xi,\eta$) coordinate system to illustrate the forcing term \tc{black}{$\hat{\bd}^{(2)}_0 \DefinedAs \cN^{(2)}_0 ( \hat{\bpsi}^{(1)}_{\pm} )$} in~\eqref{eq:weakl2}, where $\xi$ and $\eta$ denote the directions parallel and normal to the wall, respectively. Large amplification of oblique waves that result from  linearized analysis in the reattachment region triggers strongest forcing \tc{black}{$\hat{\bd}^{(2)}_{0}$} in that region. Figure~\ref{fig:vorticity_up_forcing}(b) shows the wall-normal profiles of the forcing term to the mass, momentum, and temperature equations in~\eqref{eq:weakl2} before reattachment at $x = 58$. We observe the strongest contribution of the forcing to the wall-normal and spanwise components of the momentum equations, thereby demonstrating its vortical nature. Figure~\ref{fig:vorticity_up_forcing}(c) illustrates the spatial structure of the forcing near reattachment in the ($z,\eta$) plane. The forcing term \tc{black}{$\hat{\bd}^{(2)}_{0}$} which forms counter-rotating vortices in the separated shear layer is $90^{\circ}$ out of phase relative to the induced streak response $u^\prime_s$. In contrast to the dominant vortical forcing resulting from the linearized analysis, the vortical source term that arises from weakly nonlinear interactions of oblique waves primarily lies downstream of recirculation zone.

\begin{figure}
    \centering
    {
    \begin{tabular}{cc}
    \begin{tabular}{c}
    {
   \hspace*{-0.5 cm}
    \subcaptionbox{}
    {\includegraphics[height= 5 cm, trim= 4 4 4 4, clip]{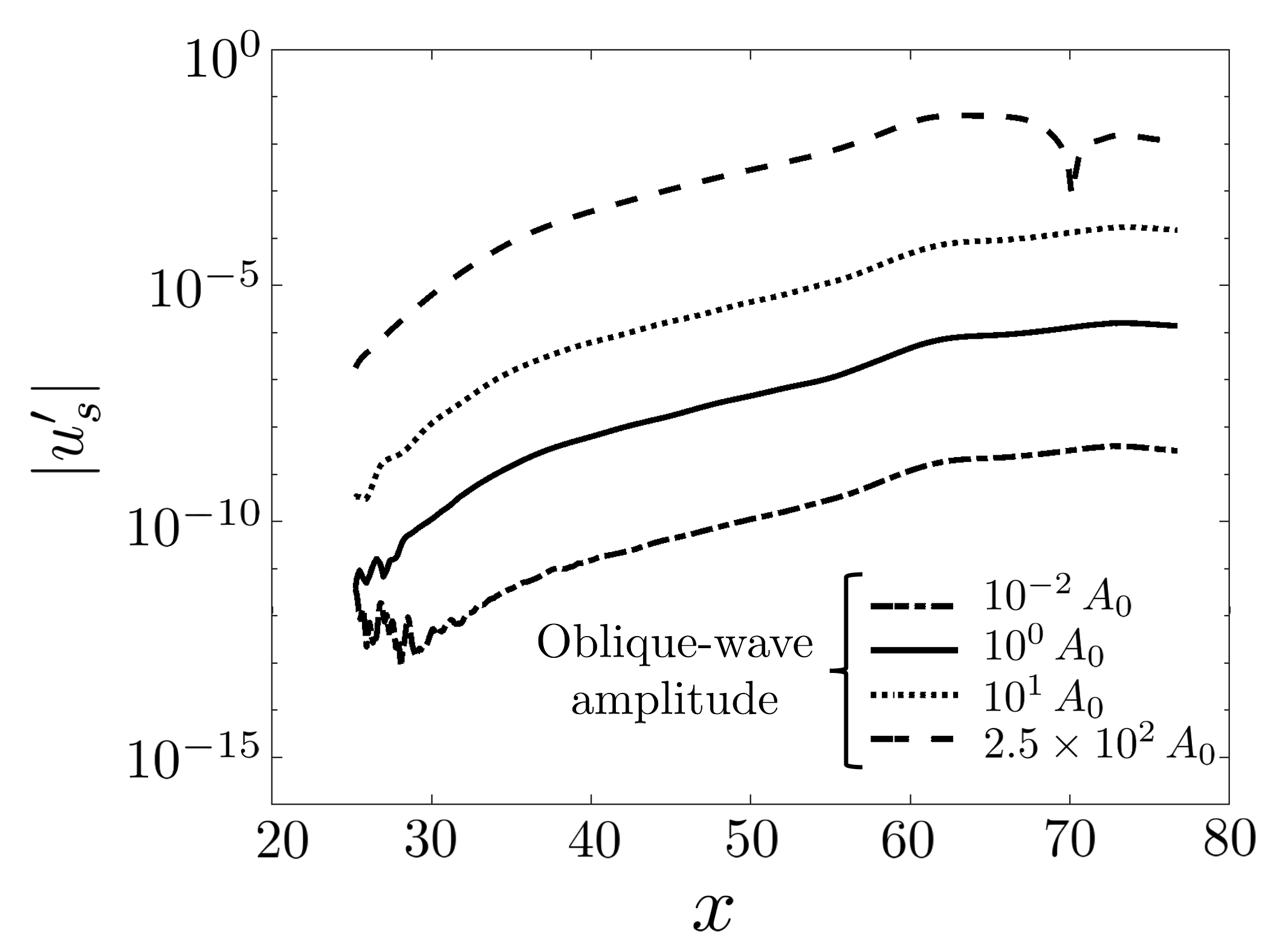}
           \label{fig.quad1}}}
    \end{tabular}
    &    
    \begin{tabular}{c}
    {
   \hspace*{-0.5cm}
    \subcaptionbox{}
    {\includegraphics[height= 5 cm, trim= 4 4 4 4, clip]{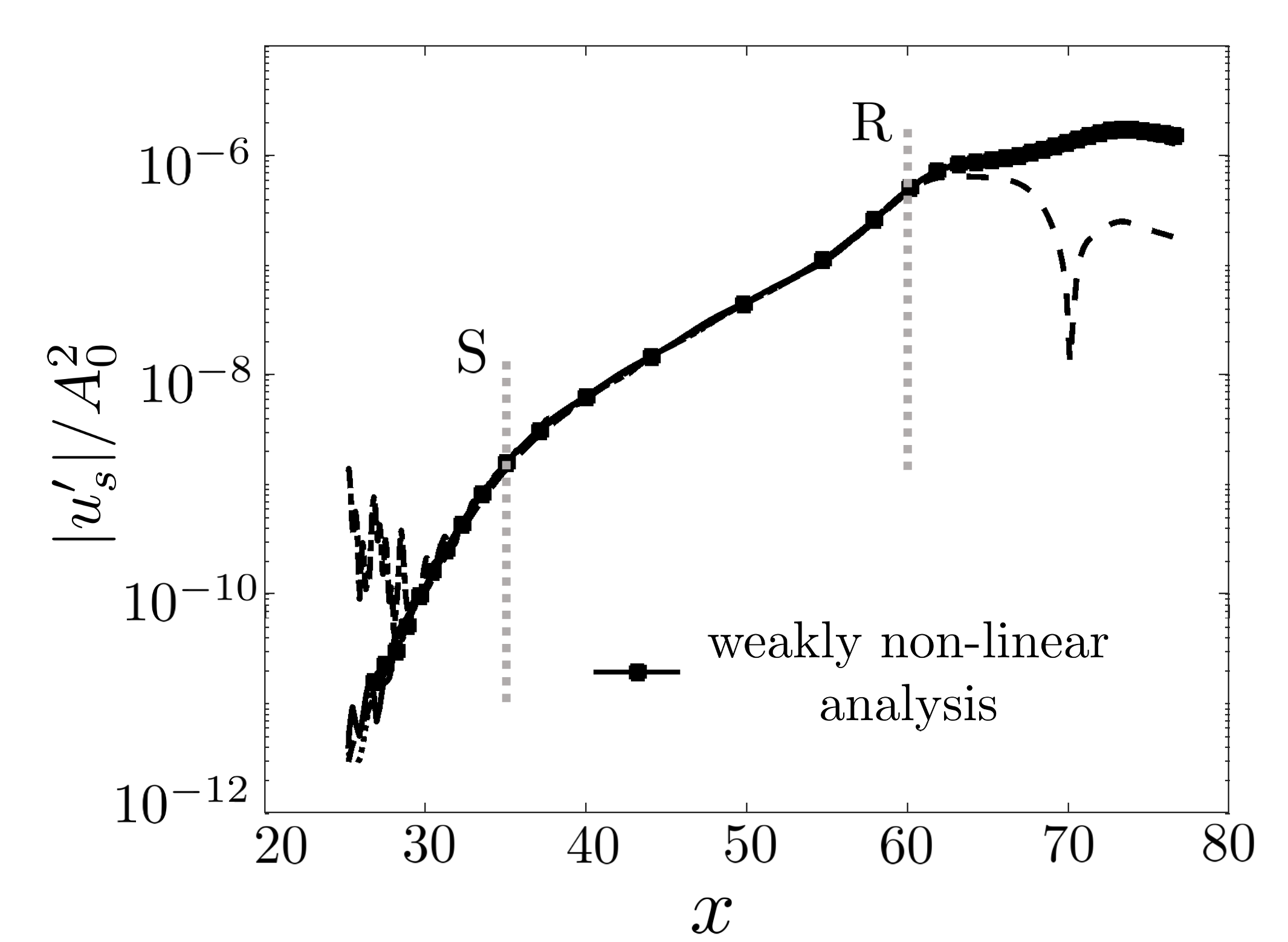}
           \label{fig.quad2}}}
    \end{tabular}
    \end{tabular}
    }    
    \caption{(a) Streamwise velocity fluctuations $u_s^{\prime}$ along the separation streamline associated with \tc{black}{$\mathcal{O}(\epsilon^{2})$} steady streaks at $\lambda_z = 1.5$; DNS results for various amplitudes of oblique disturbances with ($\omega=\pm 0.4, \lambda_{z}=3$) are shown. (b) DNS results normalized with the square of the amplitude $A_{0}$ of oblique disturbances are compared with the results of weakly nonlinear analysis.}
    \label{fig:quadratic}
    \end{figure}

\tc{black}{We utilize DNS to verify the predictions of weakly nonlinear analysis. In DNS, the input oblique modes $\hat{\bd}^{(1)}_{\pm}$ (with $\omega=\pm 0.4$, $\lambda_z = 3$, and reference amplitude $A_{0} = 1.0$) are introduced in the plane $x_\mathrm{in} = 25$. Additional details about the grid resolution and numerical implementation in our DNS study are provided in \S~\ref{sec:secinstab}. Figure~\ref{fig:quadratic} shows the spatial evolution of steady streamwise velocity fluctuations $u^\prime_s$ along the base flow separation streamline that are triggered by unsteady oblique disturbances of different amplitudes. As shown in figure~\ref{fig:quadratic}(a), steady streaks} undergo similar spatial growth across the range of forcing amplitudes. Figure~\ref{fig:quadratic}(b) illustrates that the magnitude of streamwise velocity fluctuations collapses when scaled with $A^2_0$. This demonstrates an excellent agreement between streak profiles resulting from DNS and a weakly nonlinear analysis (that leads to equation~\eqref{eq:weakl2}). Deviations from predictions of the weakly nonlinear analysis are only observed for the largest amplitude considered and they are manifested by saturation of streaks in post-reattachment region.

Collapse of spatial profiles that characterize amplification of streaks irrespective of the amplitude of oblique disturbances, which differ by $\mathcal{O}(10^4)$, shows that streaks generated via quadratic interactions of oblique waves undergo linear amplification in the separation zone. This demonstrates predictive power of the weakly nonlinear analysis across the range of forcing amplitudes. In what follows, we utilize the input-output modes obtained from the resolvent analysis to characterize evolution of \tc{black}{$\cO (\epsilon^2)$ steady streamwise streaks} and uncover corresponding amplification mechanisms.

	\vspace*{-2ex}
\subsection{\tc{black}{Resolvent mode representation of $\mathcal{O}(\epsilon^{2})$ streaks}}

As described in \S~\ref{sec:svd}, the left and right singular function of the frequency response operator provide orthonormal bases of input and output spaces that can be used to study responses of the double-wedge flow to external excitations. In particular, \tc{black}{$\cO (\epsilon^{2})$ steady} streaks resulting from weakly nonlinear interactions of \tc{black}{$\cO (\epsilon)$ unsteady} oblique waves (cf.\ equation~\eqref{eq:q^(2)}) can be represented using SVD of the frequency response operator associated with the linearized system~\eqref{eq:SScomp} at $\omega = 0$ \tc{black}{and the spanwise wavenumber $2 \beta$},
	\begin{align}
	\tc{black}{
	\hat{\bphi}^{(2)}_{0} 	(x,y)
	\; = \; 
	\left[ 
	\cH_{2 \beta} (0) \hat{\bd}^{(2)}_{0} ( \, \cdot \, , \, \cdot \, ),
	\right] (x,y)
	\; = \; 
	\displaystyle{\sum_i}
	\,
	a_i 
	\bphi_{i} (x, y).}
	\label{eq:q2-svd}
	\end{align}
\tc{black}{Here, $a_{i} \DefinedAs \sigma_i \langle \mathbf{d}_{i}, \hat{\bd}^{(2)}_{0}  \rangle_E$, with $\hat{\bd}^{(2)}_{0} \DefinedAs \cN^{(2)}_{0} (\hat{\bpsi}_{\pm}^{(1)})$, quantifies the contribution of the $i$th output mode $\bphi_{i}$ of $\cH_{2 \beta} (0)$ to $\mathcal{O} (\epsilon^{2})$ steady streaks $\hat{\bphi}^{(2)}_{0}$, $\sigma_{i}$ is the $i$th singular value, ($\mathbf{d}_{i},\bphi _{i}$) are the corresponding input-output modes of $\cH_{2 \beta} (0)$, and  for $\lambda_z = 1.5$ the inner product $\langle \cdot, \cdot \rangle_{E}$ is carried over the entire flow domain in ($x,y$).}

\begin{figure}
    \centering
    {
    \begin{tabular}{cc}
    \begin{tabular}{c}
    {
    \subcaptionbox{}
    {\includegraphics[height= 5 cm, trim= 4 4 4 4, clip]{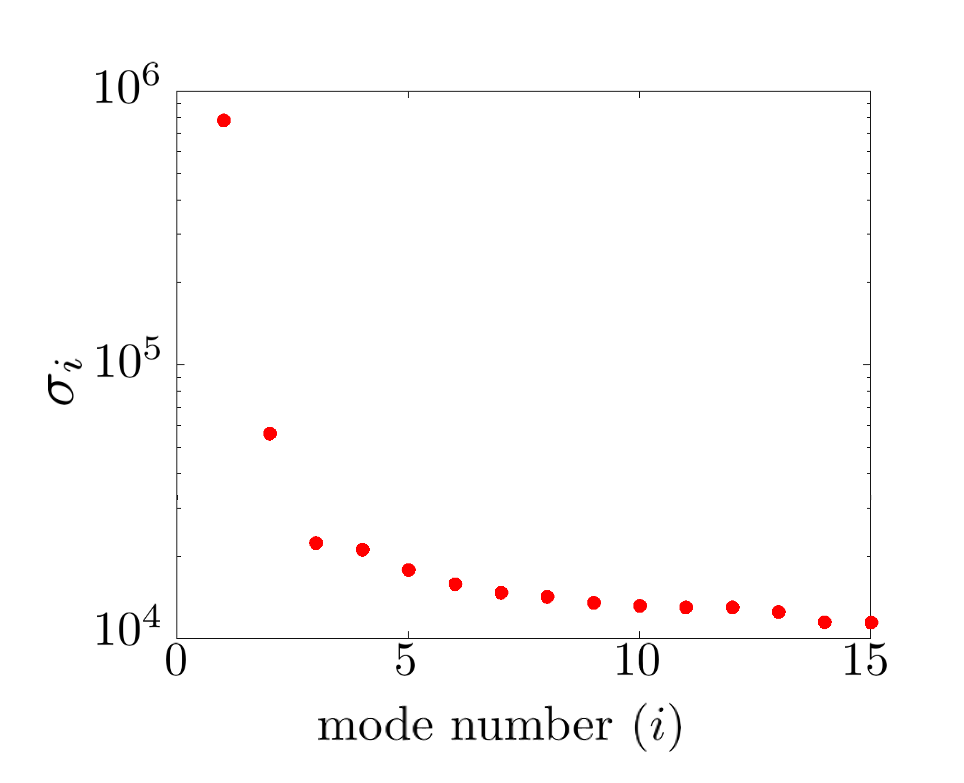}
           \label{fig.quad1}}}
    \end{tabular}
    &    
    \begin{tabular}{c}
    {
   \hspace*{-0.5cm}
    \subcaptionbox{}
    {\includegraphics[height= 5 cm, trim= 4 4 4 4, clip]{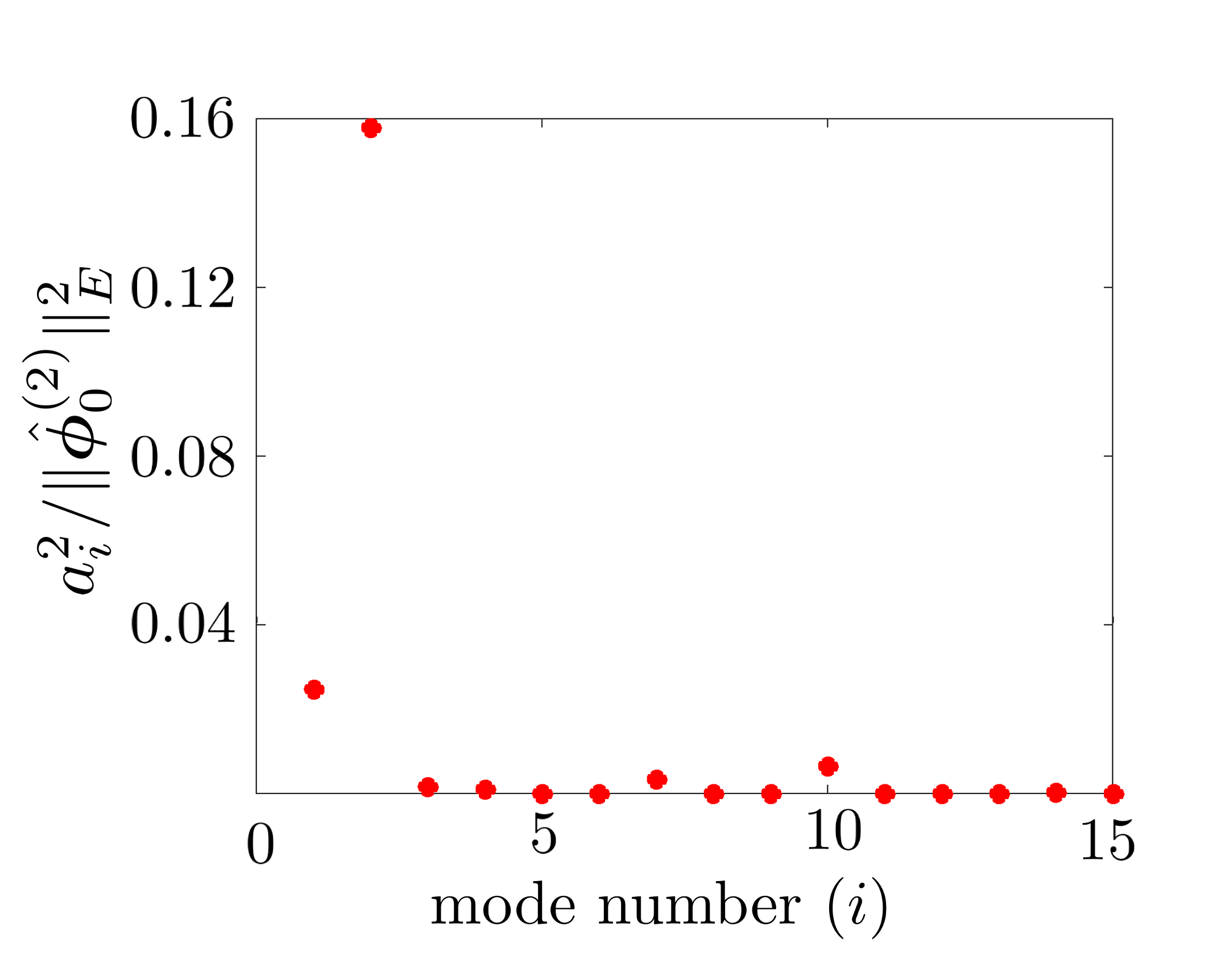}
           \label{fig.quad2}}}
    \end{tabular}
    \end{tabular}
    }    
    \caption{(a) Singular values of the resolvent operator associated with the linearized dynamics around the 2D laminar flow evaluated at $(\omega,\lambda_z) = (0,1.5)$. (b) Contribution of the $i$th output mode $\bphi_{i}$ \tc{black}{to the energy of $\mathcal{O}(\epsilon^{2})$ steady streaks \tc{black}{$\hat{\bphi}^{(2)}_{0}$}} that are triggered by weakly nonlinear interactions of \tc{black}{$\mathcal{O}(\epsilon)$} oblique waves.}
    \label{fig:siginner}
\end{figure}

\begin{figure}
    \centering
    {
    \begin{tabular}{cc}
    \begin{tabular}{c}
    {
    \subcaptionbox{}
    {\includegraphics[height= 5 cm, trim= 4 4 4 4, clip]{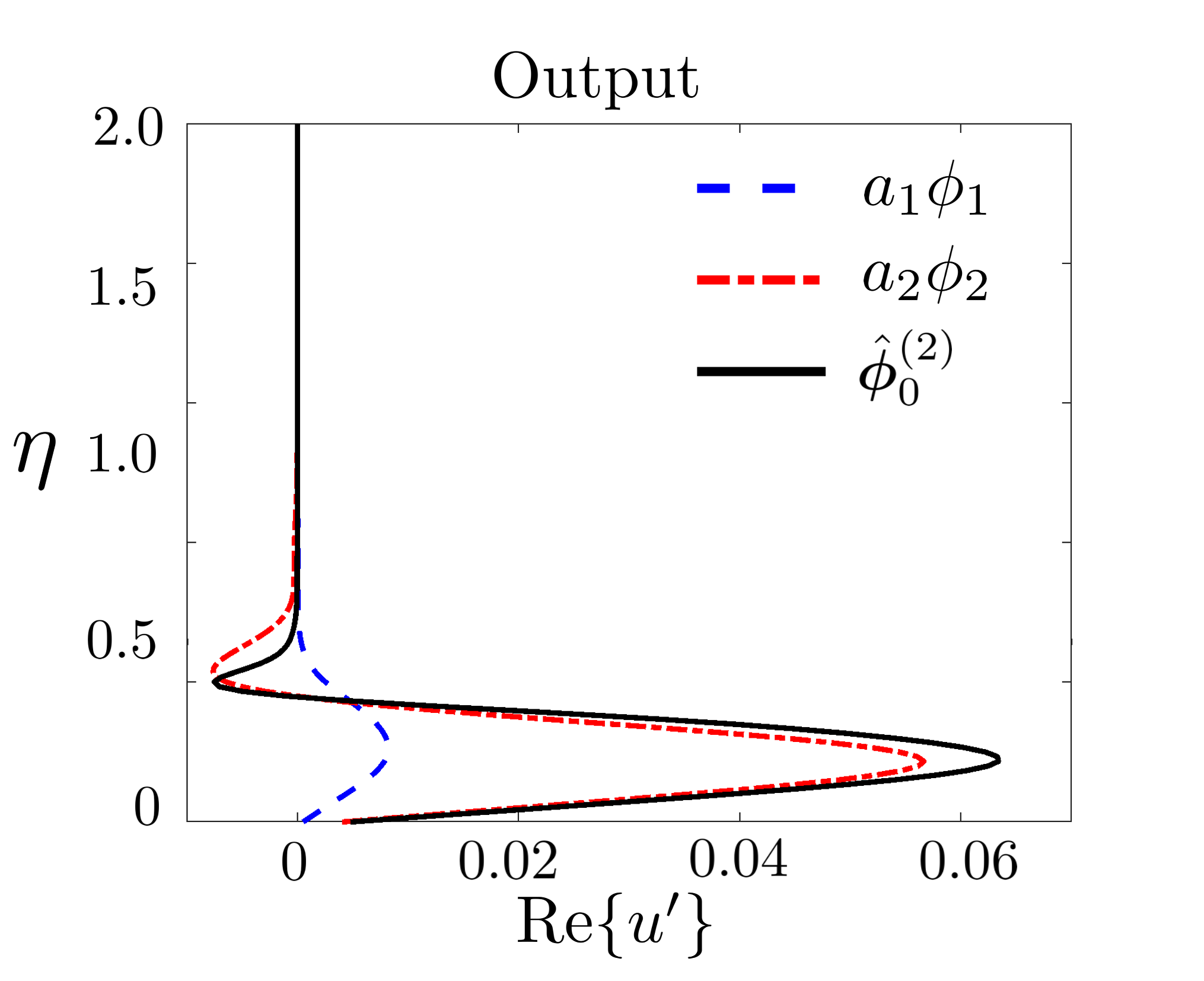}
           \label{fig.quad1}}}
    \end{tabular}
    &    
    \begin{tabular}{c}
    {
   \hspace*{-0.5cm}
    \subcaptionbox{}
    {\includegraphics[height= 5 cm, trim= 4 4 4 4, clip]{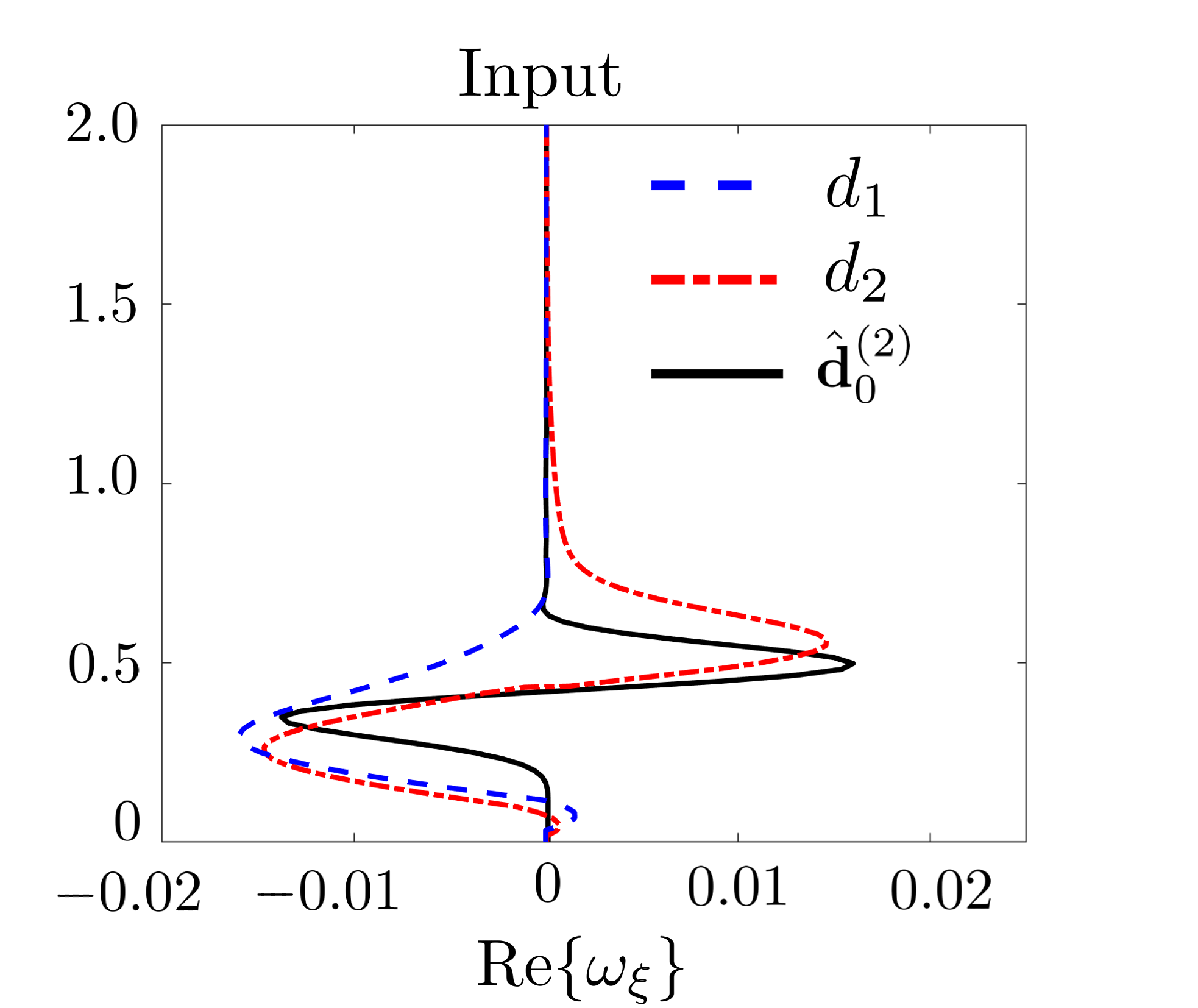}
           \label{fig.quad2}}}
    \end{tabular}
    \end{tabular}
    }    
   \caption{The wall-normal profiles of the real part of streamwise velocity fluctuations \tc{black}{resulting from weakly nonlinear interactions of oblique waves and} corresponding to (a) \tc{black}{$\mathcal{O}(\epsilon^{2})$ steady streaks $\hat{\bphi}^{(2)}_{0}$} and the first two output resolvent modes after reattachment, at $x=65$; (b) steady vortical forcing \tc{black}{$\hat{\B d}^{(2)}_{0}$} and the first two input resolvent modes in reattaching shear layer, at $x=57$.}
    \label{fig:O1O2comp}
    \end{figure}

Figure~\ref{fig:siginner}(a) shows \tc{black}{$15$} largest singular values of the resolvent for the linearized system~\eqref{eq:SScomp} with ($\omega = 0, \lambda_{z} = 1.5$). Even though the principal singular value $\sigma_1$ is an order of magnitude larger than $\sigma_2$, figure~\ref{fig:siginner}(b) demonstrates that the second output mode $\bphi_2$ contributes most to \tc{black}{$\hat{\bphi}^{(2)}_{0}$}. Figure~\ref{fig:O1O2comp}(a) shows the wall-normal profiles of the streamwise velocity component $u^{\prime}$ associated with \tc{black}{$\hat{\bphi}^{(2)}_{0}$} and the first two output modes ($\bphi_1,\bphi_2$) of the resolvent. We observe striking similarity between \tc{black}{$\cO (\epsilon^{2})$ steady streaks and $\bphi_{2}$} in the post-reattachment region, at $x = 65$. Similarly, figure~\ref{fig:O1O2comp}(b) compares the wall-normal shapes of the corresponding input modes $\bd_1$ and $\bd_2$ with the forcing $\hat{\bd}^{(2)}_0$ that arises from quadratic interactions. The input modes are visualized in the reattaching shear layer, at $x = 57$, and the streamwise vorticity component of $\bd_2$ provides a good approximation to the vortical forcing that captures interactions of unsteady oblique fluctuations. 

As demonstrated above, near reattachment, \tc{black}{$\mathcal{O}(\epsilon^{2})$ streaks} are well approximated by the second output mode of the resolvent associated with the linearized equations at ($\omega = 0, \lambda_z = 1.5$). To gain insight into the amplification mechanism that generates \tc{black}{$\mathcal{O} (\epsilon^{2})$ streaks}, we examine the dominant terms in the energy transport equation for the output mode \tc{black}{$\bphi_{2}$}. Similar to the analysis in \S~\ref{sec:oblphys}, we utilize the ($s,n,z$) coordinate system which is locally aligned with the streamlines of the base flow $(\bar{u}_{s},0,0)$. The transport of the spanwise averaged specific kinetic energy of streamwise velocity fluctuations $E_{s} = \langle {u^\prime_{s} u^{\prime}_s} \rangle$ and fluctuation shear stress ${R}_{sn} = \langle u^\prime_{s} u^{\prime}_n \rangle$ \tc{black}{associated with the second output mode is approximately governed by equation~\eqref{eq:Es-Rns} in \S~\ref{sec:oblphys}.}
	
\begin{figure}
    \centering
    {
    \begin{tabular}{cc}
    \begin{tabular}{c}
    {
   \hspace*{-0.75cm}
    \subcaptionbox{}
    {\includegraphics[height= 5.5 cm, trim= 4 4 4 4, clip]{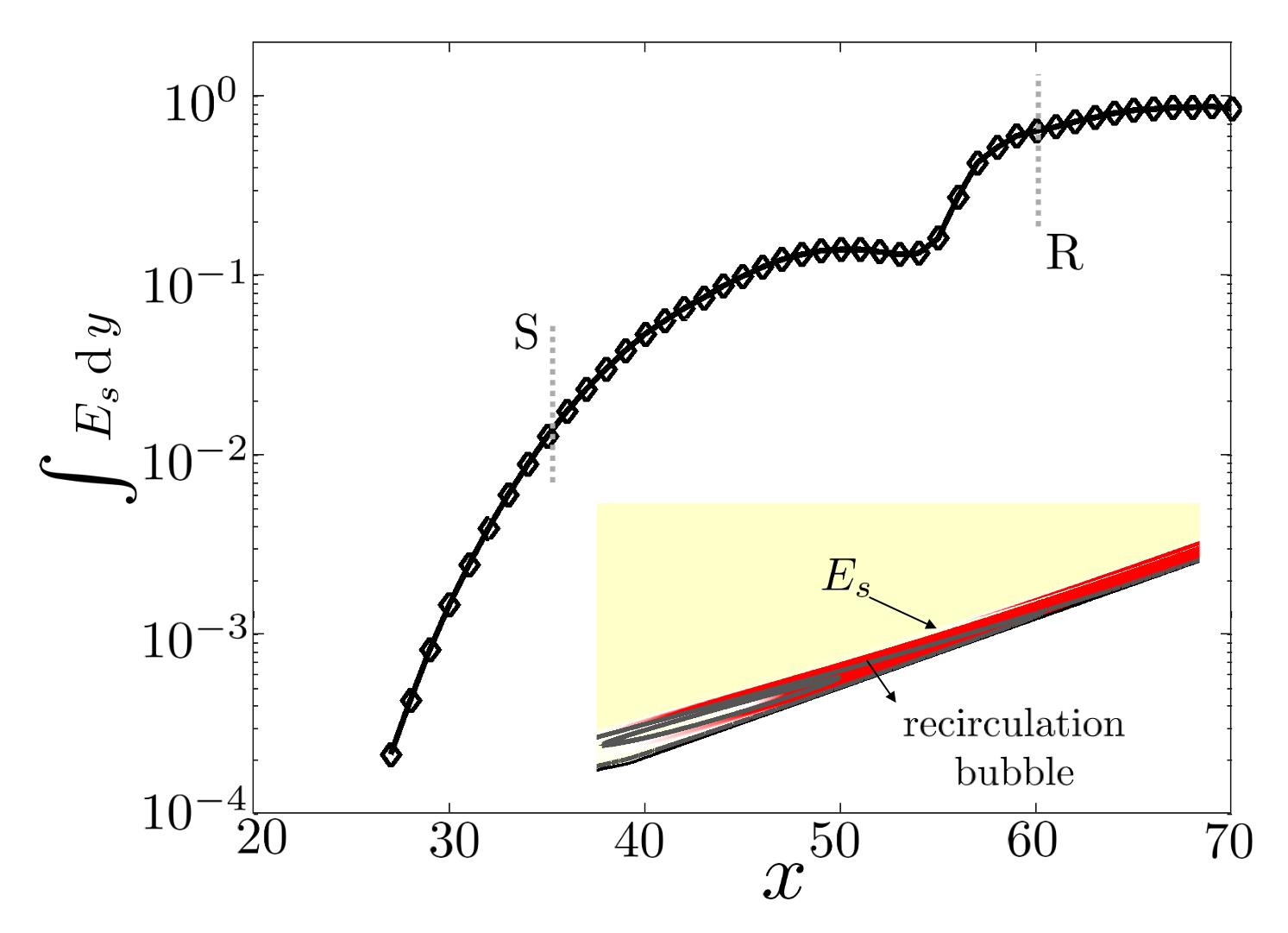}
           \label{fig.quad1}}}
    \end{tabular}
    &    
   { \begin{tabular}{c}
    \begin{tabular}{c}
    {
   \hspace*{-0.65cm}
    \subcaptionbox{}
    {\includegraphics[height= 4 cm, trim= 4 4 4 4, clip]{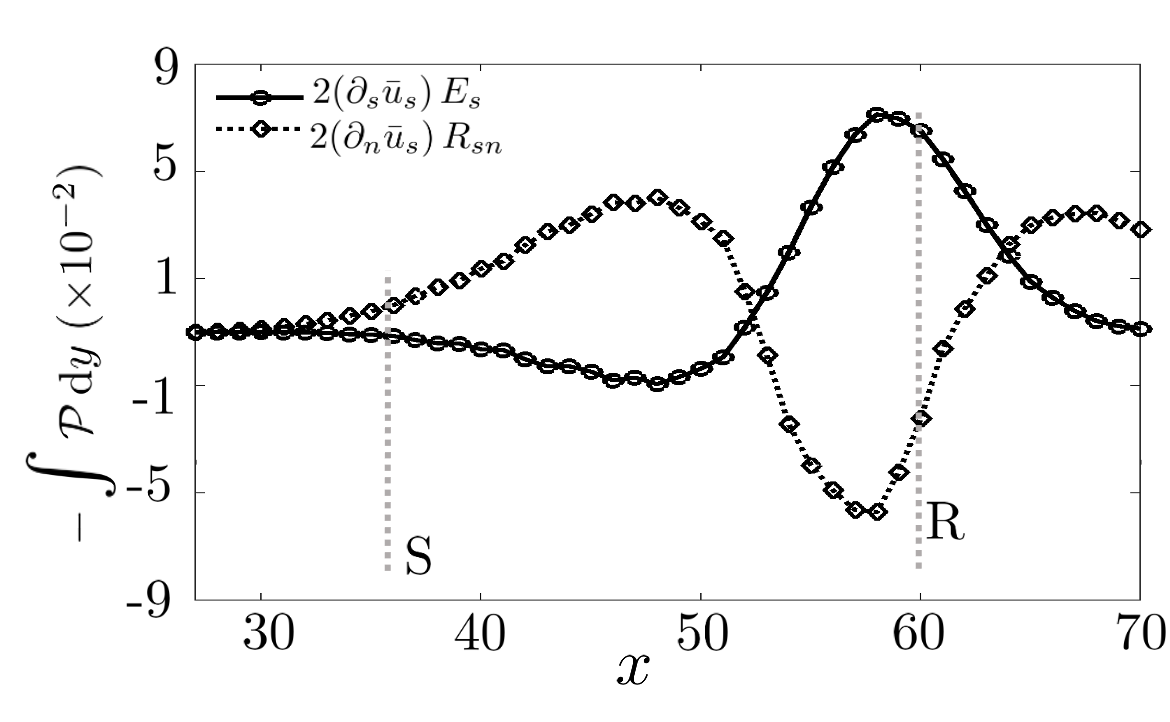}
           \label{fig.quad2}}}
    \end{tabular}
    \\
    \begin{tabular}{c}
    {
   \hspace*{-0.65cm}
    \subcaptionbox{}
    {\includegraphics[height= 3 cm, trim= 4 4 4 4, clip]{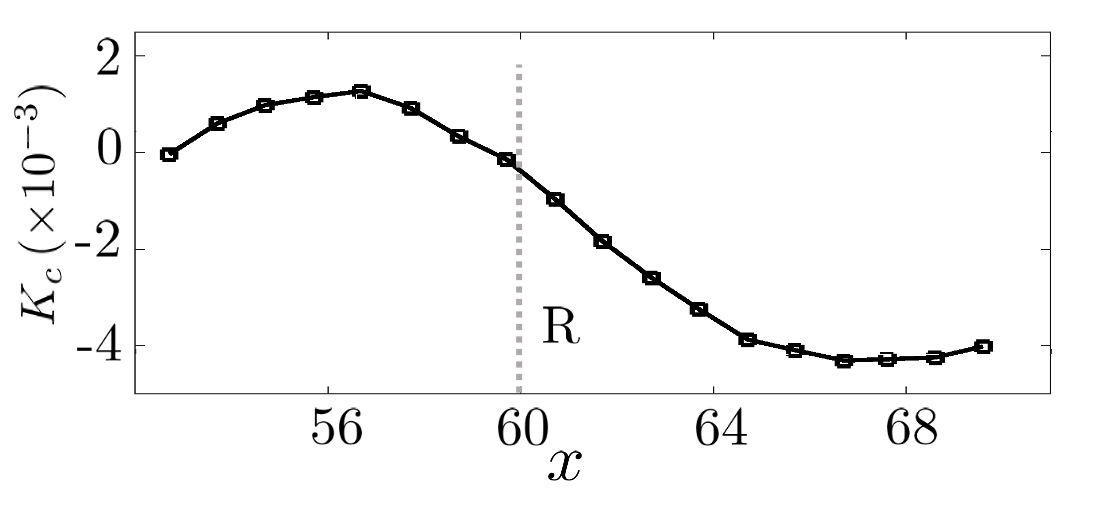}
           \label{fig.quad2}}}
    \end{tabular}
    \end{tabular}}
    \end{tabular}
    }    
   \caption{(a) Streamwise evolution of the wall-normal integral of $E_{s} = \langle {u^\prime_{s} u^{\prime}_s} \rangle$ for the second output mode associated with streaks at ($\omega = 0, \lambda_z = 1.5$) along with contours of $E_{s}$ near reattachment (inset); (b) streamwise variation of deceleration and shear components of the spanwise averaged production term $\langle \mathcal{P} \rangle$ in the transport equation for $E_{s}$; and (c) base flow curvature $K_{c}$ near reattachment.}
    \label{fig:budget}
    \end{figure}
		
Figure~\ref{fig:budget}(a) shows the streamwise evolution of the wall-normal integral of $E_{s}$ for the mode \tc{black}{$\bphi_{2}$} associated with \tc{black}{$\mathcal{O} (\epsilon^{2})$ streaks} with ($\omega = 0,\lambda_z = 1.5$). In contrast to the oblique waves  (cf.\ figure~\ref{fig:obl_spring}(a)), \tc{black}{which experience monotonic amplification throughout the separation shear layer above the recirculation bubble, we observe a non-monotonic $x$-dependence of $E_s$ for the streaks within the bubble.} To gain \tc{black}{physical} insight, we evaluate the \tc{black}{terms of the coefficient matrix in equation~\eqref{eq:Es-Rns} for steady streaks.} Figure~\ref{fig:budget}(b) shows that both streamwise deceleration (i.e., $\partial_{s} \bar{u}_{s} < 0$) and shear $\partial_{n} \bar{u}$ contribute to amplification of \tc{black}{$\mathcal{O} (\epsilon^{2})$ streaks} in different regions of the separated laminar flow. \tc{black}{Towards reattachment,} the base flow curvature $K_c$ is positive inside the recirculation bubble (cf.\ figure~\ref{fig:budget}(c)) and the streamwise deceleration term (i.e., $\partial_{s} \bar{u}_{s} < 0$) is primarily responsible for the energy $E_{s}$ amplification. \tc{black}{In this region, the transport equation for $E_{s}$ can be approximated by,}
\begin{equation}
\label{eq:Es-bub-strk}
	\tc{black}{
	\bar{u}_{s} \frac{\partial E_{s}}{\partial s}
	\; \approx \; 
	- 
	2 \, ( \partial_s \bar{u}_{s} )
	{E}_{s}.}
\end{equation}
This is in concert with the analysis of the contribution of the first (i.e., most amplified) output mode \tc{black}{$\bphi_{1}$} to the amplification of streaks in a compression ramp flow~\citep{dwisidniccanjovJFM19}, which also revealed dominance of streamwise deceleration near reattachment. On the other hand, in the post-reattachment region, \tc{black}{the shear term $\partial_{n} \bar{u}_{s} < 0$ dominates and the coupled system of equations~\eqref{eq:Es-Rns} for $E_s$ and $R_{sn}$ simplifies to the following second order equation~\eqref{eq:Es2b} for $E_s$,}
\begin{equation}
	\tc{black}{
	\bar{u}^2_{s} \, \frac{\partial^2 {E}_{s}}{\partial s^2}
	\; + \;
	4 K_c 
	( \partial_n \bar{u}_s )
	{E}_{s}
	\; \approx \; 
	0.}
	\label{eq:Es2b-strk}
\end{equation}
\tc{black}{Thus, in the post reattachment region, the concave streamline curvature of the laminar 2D base flow (i.e. $K_c < 0$) is primarily responsible for amplification of $\mathcal{O} (\epsilon^{2})$ streaks.}	

	\vspace*{-2ex}
\section{\tc{black}{Direct numerical simulations of streak breakdown}}
\label{sec:secinstab}
	
	Weakly nonlinear analysis demonstrates that small unsteady disturbances induce steady streaks in a hypersonic flow over a double-wedge. These streaks result from quadratic interactions of oblique waves and they undergo rapid amplification near reattachment. Downstream of reattachment, the streaks grow large enough to modify the time-averaged 2D flow \tc{black}{and we} utilize DNS to examine breakdown of streaks. \tc{black}{We also report instantaneous and statistical properties of the flow as it transitions to turbulence.} 
	
\vspace*{-1ex}
\subsection{\tc{black}{Numerical setup}}

\tc{black}{To study the onset of transition, we extend computational domain in the streamwise direction downstream of reattachment. We place the  inflow boundary at $x=20$ and, at this location, we interpolate the inflow profile that results from 2D base flow computations. To avoid spurious numerical errors we utilize a non-reflecting numerical sponge near the outflow. The wall is assumed to be adiabatic and periodic boundary conditions in the spanwise direction are applied.} 

\tc{black}{Computational domain of size $80 \times 13 \times 9$ in the streamwise, wall-normal, and spanwise directions is discretized using $900 \times 249 \times 384$ grid points (i.e., $86$ million cells). The grid is constructed to ensure uniform spacing in the streamwise and spanwise directions and, in the normal direction, the mesh is stretched to ensure $y^{+} < 0.22$ at the wall. {In appendix~\ref{app.grid}, we provide a grid convergence study and compare numerical resolution that we use with recent simulations of supersonic and hypersonic transitional and turbulent flows.} Our simulations utilize} low-dissipation sixth-order spatially accurate kinetic-energy-consistent (KEC) fluxes~\citep{subbareddy2009fully} for spatial discretization. The low-dissipation KEC fluxes were previously employed in high-fidelity simulations of transitional and turbulent hypersonic boundary layers~\citep{subbareddy_bartkowicz_candler_2014,subbareddy2012dns}. The time integration is carried out using the explicit third-order Runge-Kutta scheme with CFL number $0.5$. 

\vspace*{-1ex}
\subsection{\tc{black}{Secondary instability and breakdown}}
\begin{figure}
    \centering
    {
    \begin{tabular}{ccc}
    \begin{tabular}{c}
    {
   \hspace*{-0.5cm}
    \subcaptionbox{}
    {\includegraphics[height= 2.5 cm, trim= 5 5 5 5, clip]{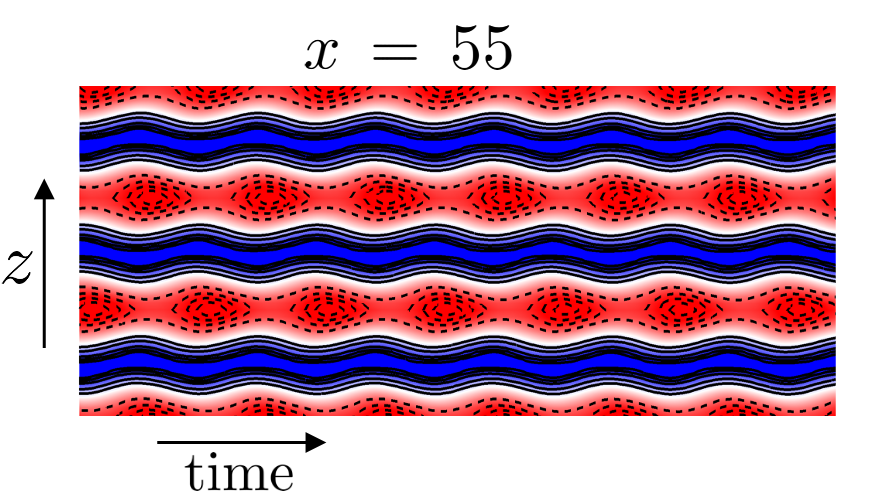}
           \label{fig.quad1}}}
    \end{tabular}
    &    
    \begin{tabular}{c}
    {
   \hspace*{-0.5cm}
    \subcaptionbox{}
    {\includegraphics[height= 2.5 cm, trim= 5 5 5 5, clip]{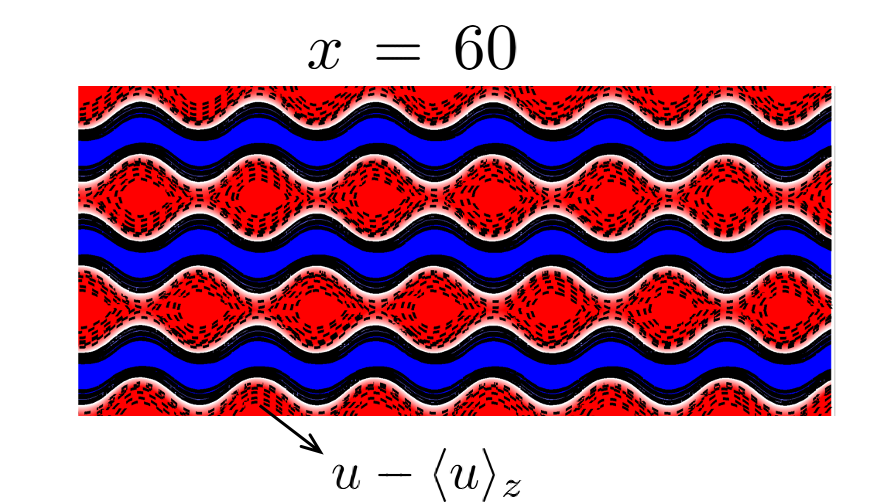}
           \label{fig.quad2}}}
    \end{tabular}
    &    
    \begin{tabular}{c}
    {
   \hspace*{-0.5cm}
    \subcaptionbox{}
    {\includegraphics[height= 2.5 cm, trim= 5 5 5 5, clip]{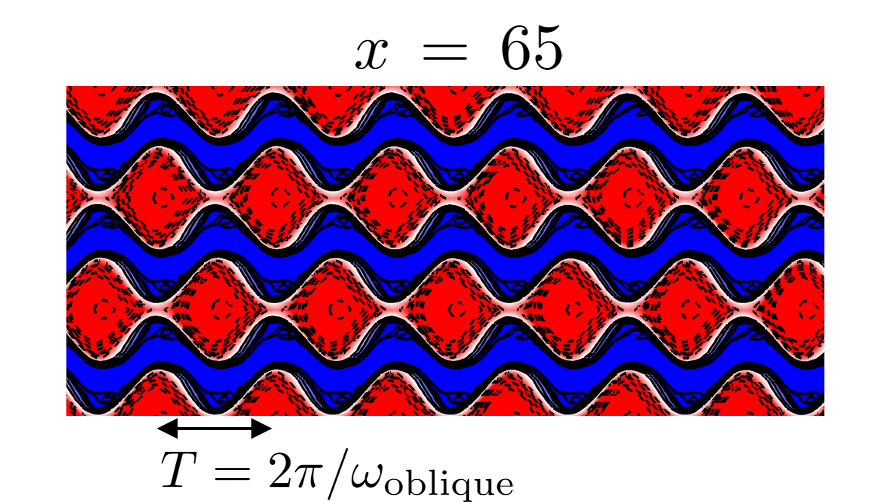}
           \label{fig.quad2}}}
    \end{tabular}
    \end{tabular}
    }    
    \caption{Streamwise streaks in the plane close to the wall, $\eta = 0.25$, at three different streamwise locations: (a) before reattachment, at $x = 55$; (b) at reattachment, at $x = 60$; and (c) after reattachment, at $x=65$. The low-speed streaks are marked in blue color (solid lines) and high-speed streaks are marked in red color (dashed lines).}
    \label{fig:sec_instab1}
    \end{figure}

\begin{figure}
    \centering
    {
    \begin{tabular}{ccc}
    \begin{tabular}{c}
    {
   \hspace*{-0.5cm}
    \subcaptionbox{}
    {\includegraphics[height= 2.35 cm, trim= 5 5 5 5, clip]{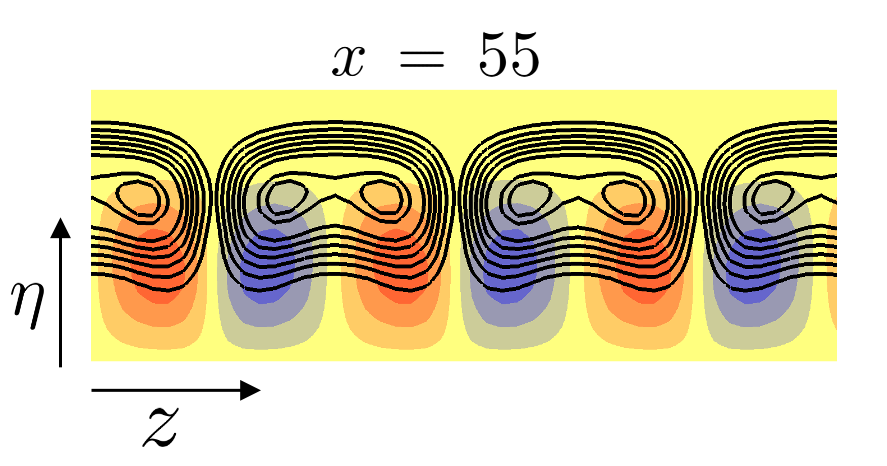}
           \label{fig.quad1}}}
    \end{tabular}
    &    
    \begin{tabular}{c}
    {
   \hspace*{-0.5cm}
    \subcaptionbox{}
    {\includegraphics[height= 2.35 cm, trim= 5 5 5 5, clip]{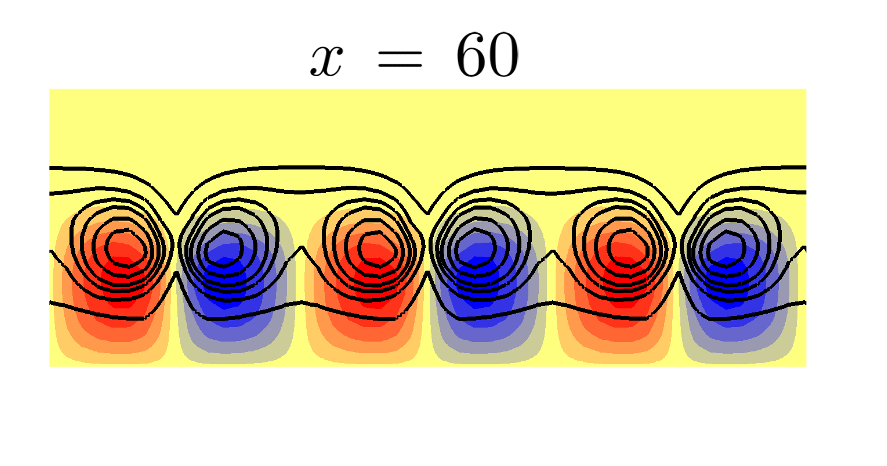}
           \label{fig.quad2}}}
    \end{tabular}
    &    
    \begin{tabular}{c}
    {
   \hspace*{-0.5cm}
    \subcaptionbox{}
    {\includegraphics[height= 2.35 cm, trim= 5 5 5 5, clip]{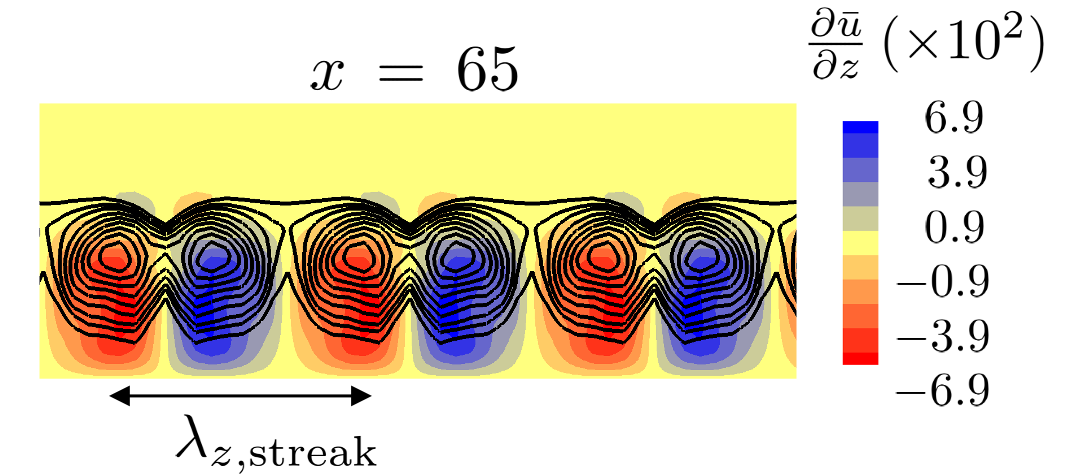}
           \label{fig.quad2}}}
    \end{tabular}
    \end{tabular}
    }    
   \caption{Color plots of the spanwise gradient of time-averaged streamwise velocity ${\partial\bar{u}}/{\partial z}$ and contour lines of $u^{\prime}_\mathrm{rms}$ at the same streamwise locations as in figure~\ref{fig:sec_instab1}.}
    \label{fig:sec_instab2}
    \end{figure}

\tc{black}{To simulate breakdown of streaks we excite the laminar 2D flow using the oblique input modes with ($\omega = \pm 0.4, \lambda_z=3$).} The disturbance amplitude is set to \tc{black}{$A_\mathrm{ob} = 2.50 \times 10^{2} {A}_{0}$} and is determined based on the results reported in figure~\ref{fig:quadratic}. With this amplitude, fluctuations grow linearly in the recirculation region but saturate nonlinearly post-reattachment (i.e., beyond $x=60$). 

In figure~\ref{fig:sec_instab1}, we show time evolution of streaks in the plane close to the wall, at $\eta = 0.25$, for three different values of $x$. These plots demonstrate that, in the presence of unsteady oblique disturbances, \tc{black}{$\mathcal{O}(\epsilon^{2})$} steady streaks undergo spanwise motion close to reattachment. The identified spanwise oscillations become stronger as we progress downstream and their period corresponds to the time period of oblique wave inputs. We note the presence of a ``sinuous-subharmonic'' motion, where two adjacent low-speed streaks oscillate out of phase, and observe that  the amplitude of spanwise oscillations almost doubles as we move from $x=55$ to $x = 65$.

In figure~\ref{fig:sec_instab2}, we illustrate the effect of streak oscillations on the mean flow by visualizing the root-mean-square of temporal fluctuations in the streamwise velocity.  At different streamwise locations, fluctuations are plotted against $\partial \bar{u}/\partial z$, which characterizes steady spanwise variations of the mean flow. Initially, unsteadiness is restricted to the oblique wave pair, which is located further away from the wall relative to streaks. However, at $x=65$, unsteadiness is observed in the region of the largest spanwise shear because of the presence of streaks. This ``locking-in'' of $u_\mathrm{rms}$ with streak oscillations identifies late stages of the streak evolution just before smaller-scales \mbox{(associated with higher frequencies) set in.} 

\begin{figure}
    \centering
    {
    \begin{tabular}{cc}
    \begin{tabular}{c}
    {
   \hspace*{-0.5cm}
    \subcaptionbox{}
    {\includegraphics[height= 3 cm, trim= 4 4 4 4, clip]{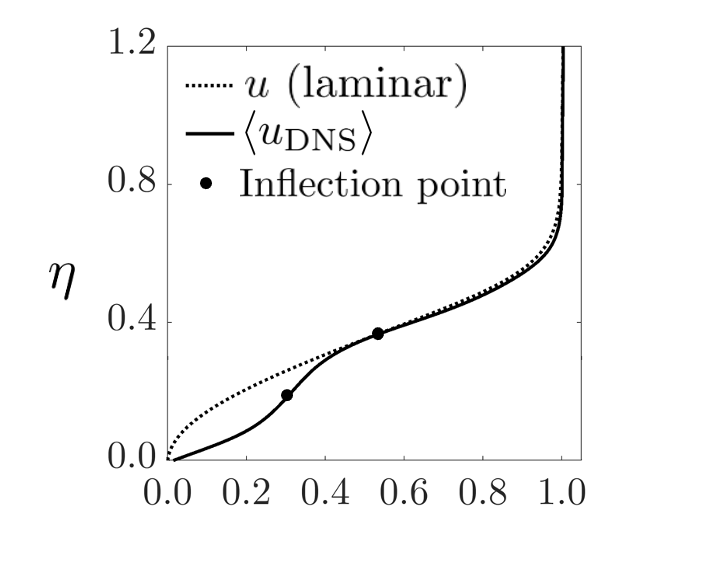}
           \label{fig.quad1}}}
    \end{tabular}
    &    
    \begin{tabular}{c}
    {
   \hspace*{-0.5cm}
    \subcaptionbox{}
    {\includegraphics[height= 3 cm, trim= 5 5 5 5, clip]{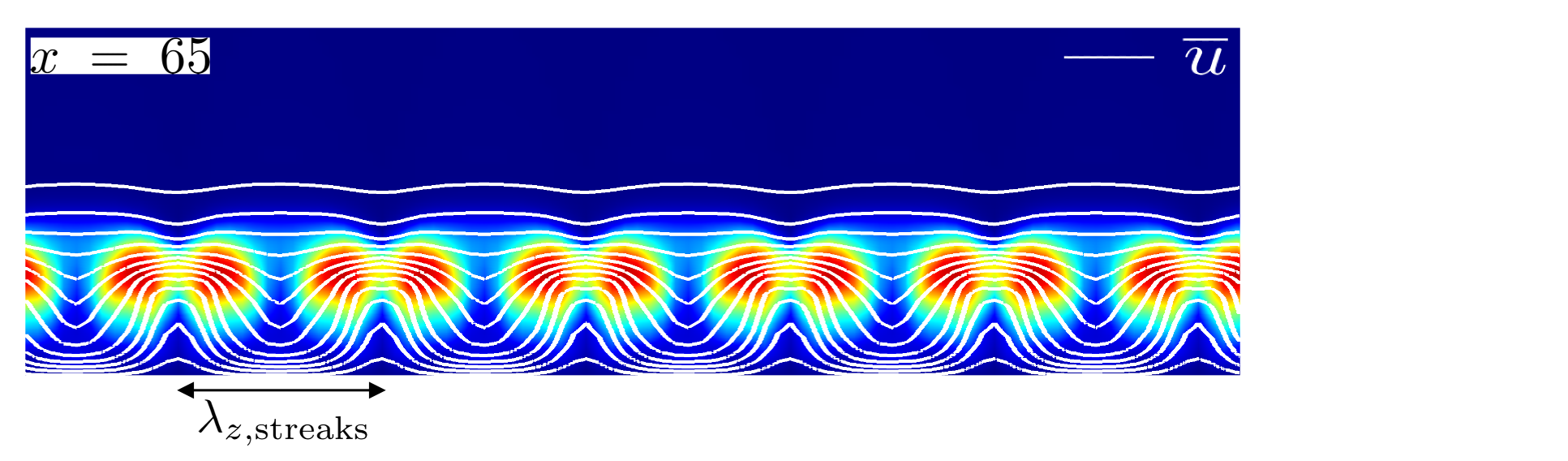}
           \label{fig.quad2}}}
    \end{tabular}
    \\    
    \begin{tabular}{c}
    {
   \hspace*{-0.5cm}
    \subcaptionbox{}
    {\includegraphics[height= 3 cm, trim= 4 4 4 4, clip]{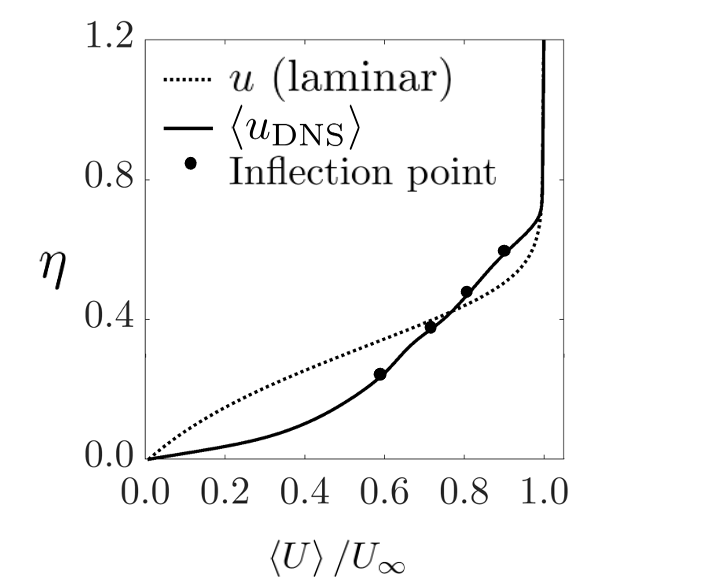}
           \label{fig.quad2}}}
    \end{tabular}
    &    
    \begin{tabular}{c}
    {
   \hspace*{-0.5cm}
    \subcaptionbox{}
    {\includegraphics[height= 3 cm, trim= 5 5 5 5, clip]{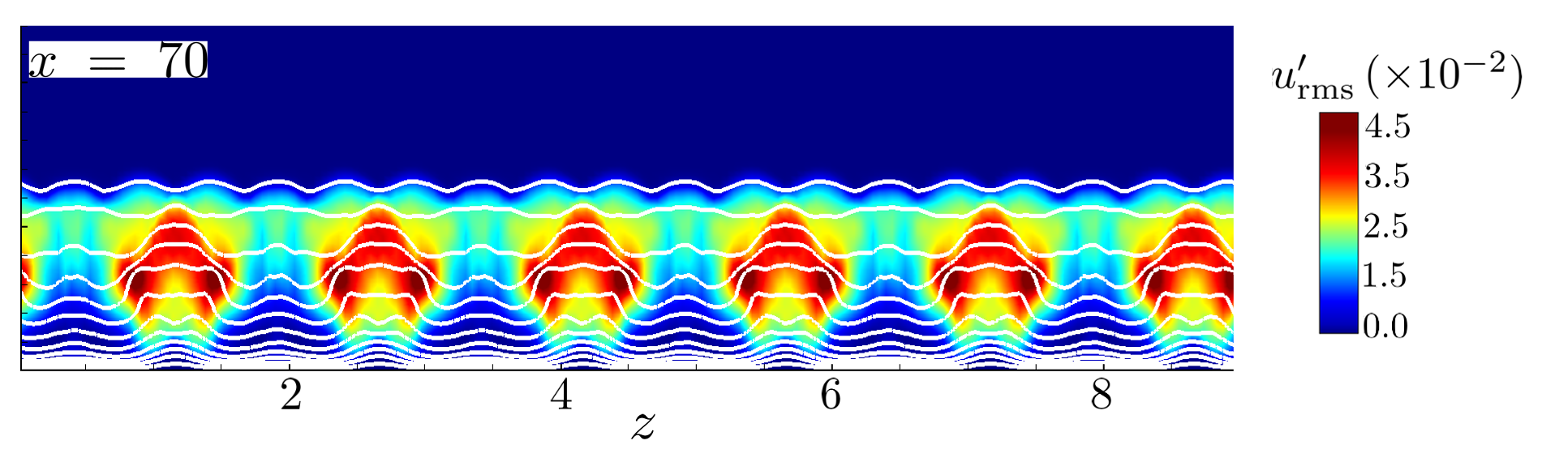}
           \label{fig.quad2}}}
    \end{tabular}
    \end{tabular}
    }    
   \caption{(a) Wall-normal profiles of laminar and mean streamwise velocity components at $x = 70$. (b) Color plots of $u^{\prime}_\mathrm{rms}$ along with contour lines of time-averaged streamwise velocity.}
    \label{fig:u_inflex}
    \end{figure}

	Amplification of streaks and their unsteadiness induce rapid steepening of spanwise and wall-normal mean flow gradients, thereby leading to the emergence of inflection points in the mean (time- and spanwise-averaged) flow. Figure~\ref{fig:u_inflex}(a) shows the resulting inflectional mean flow profile and figure~\ref{fig:u_inflex}(b) shows the location of unsteady fluctuations with respect to the streaks. As we move from $x=65$ to $x=70$, we note the appearance of higher spanwise wavenumbers in the streaks as well as in the unsteady fluctuations. As discussed by~\citet{hall_horseman_1991, yu1994mechanism}, inflectional points in the mean velocity serve as an indicator of its susceptibility to the growth of broadband fluctuations. Amplification of high-frequency harmonics is also observed in the temporal spectra of the fluctuations. Therefore, at $x=70$, flow is well within fully nonlinear stages of transition. We also note strong spatial correlation of unsteady fluctuations with the spanwise shear associated with the streaks, even at this nonlinear stage. 

\begin{figure}
    \centering
    {
    \begin{tabular}{cc}
    \begin{tabular}{c}
    {
   \hspace*{-0.5cm}
    \subcaptionbox{}
    {\includegraphics[height= 5.5 cm, trim= 5 7 5 5, clip]{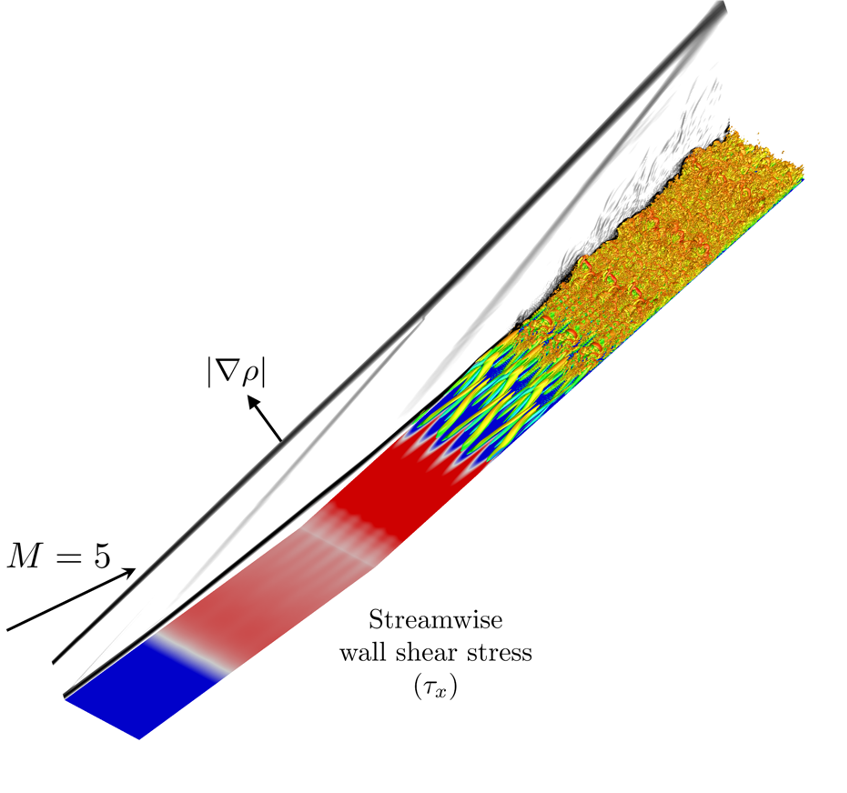}
           \label{fig.quad1}}}
    \end{tabular}
    &    
    \begin{tabular}{c}
    {
   \hspace*{-0.5cm}
    \subcaptionbox{}
    {\includegraphics[height= 5.5 cm]{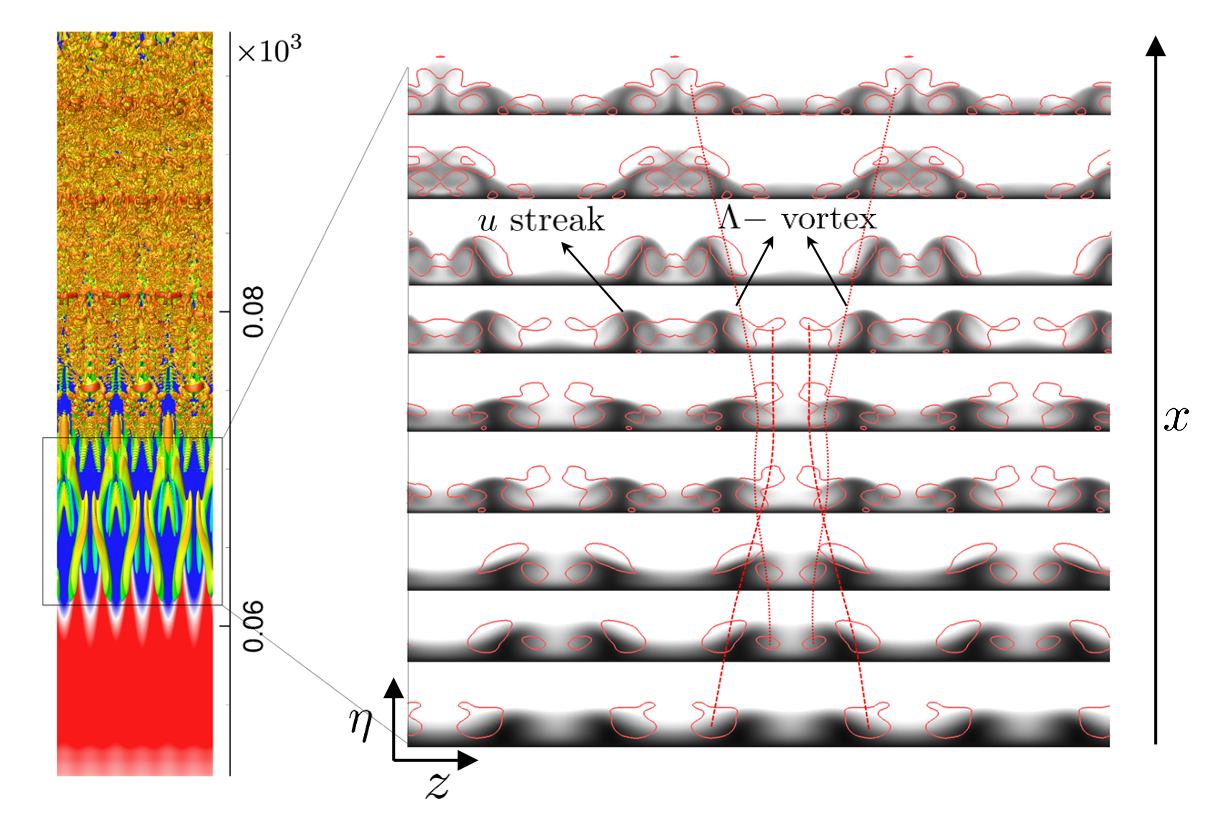}
           \label{fig.quad2}}}
    \end{tabular}
    \end{tabular}
    }    
    \caption{(a) Instantaneous isosurface of the $Q$-criterion {colored by contours of the streamwise velocity;} (b) contour plot of the wall-shear stress show the formation of streaks before transition and $(\eta,z)$ slices of instantaneous streamwise velocity contours along with contour lines of the $Q$-criteria.}
    \label{fig:Qiso}
    \end{figure}

	In figure~\ref{fig:Qiso}, we plot $Q$-isosurfaces of the instantaneous flow-field. These visualize vortical structures that arise from sinuous-subharmonic motion of the streaks, prior to breakdown of the flow. We note the appearance of staggered lambda vortices, similar to the ones identified in transition induced by oblique waves in incompressible boundary layers~\citep{berlin1999numerical}. These vortical structures are also sometimes referred to as $X$-vortices, for reasons illustrated in the ($\eta,z$) slice array plot in figure~\ref{fig:Qiso}. Initially, at $x \approx 62$, we observe that the interaction with the oblique waves causes the roll-up of the low-speed streak as they come together due to sinuous-subharmonic motion. Further downstream, as the flow structures associated with the $X$-vortices spread apart, we see that wall normal jets of low-speed flow form mushroom structures. As these low-speed regions oscillate further, they interact to generate fluctuations with higher spanwise wavenumbers. Finally, at $x\approx70$, the shear introduced by the upward jets of low-speed streaks becomes significant enough to cause a local Kelvin-Helmholtz instability that induces streamwise rollers~\citep{reddy_schmid_baggett_henningson_1998}. At this stage, the laminar boundary layer flow has started to break down to turbulence.

	\vspace*{-2ex}
\section{Transition to turbulence}
\label{sec:transition}
Our DNS study demonstrates that amplification of steady streaks, that result from quadratic interactions of oblique waves, leads to a formation of a 3D inflectional boundary layer profile. During this process, fluctuations with multiple spatial and temporal scales develop \tc{black}{and, despite periodic nature of upstream forcing, the flow downstream of reattachment becomes turbulent. In this section, we utilize wall and boundary layer statistics to illustrate the onset of turbulence on the second wedge.} 

	\vspace*{-1ex}
\subsection{\tc{black}{Wall statistics: skin friction}}

\begin{figure}
    \centering
    {
    \begin{tabular}{cc}
    {\begin{tabular}{c}
    \begin{tabular}{c}
    {
   \hspace*{-0.55cm}
    \subcaptionbox{}
    {\includegraphics[height= 2.65 cm, trim= 4 4 4 4, clip]{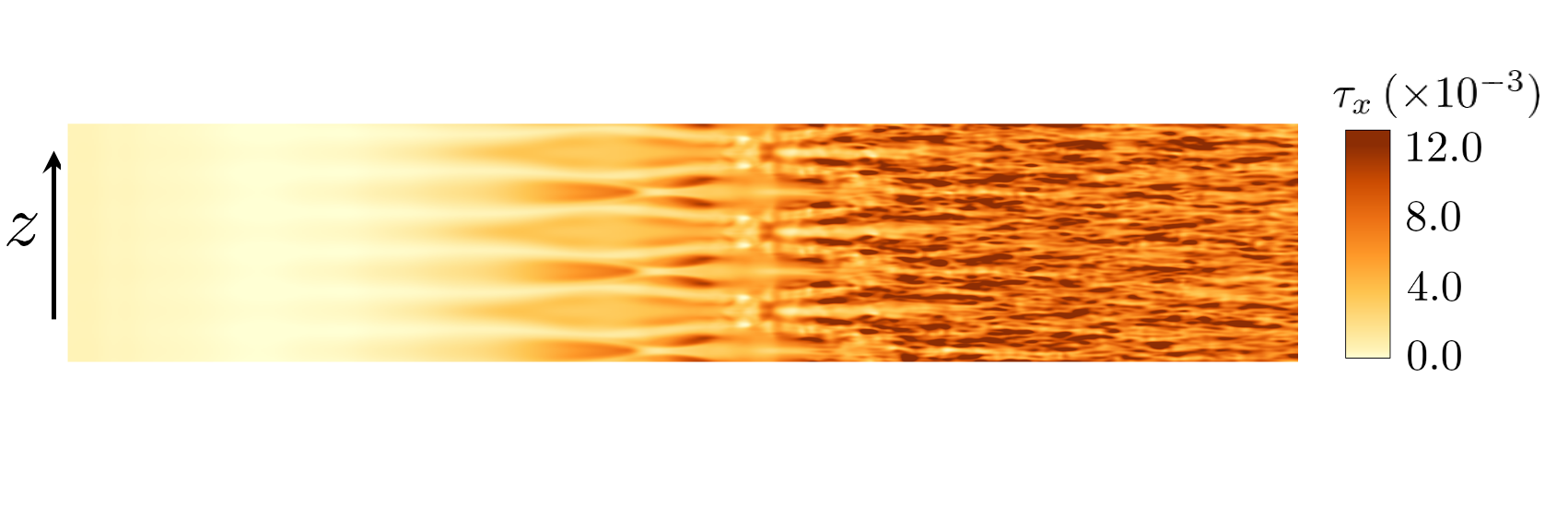}
           \label{fig.quad1}}}
    \end{tabular}
    \\
      \begin{tabular}{c}
    {
   \hspace*{-0.65cm}
    \subcaptionbox{}
    {\includegraphics[height= 2.65 cm, trim= 5 5 5 5, clip]{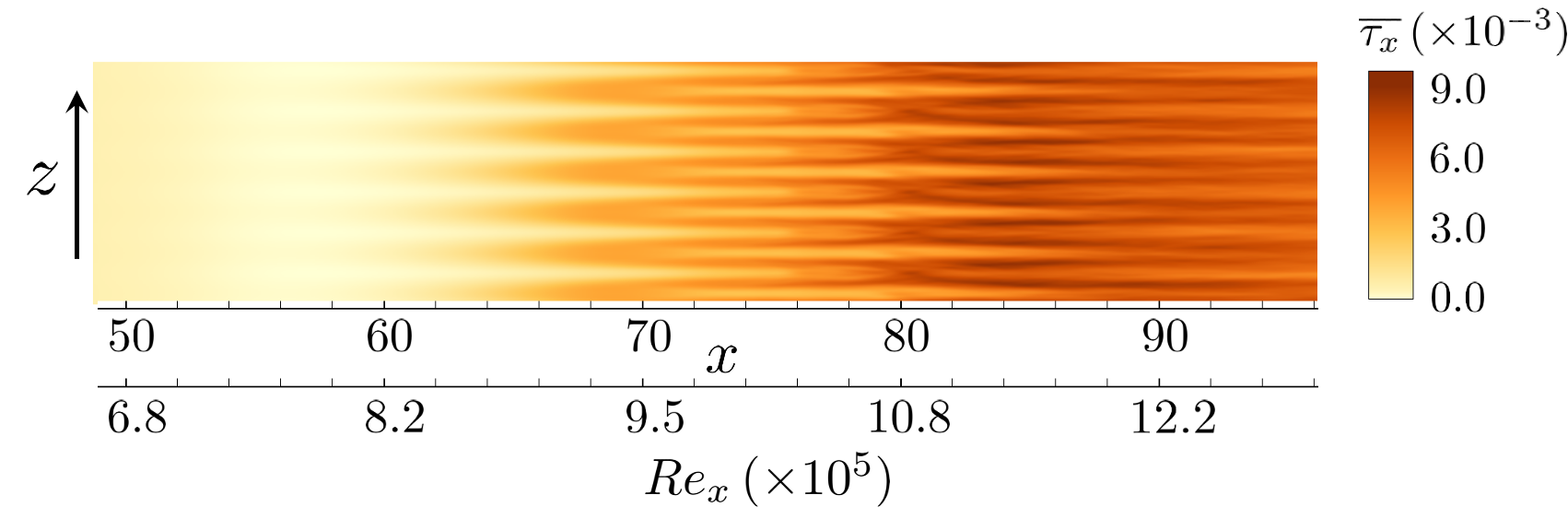}
           \label{fig.quad2}}}
    \end{tabular}
    \end{tabular}}
    &    
   {\begin{tabular}{c}
    {
   \hspace*{-0.65cm}
    \subcaptionbox{}
    {\includegraphics[height= 4.75 cm, trim= 4 4 4 4, clip]{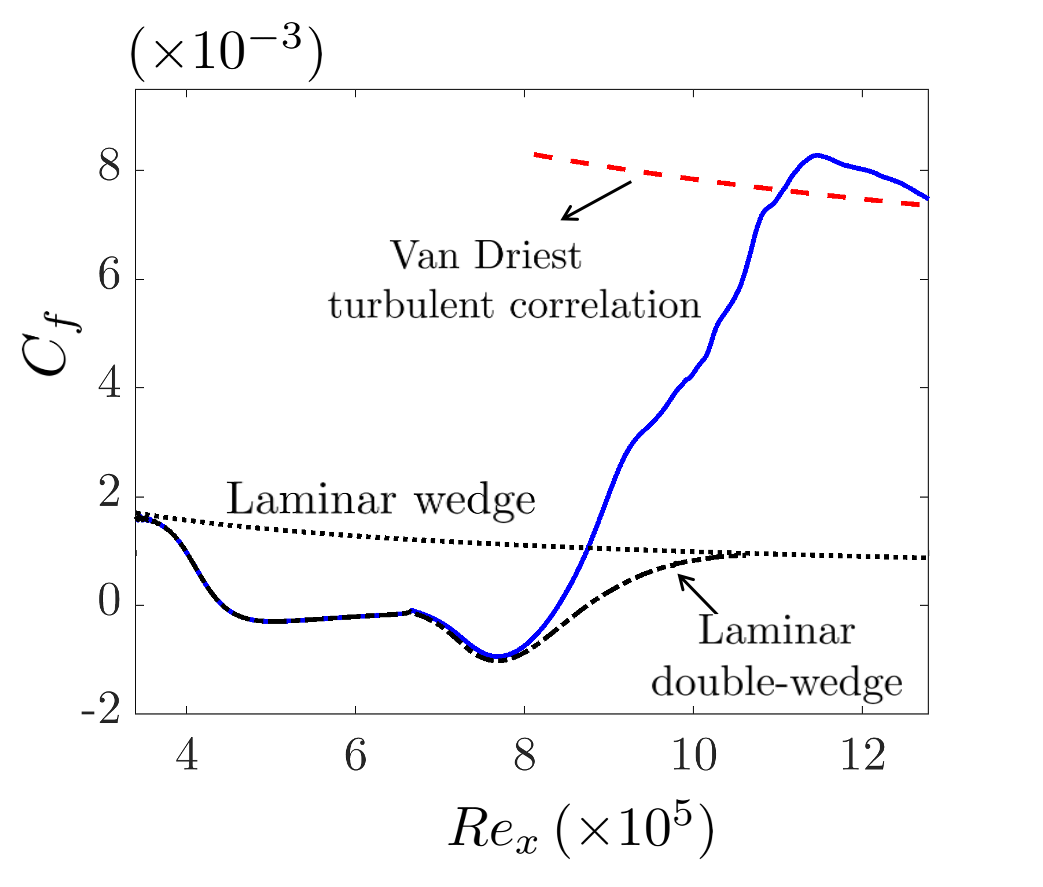}
           \label{fig.quad2}}}
    \end{tabular}}
    \end{tabular}
    }    
  \caption{Streamwise variations of (a) instantaneous and (b) time-averaged streamwise shear stresses as well as (c) wall skin friction coefficient.}
    \label{fig:meanCf}
    \end{figure}

	Boundary layer transition is characterized by a rapid increase in wall friction. Figure~\ref{fig:meanCf}(a) utilizes instantaneous wall-shear distribution to identify transitional and turbulent regions. The wall-shear stress experiences sinuous-subharmonic modulation downstream of the reattachment (at $x=60$, i.e., $Re_{x} = 8.2 \times 10^{5}$) and finer spanwise scales emerge after $x=80$ (i.e., $Re_{x} \approx 11 \times 10^{5}$). As shown in figure~\ref{fig:meanCf}(b), this location coincides with the highest value of time-averaged wall shear. Even though significant attenuation of spanwise variations of the wall-shear stress associated with the initial streaks takes place by $x = 80$, nonlinear interactions within the transition zone lead to the appearance of new streaks further downstream.

The spatial extent of transition zone is visualized in figure~\ref{fig:meanCf}(c) by showing streamwise development of the skin friction coefficient, 
\begin{align}
    C_{f} 
    \; = \; 
    \frac{1}{T}\frac{1}{L_{z}}\int_{0}^{T} \int_{0}^{L_{z}}
    \frac{\tau^{*}_{w}}{\frac{1}{2} \rho^*_{e} (U_{e}^{*})^2} 
    \; \mathrm{d} z \, \mathrm{d} t.
    \label{eq:meanCf}
\end{align}
Here, $\tau^*_{w}$ is the dimensional shear stress at the wall, $\rho^*_{e}$ and $u^*_{e}$ are the dimensional density and streamwise velocity at the boundary layer edge, $L_{z}$ is the spanwise extent of the computational domain, and $T = 4 L_{x}/u^*_{\infty}$. The values of $C_{f}$ in laminar flows over a $12$ degree wedge and over the double-wedge are also plotted for comparison. The skin friction coefficient drops because of flow separation but it grows again after reattachment. Near and downstream of reattachment, we observe a significant difference between laminar and turbulent values of $C_f$. After $Re_{x} = 11.5 \times 10^{5}$, when $C_f$ starts to decay with $x$, the wall-friction coefficient is about seven times larger than its laminar counterpart. A comparison with the Van-Driest turbulent correlation for standard compressible boundary layers demonstrates that the flow on the second wedge is approaching fully developed turbulent values towards the end of the computational domain. \tc{black}{In addition to skin friction, transition also has a significant impact on wall temperature. Additional details about the mean temperature and the wall statistics are included in appendix~\ref{app.temp}.}

	\vspace*{-1ex} 
\subsection{Towards a turbulent compressible boundary layer}

\begin{figure}
    \centering
    {
    \begin{tabular}{cc}
    \begin{tabular}{c}
    {
   \hspace*{-0.5cm}
    \subcaptionbox{}
    {\includegraphics[height= 5.0 cm, trim= 4 4 4 4, clip]{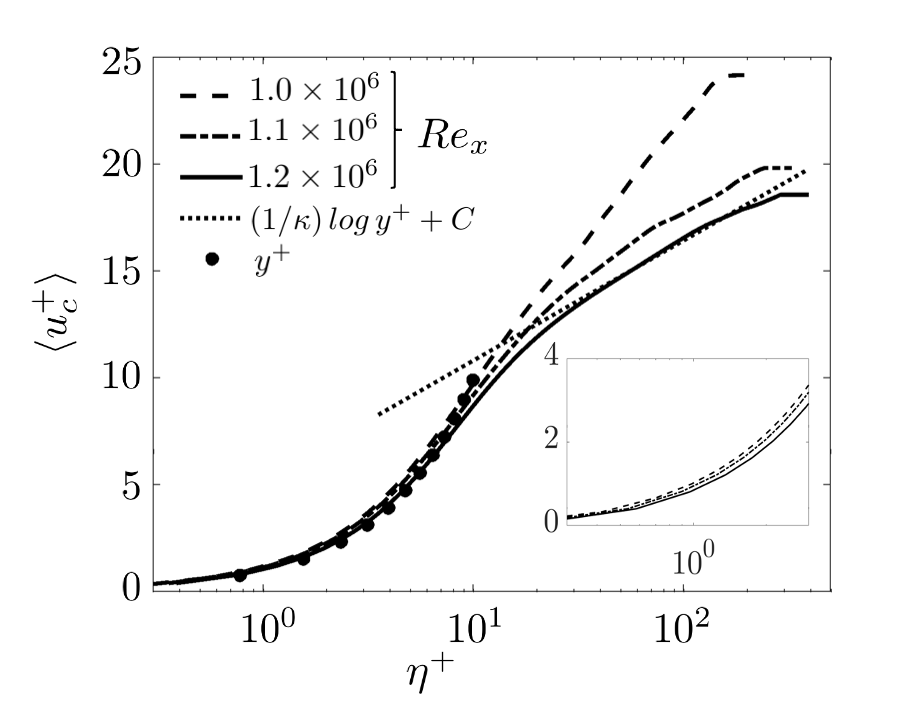}
           \label{fig.quad1}}}
    \end{tabular}
    &    
    \begin{tabular}{c}
    {
   \hspace*{-0.5cm}
    \subcaptionbox{}
    {\includegraphics[height= 5.0 cm, trim= 4 4 4 4, clip]{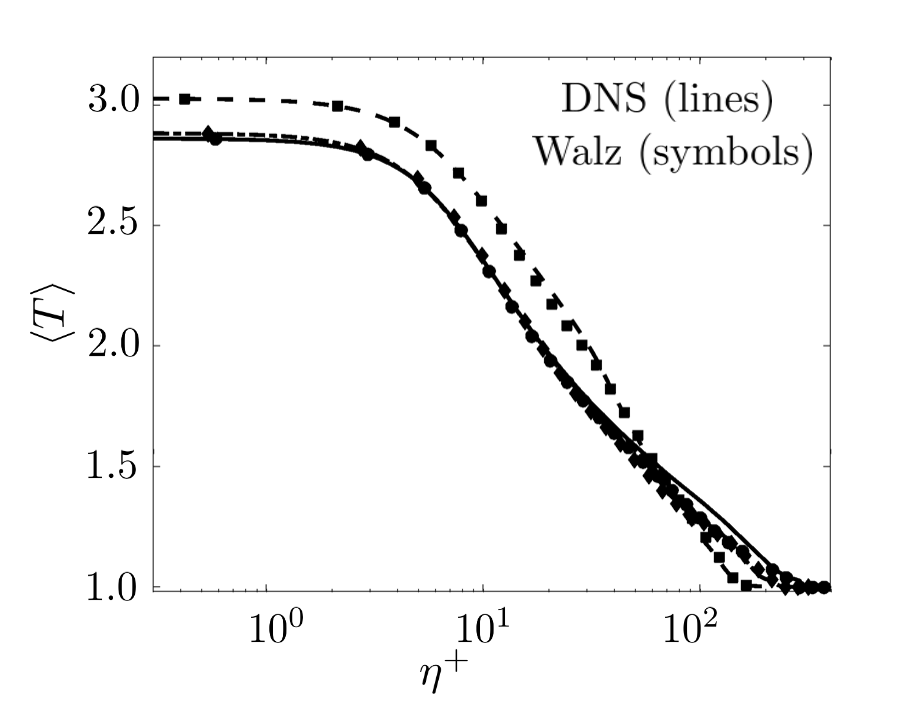}
           \label{fig.quad2}}}
    \end{tabular}
    \end{tabular}
    }    
    \caption{The wall-normal profiles of mean (a) streamwise velocity and (b) temperature.}
    \label{fig:turb_mean}
\end{figure}

To illustrate the onset of turbulence, we also examine first- and second-order statistics in the latter stages of transition. At a given streamwise location, we report statistics in terms of the inner coordinate $\eta^{+}$ which is obtained by non-dimensionalizing the wall-normal coordinate with the local viscous length $\delta_{\nu}$. Figure~\ref{fig:turb_mean}(a) shows the mean streamwise velocity $u^{+}$ non-dimensionalized by the local friction velocity $u_{\tau}$. The mean profile is obtained by averaging in time over $2 L_{x}/u_{\infty}$ and in the spanwise direction over the extent of the computational domain $L_{z}$. In the presence of density variation along the compressible boundary layer, we utilize the Van-Driest transformation,
\begin{align}
	u_{c}(\eta)
	\; = \; 
	\int_{0}^{\braket{u}} 
	\sqrt{\dfrac{\braket{\rho}}{\braket{\rho}_{w}}}
	\; \mathrm{d} \hat{u}
	\; = \; 
	\int_{0}^{\eta} 
	\sqrt{\dfrac{\braket{\rho}}{\braket{\rho}_{w}}} 
	\, 
	\frac{\partial \! \braket{u}}{\partial \hat{\eta}}
	 \, 
	 \mathrm{d}\hat{\eta},
\label{eq:vandriest}    
\end{align}
to compare the transitional velocity profiles with the `log-law' observed in incompressible turbulent boundary layers~\citep{white2006viscous}, where $\braket{\cdot}$ denotes averaging over time and spanwise direction, and $\braket{\rho}_{w}$ is the mean density at the wall. A fully developed turbulent boundary layer is characterized by the presence of the viscous sublayer close to the wall ($\eta^{+} < 10$), where ${u^+_{c}} = \eta^+$, and further away from the wall the mean velocity obeys the log-law,
\begin{align}
	{u^+_{c}} 
	\; = \; 
	(1/\kappa) \operatorname{ln}(\eta^{+}) \; + \; C, 
\label{eq:loglaw}    
\end{align}
where $\kappa = 0.41$ and $C = 5.2$.

	Figure~\ref{fig:turb_mean}(a) shows that the mean velocity in the boundary layer approaches the fully turbulent profile at $Re_{x} = 1.2\times 10^6$. Upstream of this location, within the transition zone, the boundary layer has a significantly greater momentum and most of it lies away from the wall. Furthermore, closer to the wall, the slope of the streamwise velocity profile is larger than the slope obtained using viscous sublayer profile of a fully turbulent flow. This observation is consistent with the overshoot in skin-friction coefficient within the transition zone; cf.\ figure~\ref{fig:meanCf}(b).

	In addition to the mean velocity, the mean temperature profile plays an important role in heat transfer and material response analysis of hypersonic flows. Walz's modified Crocco-Busemann relation, 
\begin{align}
    \frac{\braket{T}}{\braket{T_{e}}} 
    \; = \;
    \frac{\braket{T_{w}}}{\braket{T_{e}}} 
    \; + \;  
    \frac{\braket{T_{r}}-\braket{T_{w}}}{\braket{T_{e}}} \frac{\braket{u}}{\braket{u_{e}}}
    \; + \; 
    \frac{\braket{T_{e}}-\braket{T_{r}}}{\braket{T_{e}}} 
    \left( \frac{\braket{u}}{\braket{u_{e}}} \right)^{2},
    	\label{eq.cb}
\end{align}
is commonly used to describe the relation between temperature and velocity in a zero pressure gradient turbulent boundary layer. Here, $\braket{u_{e}}$ and $\braket{T_{e}}$ denote the mean boundary layer edge velocity and temperature, respectively, $\braket{T_{r}}$ is the mean recovery temperature and, since the wall is adiabatic, the mean wall temperature is determined by $\braket{T_{w}} = \braket{T_{r}}$. In spite of pressure variations post-reattachment, figure~\ref{fig:turb_mean}(b) demonstrates that the quadratic relation in equation~\eqref{eq.cb} holds throughout the transition zone. As the flow approaches a fully turbulent profile, we see a slight deviation from this relation in the outer region away from the wall; this observation is in agreement with prior studies of fully turbulent compressible boundary layers~\citep{duan_beekman_martin_2011,franko2013breakdown}. 

\begin{figure}
    \centering
    {
    \begin{tabular}{cc}
    \begin{tabular}{c}
    {
   \hspace*{-0.5cm}
    \subcaptionbox{}
    {\includegraphics[height= 5.0 cm, trim= 4 4 4 4, clip]{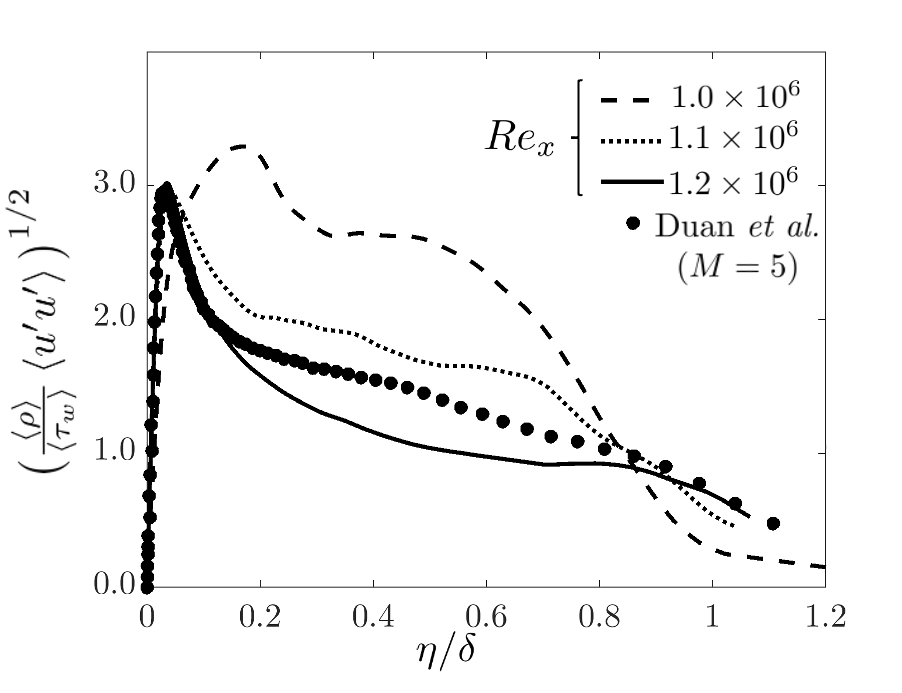}
           \label{fig.quad1}}}
    \end{tabular}
    &    
    \begin{tabular}{c}
    {
   \hspace*{-0.5cm}
    \subcaptionbox{}
    {\includegraphics[height= 5.0 cm, trim= 4 4 4 4, clip]{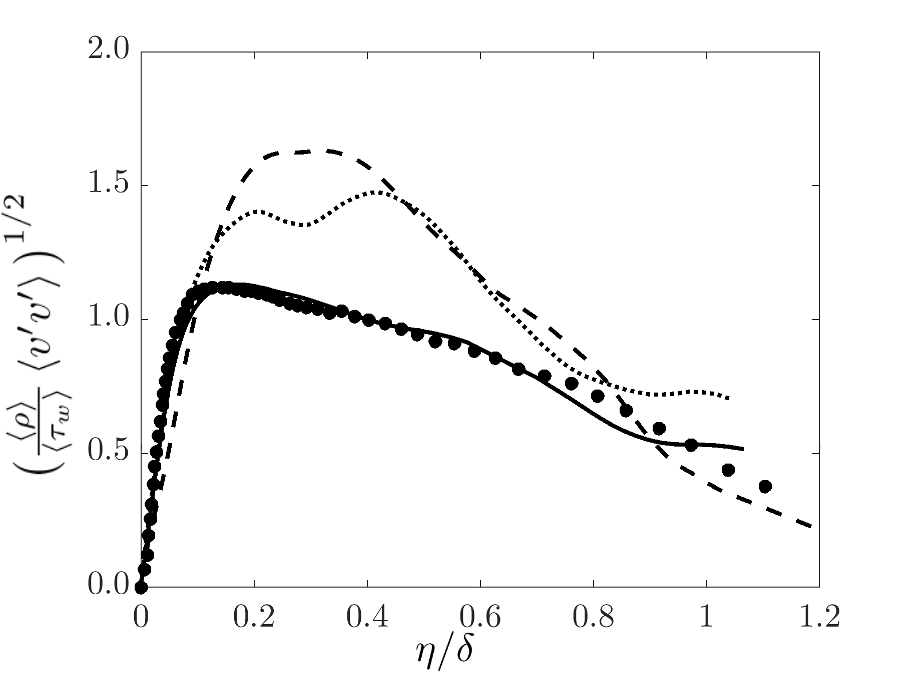}
           \label{fig.quad2}}}
    \end{tabular}
    \\
    \begin{tabular}{c}
    {
   \hspace*{-0.5cm}
    \subcaptionbox{}
    {\includegraphics[height= 5.0 cm, trim= 4 4 4 4, clip]{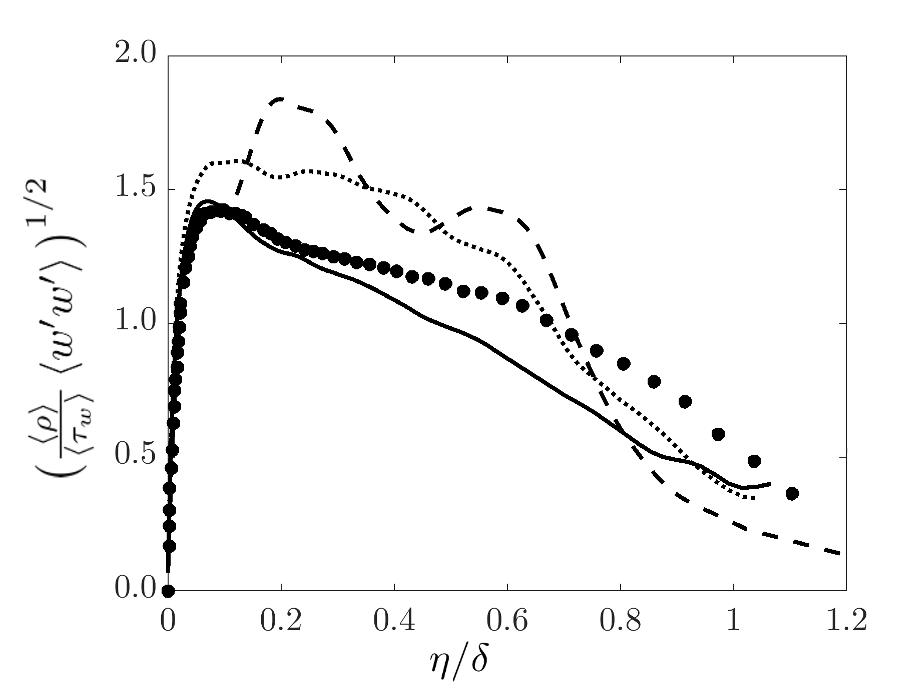}
           \label{fig.quad1}}}
    \end{tabular}
    &    
    \begin{tabular}{c}
    {
   \hspace*{-0.5cm}
    \subcaptionbox{}
    {\includegraphics[height= 5.0 cm, trim= 4 4 4 4, clip]{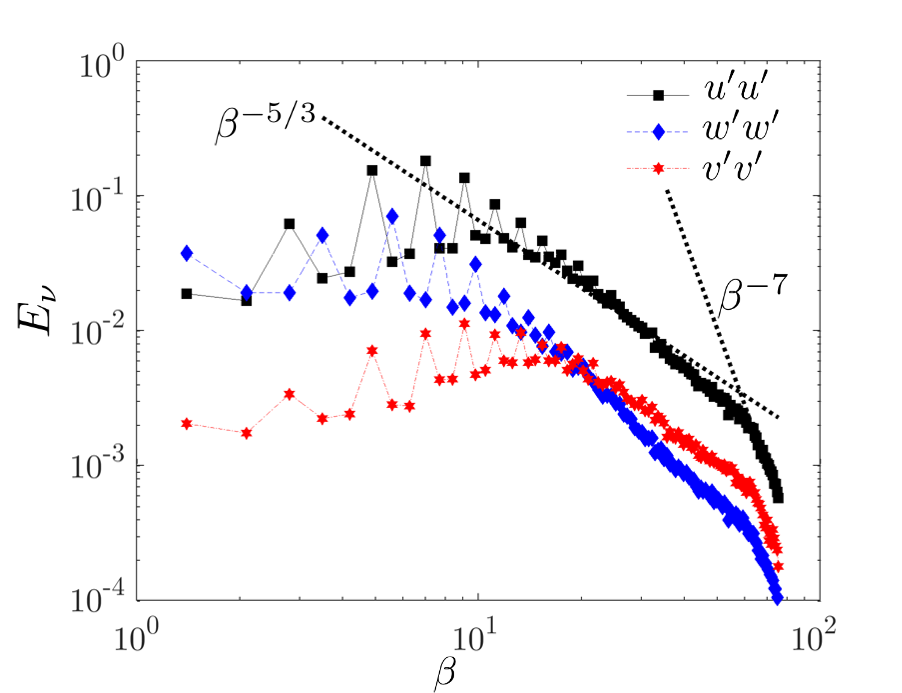}
           \label{fig.quad2}}}
    \end{tabular}
    \end{tabular}
    }    
    \caption{The wall-normal profiles of density weighted RMS values of (a) streamwise; (b) wall-normal; and (c) spanwise velocity fluctuations at different Reynolds numbers. (d) Energy spectrum as a function of spanwise wavenumber $\beta$ at $Re_{x} = 1.2 \times 10^6$.}
    \label{fig:turb_spec}
\end{figure}

We next examine spatial development of flow fluctuations by evaluating second-order statistics and comparing them with those observed in a Mach 5 fully developed turbulent flat plate boundary layer~\citep{duan_beekman_martin_2011}. Figures~\ref{fig:turb_spec}(a)-(c) show the streamwise variation of the density-weighted root-mean-square (RMS) values of the streamwise, wall-normal, and spanwise velocity fluctuation components. In the transition zone, we observe large values of fluctuations away from the wall in all three plots. Further downstream, at $Re_{x} = 1.2 \times 10^6$, the RMS values closer to the wall are in agreement with fully turbulent values. However, away from the wall, the RMS values of $u'$ and $w'$ deviate from the flat plate profiles. We attribute this deviation to the presence of 3D oblique waves and streaks that persist downstream because of continuous upstream forcing in our setup.

\begin{figure}
    \centering
    {
    \begin{tabular}{ccc}
    \begin{tabular}{c}
    {
   \hspace*{-0.5cm}
    \subcaptionbox{}
    {\includegraphics[height= 3.5 cm, trim = 5 5 5 5, clip]{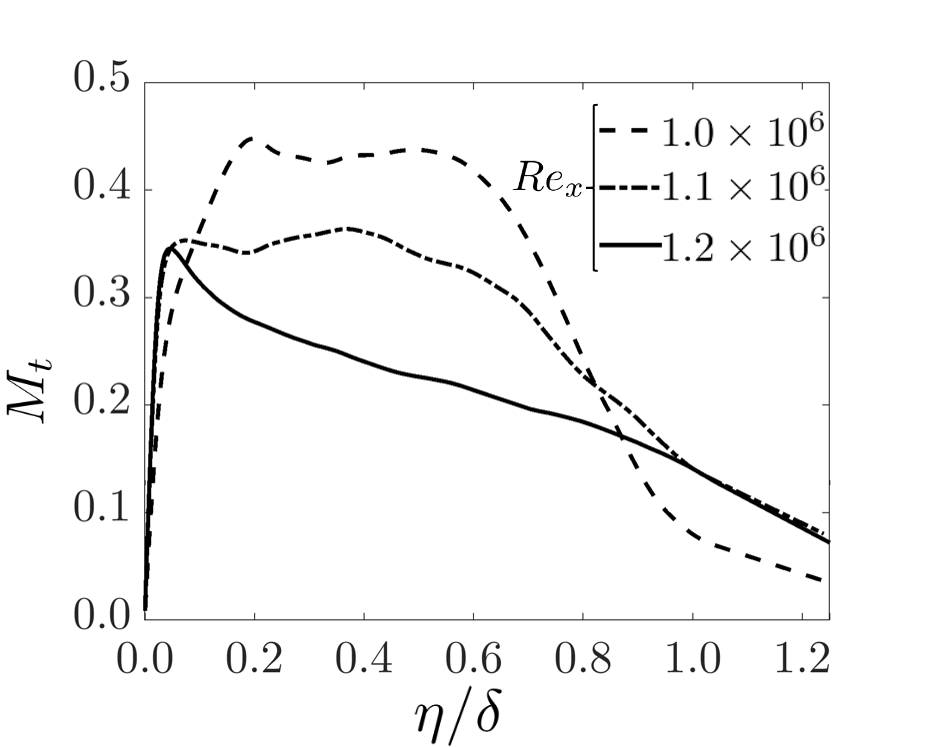}
           \label{fig.quad1}}}
    \end{tabular}
    &    
    \begin{tabular}{c}
    {
   \hspace*{-0.5cm}
    \subcaptionbox{}
    {\includegraphics[height= 3.5 cm, trim = 5 5 5 5, clip]{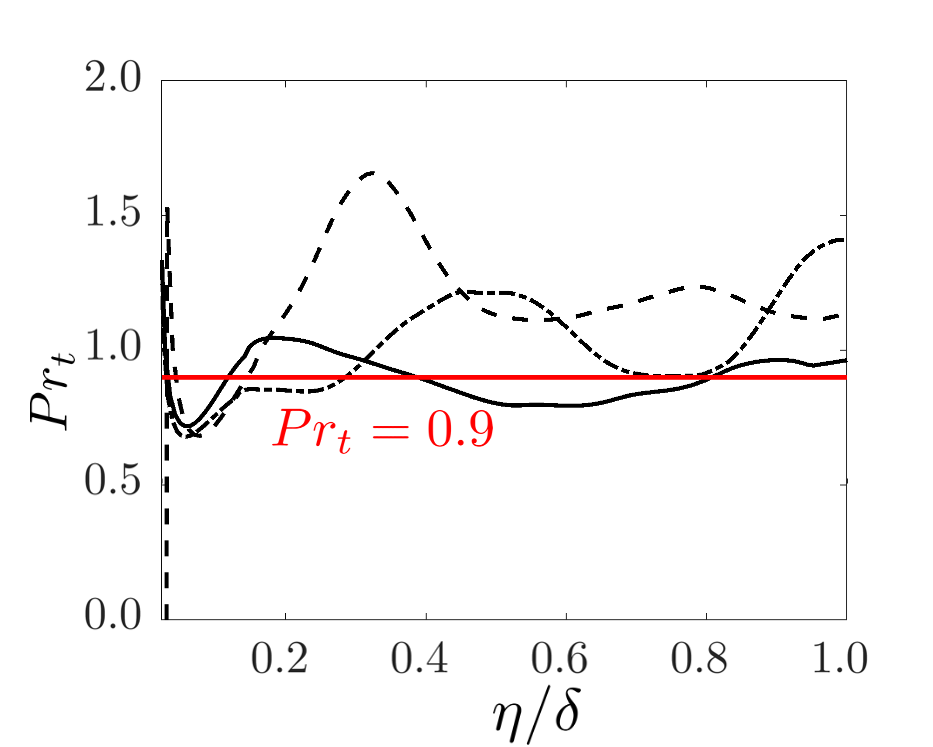}
           \label{fig.quad2}}}
    \end{tabular}
     &    
    \begin{tabular}{c}
    {
   \hspace*{-0.5cm}
    \subcaptionbox{}
    {\includegraphics[height= 3.5 cm, trim = 5 5 5 5, clip]{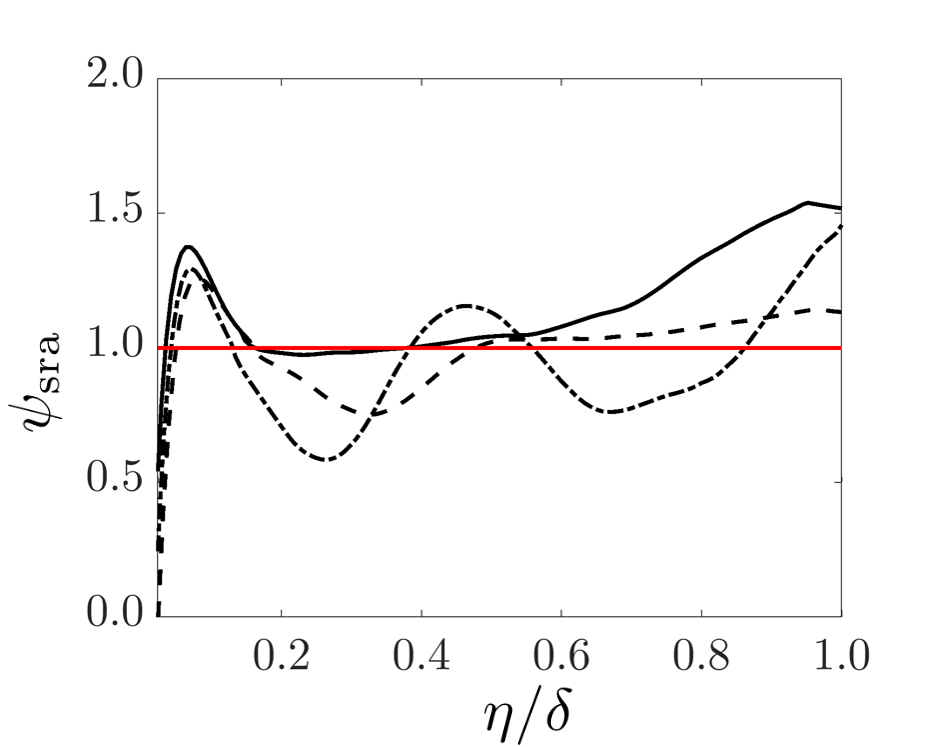}
           \label{fig.quad2}}}
    \end{tabular}
    \end{tabular}
    }    
    \caption{The wall-normal profiles of (a) fluctuation Mach number, $M_t$; (b) fluctuation Prandtl number, $Pr_t$; and (c) Huang's modified strong Reynolds analogy \mbox{parameter,~$\psi_\mathrm{sra}$.}}
    \label{fig:hsra}
\end{figure}

To illustrate the broadband nature of velocity fluctuations in the later stages of transition, in figure~\ref{fig:turb_spec}(d) we evaluate the spanwise wavenumber dependence of the individual contributions of velocity fluctuations to the 1D energy spectrum at $Re_{x} = 1.2 \times 10^{6}$ and $\eta^{+}=40$. At this location, the energy spectrum of streamwise velocity fluctuations clearly exhibits an inertial and a dissipative subrange features, indicating that the flow is approaching a fully turbulent stage.

	Our DNS also provides data for evaluating non-dimensional parameters which can be utilized for low-complexity modeling of high-speed compressible turbulent flows using time-averaged NS equations~\citep{hirsch2007numerical}. In particular, we examine the spatial evolution of fluctuation Mach number, $M_{t}$, fluctuation Prandtl number, $Pr_t$, and Huang's modified strong Reynolds analogy parameter~\citep{huang_coleman_bradshaw_1995}, $\psi_\mathrm{sra}$, 
	\begin{equation}
    M_t 
    \, = \, 
    \sqrt{\frac{{\braket{u_i^\prime u_i^\prime}}}{\gamma \,R \braket{T}}}, \;\; 
    Pr_t 
    \, = \, 
    \frac{\braket{ u^\prime v^\prime}}
    {\braket{ v^\prime T^\prime}}
    \frac{\partial_{\eta} \! \braket{T}}{\partial_{\eta} \! \braket{u}},\;\;\;
    \psi_{\mathrm{sra}} 
    \, = \, 
    \frac{(1-\frac{\partial \braket{T_{0}}}{\partial \braket{T}}) \, Pr_t}{(\gamma-1)\braket{M}^2}\,\frac{\braket{u}}{\braket{T}}\,\frac{T'_\mathrm{rms}}{u'_\mathrm{rms}}.
    \end{equation} 
Here, $\braket{u_i^\prime u_i^\prime} = \braket{u^\prime u^\prime} + \braket{v^\prime v^\prime} + \braket{w^\prime w^\prime}$, $\gamma$ and $R$ denote the specific heat ratio and the gas constant, respectively, and $T_{0}$ is the stagnation temperature. In a fully developed turbulent boundary layer, $\psi_\mathrm{sra}$ provides a measure of correlation between velocity and temperature fluctuations~\citep{huang_coleman_bradshaw_1995}. 

Figure~\ref{fig:hsra} shows that towards the end of the computational domain, at $Re_x= 1.2\times 10^{6}$, these parameters are close to those observed in canonical hypersonic boundary layers, i.e., $M_t \approx 0.2-0.3$, $Pr_{t} \approx 0.9$, and $\psi_{\mathrm{sra}} \approx 1$~\citep{pirozzoli2011turbulence}. However, upstream of the breakdown region, at $Re_x= 1.0\times 10^{6}$, there is a significant deviation compared to these canonical values. Here, $M_t$ becomes as high as $0.45$ which suggests that compressibility effects on flow fluctuations cannot be neglected in the transition zone. Similarly, $Pr_t$ can reach values close to $1.5$ which correspond to decreased fluctuation temperature fluxes in the breakdown region. Furthermore, $Pr_t$ exhibits large variations away from the wall, which is in contrast to the observations in flat-plate turbulent boundary layers, where $Pr_t$ has value of $0.9$ throughout the boundary layer~\citep{saffman1974turbulence, smith1993simultaneous, pirozzoli2011turbulence}. Similar deviations from canonical values in the outer region of boundary layer are also observed in $\psi_\mathrm{sra}$. We conjecture that, in the presence of persistent upstream excitations, the resulting unsteady fluctuations are primarily responsible for these discrepancies.

		\vspace*{-2ex}
\section{\tc{black}{Concluding remarks}} 
\label{sec:discuss}

\begin{figure}
    \centering
    {
    \begin{tabular}{ccc}
    \begin{tabular}{c}
    {
   \hspace*{-0.5cm}
    \subcaptionbox{Experiments}
    {\includegraphics[height= 2.9 cm , trim = 5 5 5 5, clip]{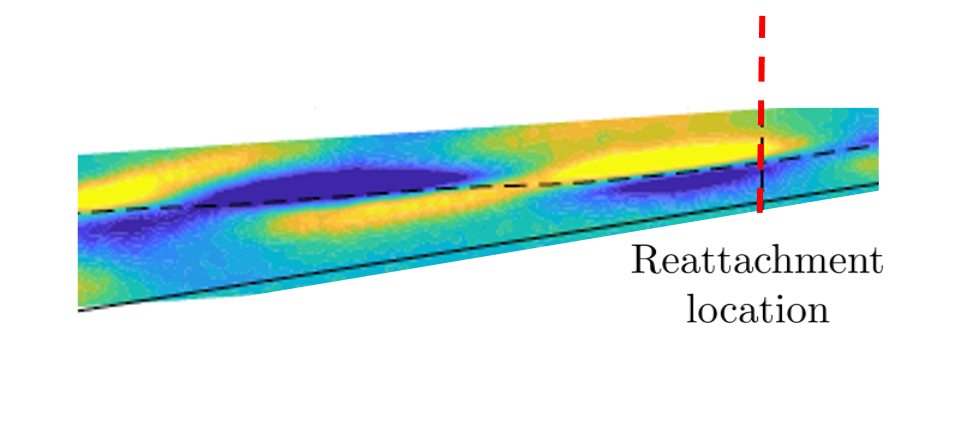}
           \label{fig.quad1}}}
    \end{tabular}
    &    
    \begin{tabular}{c}
    {
   \hspace*{-0.5cm}
    \subcaptionbox{Input-output analysis}
    {\includegraphics[height= 2.9 cm , trim = 5 5 5 5, clip]{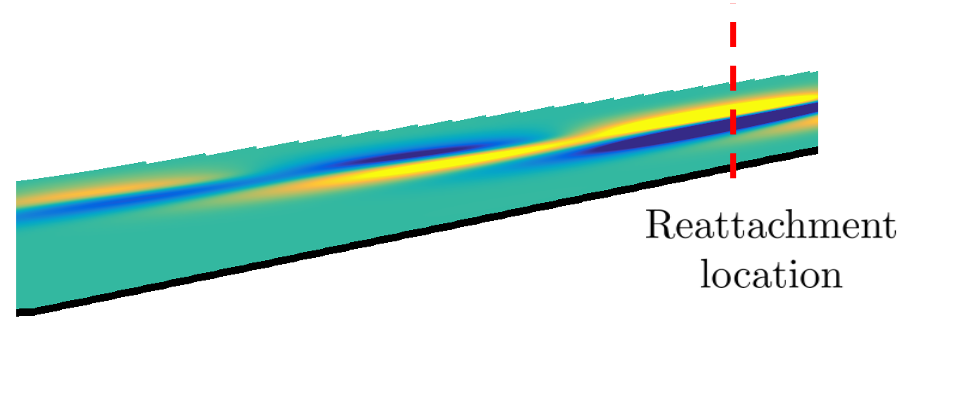}
           \label{fig.quad2}}}
    \end{tabular}
    \end{tabular}
    }    
    \caption{Qualitative comparison of spatial structures of (a) unsteady fluctuations observed using schlieren measurements in Mach $6$ reattaching flow on axisymmetric cone-flare at $U_{\infty}/\nu_{\infty} = 3 \times 10^{6} \, \mathrm{m}^{-1}$~\citep{butler_laurence_2021}; and (b) dominant oblique density fluctuations resulting from input-output analysis of linearization around laminar 2D Mach $5$ reattaching flow at $U_{\infty}/\nu_{\infty} = 13.6 \times 10^{6} \, \mathrm{m}^{-1}$.} 
    \label{fig:expdrdy}
\end{figure}

\tc{black}{Axisymmetric cone-flare experiments~\citep{benitez2020instability,butler_laurence_2021} identified unsteady fluctuations in the separation zone and it is believed that these play an important role in initiating hypersonic flow transition.} As demonstrated in figure~\ref{fig:expdrdy}, we observe strong qualitative similarity between spatial structures of fluctuations observed using schlieren measurements in Mach $6$ reattaching flow on axisymmetric cone-flare~\citep{butler_laurence_2021} and the dominant oblique density fluctuations \tc{black}{that we identify via resolvent} analysis of separated flow over a slender double-wedge. Inspired by these observations, we have examined transition mechanisms in a Mach $5$ hypersonic flow over a slender double-wedge subject to unsteady upstream disturbances.

To investigate the early stages of transition, \tc{black}{we employ resolvent analysis to evaluate responses of the laminar 2D base flow to exogenous time-periodic inputs. This allows us to identify prevailing spatio-temporal scales, the spatial structure of disturbances that most effectively excite the double-wedge flow, as well as the spatial structure of the resulting responses {and the underlying physical mechanisms.} In the presence of flow separation, our analysis shows that two types of disturbances are strongly amplified by the linearized dynamics: steady streamwise vortices and unsteady oblique waves. While amplification of steady upstream vortical disturbances has been studied in~\citet{dwisidniccanjovJFM19},  oblique waves that amplify within separation-reattachment zone have not been investigated. Recently,~\citet{lugrin_beneddine_leclercq_garnier_bur_2021} examined the growth of unsteady perturbations that arise from the ``first mode'' instability of an axisymmetric boundary layer over cylinder-flare geometry. In the presence of stochastic disturbances in the inlet of the computational domain, DNS was used to demonstrate that oblique waves that emerge in the upstream boundary layer (i.e., before separation) can undergo nonlinear interactions similar to those observed in attached compressible boundary layers~\citep{chang_malik_1994,mayer2011direct} and cause transition in separated high-speed flows~\citep{lugrin_beneddine_leclercq_garnier_bur_2021}. In contrast, we show that unsteady disturbances that are localized upstream of the corner trigger oblique waves downstream of the corner even in the absence of  local or global boundary layer instabilities. These oblique waves experience significant amplification within separation-reattachment zone and their role in initiating transition in flows with SWBLI has not been studied before.}

\tc{black}{By carrying out resolvent analysis of the linearized flow equations subject to disturbances that are introduced in a plane immediately upstream of separation, we identify the physical mechanism responsible for non-modal amplification of oblique waves in presence of flow separation. The subsequent nonlinear interaction of dominant unsteady oblique fluctuations is examined using a weakly nonlinear analysis to demonstrate the emergence of steady reattachment streaks inside the recirculation bubble. DNS confirms the predictions of our analysis and provides a detailed characterization of transition initiated by time-periodic upstream oblique disturbances.}

	\vspace*{1ex}
	
\tc{black}{We next briefly summarize our main contributions:}

	\vspace*{1ex}
\begin{enumerate}

\item{\bf \tc{black}{Amplification of oblique waves by base flow curvature.}} 
We analyze  fluctuations' kinetic energy in a streamline-aligned orthogonal coordinate system  to identify physical mechanisms responsible for amplification of oblique waves in the separation zone. Large energy amplification arises from the growth of the fluctuation shear stress due to streamline curvature in the separated shear layer. This is in contrast to the attached boundary layers, where no such mechanism exists. To compare separated and attached boundary layers, we also conduct resolvent analysis of the flow over a wedge that does not contain the compression corner (this wedge is identical to the first wedge in the double-wedge configuration analyzed in the paper). Figure~\ref{fig:fpVssbli} demonstrates that the presence of a recirculation zone in the double-wedge flow significantly increases amplification relative to the single-wedge flow. The amplification profiles of the fluctuation shear stresses differ in these two cases because of fundamentally different physical mechanisms.

\begin{figure}
    \centering
    {
    \begin{tabular}{cc}
    \begin{tabular}{c}
    {
   \hspace*{-0.5cm}
    \subcaptionbox{}
    {\includegraphics[height= 5 cm,  trim = 5 5 5 5, clip]{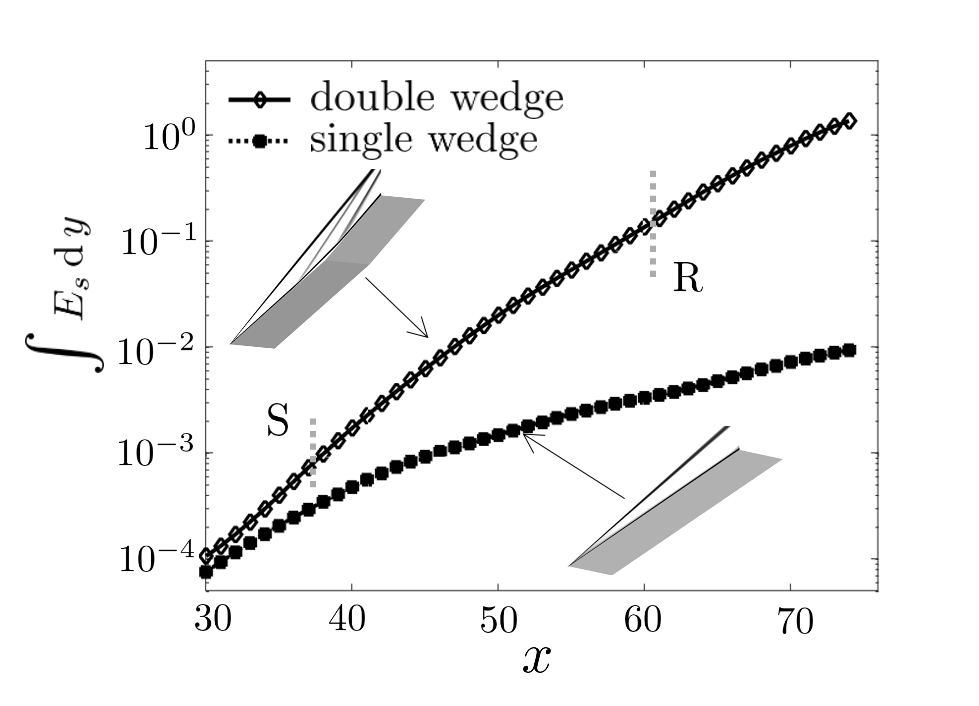}
           \label{fig.quad1}}}
    \end{tabular}
    &    
    \begin{tabular}{c}
    {
   \hspace*{-0.65cm}
    \subcaptionbox{}
    {\includegraphics[height= 5 cm,  trim = 5 5 5 5, clip]{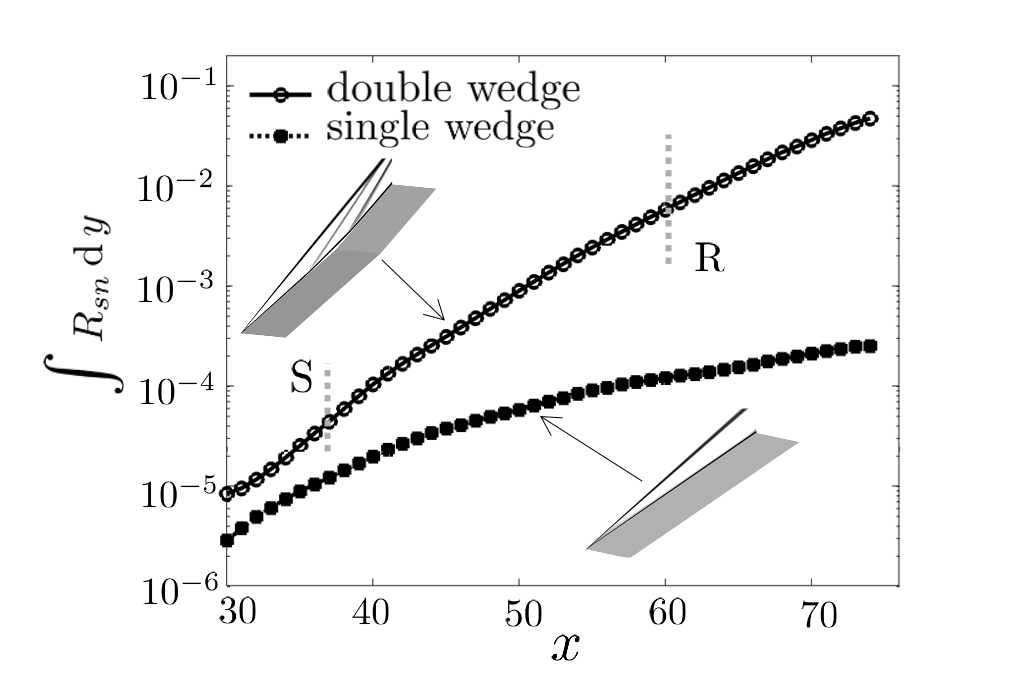}
           \label{fig.quad2}}}
    \end{tabular}
    \end{tabular}
    }    
   \caption{The streamwise evolution of (a) streamwise specific energy and (b) fluctuation shear stress of the dominant output mode $\bphi_1$ of the resolvent associated with linearization around laminar flows over double- and single-wedges with ($\omega = 0.4,\lambda_{z} = 3.0$). }
    \label{fig:fpVssbli}
    \end{figure}

		\vspace*{1ex}
\item{\bf \tc{black}{Steady reattachment streaks through base flow deceleration.}} 
We utilize a weakly nonlinear analysis to show that the resolvent operator associated with the linearized dynamics governs the evolution of steady streaks that arise from quadratic interactions of unsteady oblique waves. \tc{black}{Vortical excitations in the reattaching shear layer generated by these interactions lead to the formation of streaks in the recirculation zone and their subsequent amplification downstream.} Additionally, we use SVD of the resolvent operator to demonstrate that \tc{black}{secondary reattachment} streaks are well approximated by the second output \tc{black}{resolvent} mode. Similar to the most amplified steady output \tc{black}{in compression ramp flow}~\citep{dwisidniccanjovJFM19}, our analysis of the energy budget shows that deceleration of the laminar base flow near reattachment is responsible for amplification of reattachment streaks associated with this sub-optimal mode.

		\vspace*{1ex}
\item{\tc{black}{\bf Transition to turbulence.}} \tc{black}{We use DNS to examine nonlinear stages of the evolution of flow fluctuations.} In the presence of strong upstream oblique disturbances, steady streaks saturate after reattachment and experience sinuous sub-harmonic oscillations.  \tc{black}{The resulting 3D boundary layer 
breaks down further downstream and the observed flow structures are similar to those in other canonical configurations: nonlinear interactions of streaks with oblique waves lead to the development of staggered patterns of lambda vortices, followed by a rapid emergence of higher harmonics in fluctuations and multiple inflection points in the mean velocity profile before breakdown to turbulence~\citep[e.g., see][]{hall_horseman_1991,yu1994mechanism,reddy_schmid_baggett_henningson_1998}.} As the flow transitions to turbulence, the wall friction increases rapidly before settling to the values predicted by turbulent correlations. \tc{black}{Within the transitional zone, non-dimensional parameters that are critical for modeling temperature and compressibility effects in high-speed turbulent flows~\citep{hirsch2007numerical}, e.g., the fluctuation Prandtl and Mach numbers, significantly differ from their fully developed turbulent values.} Post-breakdown, the boundary layer develops mean and fluctuation statistics that agree well with  observations made in attached hypersonic turbulent boundary layers, \tc{black}{thereby demonstrating the efficacy of unsteady oblique waves in triggering transition in separated high-speed boundary layer flows.}
\end{enumerate}

		\vspace*{1ex}
		Unsteady disturbances in hypersonic boundary layers can arise from free-stream turbulence in wind tunnel experiments~\citep{schneider2001, schneider2015developing}, \tc{black}{interactions of unsteady free-stream disturbances with surface roughness}~\citep{wu2001,Gonzalez2019}, atmospheric particulates associated with ice-clouds and volcanic dust~\citep{turco1992upper,chuvakhov2019numerical}, and atmospheric turbulence~\citep{bushnell1990notes}. Novel physical mechanisms that we identify are unique to high-speed boundary layers with a separation-reattachment zone. \tc{black}{We expect that insights about transition mechanisms that we provide using a combination of resolvent and weakly nonlinear analyses with DNS will motivate a systematic evaluation of specific disturbance environments that appear in wind tunnels or free flights and guide the development of low-complexity models for fast and accurate prediction of transition in hypersonic flows under realistic in-flight conditions.}

	\vspace*{-2ex}
\section*{Acknowledgments}

We would like to thank Prof.\ Graham V.\ Candler  for providing access to the US3D solver and computational facilities at the University of Minnesota and the Air Force Office of Scientific Research for financial support under Award FA9550-18-1-0422. 

	\vspace*{-2ex}
\section*{Declaration of interests}

The authors report no conflict of interest.

	\newpage
\appendix
	
	\vspace*{-2ex}
\section{Nonlinear terms at $\mathcal{O}(\epsilon^2)$}
\label{app.nonlin}
	\tc{black}{As shown in \S~\ref{ref:wnla}, $\hat{\bd}_{0}^{(2)} \DefinedAs \mathcal{N}_{0}^{(2)}(\hat{\bpsi}^{(1)}_{\pm})$ accounts for quadratic interactions between $\hat{\bpsi}^{(1)}_{+}$ and $\hat{\bpsi}^{(1)}_{-}$ at $\mathcal{O}(\epsilon^2)$. For steady streaks $\hat{\bpsi}^{(2)}_{0}$, $\hat{\bd}^{(2)}_{0,\rho} = 0$ and the contributions to the equations for the momentum and total energy fluctuations are given by, 
	\begin{equation}
\begin{array}{rcl}
	\hat{\bd}^{(2)}_{0,\rho u} 
	& \!\!\! = \!\!\! &
	\left(\gamma-1\right) \dfrac{\partial}{\partial x} \bar{\rho} \left(\dfrac{\gamma - 3}{\gamma - 1} (u_{+}^{(1)}u_{-}^{(1)})\,+\, (v_{+}^{(1)} v_{-}^{(1)}) + (w_{+}^{(1)} w_{-}^{(1)}) \right)  
	~ -
	\\[0.cm]
	& \!\!\! \!\!\! &
	\dfrac{\partial}{\partial y} \bar{\rho} \left(u_{-}^{(1)}v_{+}^{(1)}\,+\, u_{+}^{(1)} v_{-}^{(1)}  \right) \,-\, \mri \beta \bar{\rho} \left(u_{-}^{(1)}w_{+}^{(1)}\,+\, u_{+}^{(1)} w_{-}^{(1)}  \right),
	\\[0.cm]
	\hat{\bd}^{(2)}_{0,\rho v}  
	& \!\!\! = \!\!\! &
	\left(\gamma-1\right) \dfrac{\partial}{\partial y} \bar{\rho} \left(\dfrac{\gamma - 3}{\gamma - 1} (v_{+}^{(1)}v_{-}^{(1)})\,+\, (u_{+}^{(1)} u_{-}^{(1)}) + (w_{+}^{(1)} w_{-}^{(1)}) \right) 
	~ -
	\\[0.cm]
	& \!\!\! \!\!\! &
	\dfrac{\partial}{\partial x} \bar{\rho} \left(u_{-}^{(1)}v_{+}^{(1)}\,+\, u_{+}^{(1)} v_{-}^{(1)}  \right) \,-\, \mri \beta \bar{\rho} \left(v_{-}^{(1)}w_{+}^{(1)}\,+\, v_{+}^{(1)} w_{-}^{(1)}  \right),
	\\[0.cm]
	\hat{\bd}^{(2)}_{0,\rho w}  
	& \!\!\! = \!\!\! &
	\mri \beta \bar{\rho} \left(\gamma-1\right) \left(\dfrac{\gamma - 3}{\gamma - 1} (w_{+}^{(1)}w_{-}^{(1)})\,+\, (u_{+}^{(1)} u_{-}^{(1)}) + (v_{+}^{(1)} v_{-}^{(1)}) \right)  
	~ -
	\\[0.cm]
	& \!\!\! \!\!\! &
	\dfrac{\partial}{\partial x} \bar{\rho} \left(u_{-}^{(1)}w_{+}^{(1)}\,+\, u_{+}^{(1)} w_{-}^{(1)}  \right) \,-\, \dfrac{\partial}{\partial y} \bar{\rho} \left(v_{-}^{(1)}w_{+}^{(1)}\,+\, v_{+}^{(1)} w_{-}^{(1)}  \right),
	\\
	\hat{\bd}^{(2)}_{0,E_{t}}
	 & \!\!\! = \!\!\! & 
	- \mri \beta \bar{\rho}\, w_{+}^{(1)} \left(\gamma{C_v} T_{-}^{(1)} + \bar{u} u_{-}^{(1)} + \bar{v} v_{-}^{(1)}\right) 
	\, - 
	\\[0.cm]
	& \!\!\! \!\!\! & 
	\phantom{-}
	\mri \beta \bar{\rho}\, w_{-}^{(1)} \left(\gamma{C_v} T_{+}^{(1)} + \bar{u} u_{+}^{(1)} + \bar{v} v_{+}^{(1)}\right)
	\, -
	\\[0.cm]
	& \!\!\! \!\!\! & 
	\left(\gamma-1\right) \dfrac{\partial}{\partial x} \bar{\rho} \bar{u} \left(\dfrac{\gamma - 3}{\gamma - 1} (u_{+}^{(1)}u_{-}^{(1)})\,+\, (v_{+}^{(1)} v_{-}^{(1)}) + (w_{+}^{(1)} w_{-}^{(1)}) \right)
	~ + 
	\\[0.cm]
	& \!\!\! \!\!\! & 
	\left(\gamma-1\right) \dfrac{\partial}{\partial y} \bar{\rho} \bar{v} \left(\dfrac{\gamma - 3}{\gamma - 1} (v_{+}^{(1)}v_{-}^{(1)})\,+\, (u_{+}^{(1)} u_{-}^{(1)}) + (w_{+}^{(1)} w_{-}^{(1)}) \right)
	~ -
	\\[0.cm]
	& \!\!\! \!\!\! & 	
	\dfrac{\partial}{\partial x}\bar{\rho}\, u_{+}^{(1)} \left(\gamma{C_v} T_{-}^{(1)} + \bar{v} v_{-}^{(1)}\right) \,-\, \dfrac{\partial}{\partial x}\bar{\rho}\, u_{-}^{(1)} \left(\gamma{C_v} T_{+}^{(1)} + \bar{v} v_{+}^{(1)}\right) 
	~ -
	\\[0.cm]
	& \!\!\! \!\!\! &
	\dfrac{\partial}{\partial y}\bar{\rho}\, v_{+}^{(1)} \left(\gamma{C_v} T_{-}^{(1)} + \bar{u} u_{-}^{(1)}\right) \,-\, \dfrac{\partial}{\partial y}\bar{\rho}\, v_{-}^{(1)} \left(\gamma{C_v} T_{+}^{(1)} + \bar{u} u_{+}^{(1)}\right) 
	~ -
	\\[0.cm]
	& \!\!\! \!\!\! & 
	\mri \beta \lambda \left( w^{(1)}_{+} \dfrac{\partial}{\partial x}  u^{(1)}_{-} +  w^{(1)}_{-} \dfrac{\partial}{\partial x}  u^{(1)}_{+}\right) \,+\,  \mri \beta \mu \left( u^{(1)}_{+} \dfrac{\partial}{\partial x}  w^{(1)}_{-} +  u^{(1)}_{-} \dfrac{\partial}{\partial x}  w^{(1)}_{+}\right)
	~ -
	\\[0.cm]
	& \!\!\! \!\!\! & 
	\mri \beta \lambda \left( w^{(1)}_{+} \dfrac{\partial}{\partial y}  y^{(1)}_{-} +  w^{(1)}_{-} \dfrac{\partial}{\partial y}  v^{(1)}_{+}\right) \,+\,  \mri \beta \mu \left( v^{(1)}_{+} \dfrac{\partial}{\partial y}  w^{(1)}_{-} +  v^{(1)}_{-} \dfrac{\partial}{\partial y}  w^{(1)}_{+}\right)
	~ -
	\\[0.cm]
	& \!\!\! \!\!\! & 
	2 \beta^{2} \mu \left( u^{(1)}_{+} u^{(1)}_{-} + v^{(1)}_{+} v^{(1)}_{-} + \dfrac{\lambda+2\mu}{\mu} w^{(1)}_{+} w^{(1)}_{-} \right) 
	~+ 
	\\[0.cm]
	& \!\!\! \!\!\! &
	\mri 2 \beta \dfrac{\partial}{\partial y} \left((\mu+\lambda) (w^{(1)}_{+} v^{(1)}_{-} + w^{(1)}_{-} v^{(1)}_{+}) \right)
	 ~+
	 \\[0.cm]
	& \!\!\! \!\!\! &
	 \mri 2 \beta \dfrac{\partial}{\partial x} \left((\mu+\lambda) (w^{(1)}_{+} u^{(1)}_{-} + w^{(1)}_{-} u^{(1)}_{+}) \right)
	 ~+
	 \\[0.cm]
	& \!\!\! \!\!\! &
	\dfrac{\partial}{\partial x} \mu \left(v^{(1)}_{+} \dfrac{\partial u^{(1)}_{-}}{\partial y}  +  v^{(1)}_{-} \dfrac{\partial u^{(1)}_{+}}{\partial y} \right) \,+\, \dfrac{\partial}{\partial x} \lambda \left(u^{(1)}_{+} \dfrac{\partial v^{(1)}_{-}}{\partial y}  +  u^{(1)}_{-} \dfrac{\partial v^{(1)}_{+}}{\partial y} \right)
	 ~+
	 \\[0.cm]
	& \!\!\! \!\!\! &
	\dfrac{\partial}{\partial y} \mu \left(u^{(1)}_{+} \dfrac{\partial v^{(1)}_{-}}{\partial x}  +  u^{(1)}_{-} \dfrac{\partial v^{(1)}_{+}}{\partial x} \right) \,+\, \dfrac{\partial}{\partial y} \lambda \left(v^{(1)}_{+} \dfrac{\partial u^{(1)}_{-}}{\partial x}  +  v^{(1)}_{-} \dfrac{\partial u^{(1)}_{+}}{\partial x} \right)
	~ +
	\\[0.cm]
	& \!\!\! \!\!\! & 
	\dfrac{\partial}{\partial x} \mu \left(\dfrac{\partial v^{(1)}_{+} v^{(1)}_{-}}{\partial x} + \dfrac{\partial v^{(1)}_{+} v^{(1)}_{-}}{\partial x} + \dfrac{\lambda+2\mu}{\mu} \dfrac{\partial u^{(1)}_{+} u^{(1)}_{-}}{\partial x} \right) 
	~ +
	\\[0.cm]
	& \!\!\! \!\!\! & 
	\dfrac{\partial}{\partial y} \mu \left(\dfrac{\partial u^{(1)}_{+} u^{(1)}_{-}}{\partial y} + \dfrac{\partial w^{(1)}_{+} w^{(1)}_{-}}{\partial y} + \dfrac{\lambda+2\mu}{\mu} \dfrac{\partial v^{(1)}_{+} v^{(1)}_{-}}{\partial y} \right),
\end{array} 
	\label{eq.N2}
\end{equation}
where we drop the caret notation on the right-hand side of equation~\eqref{eq.N2} for brevity. Here, $C_{v} = R/(\gamma -1)$, where $R$ is the gas constant, $\gamma = 1.4$ is the ratio of the specific heat capacities, and $\mu(x,y)$ and $\lambda(x,y)$ are the coefficients of viscosity and bulk viscosity, respectively. We utilize Sutherland's law for computing viscosity and assume $\lambda=-{2 \mu}/{3}$.}

\vspace*{-2ex}
{\color{black}\section{Relation between the state $\bPsi$ and the output $\bPhi$}
\label{app.conv2Nconv}

Herein, we utilize a weakly nonlinear expansion to establish the relation between the components of the state vector $\bPsi = (\Psi_1, \bPsi_{2}, \Psi_3) \DefinedAs (\rho, \rho \B u, E_t)$ in conserved variables and the components of the vector $\bPhi = (\Phi_1, \bPhi_{2}, \Phi_3) \DefinedAs (\rho, \bu, T)$ in primitive variables. Here, $E_t$ is the total energy per unit volume of the gas,
 	\begin{equation}
	E_t 
	\; = \;
	C_v \rho T \; + \; \dfrac{1}{2} \, \rho | \bu |^2,
	\end{equation}
$C_v$ is the specific heat at constant volume, and $| \bu |^2 \DefinedAs \bu^T \bu$. Within the weakly nonlinear framework, we can decompose $\bPsi$ and $\bPhi$ into the sums of base and fluctuating components, 
	\begin{equation}
	\begin{array}{rcl}
	\bPsi 
	& \!\! = \!\! & 
	\bar{\bPsi}
	\; + \, 
	\bpsi
	\; = \;
	\bar{\bPsi}
	\; + \, 
	\epsilon \bpsi^{(1)}
	\; + \; 
	\epsilon^2 \bpsi^{(2)}
	\; + \; 
	{\cal O} (\epsilon^3),
	\\[0.1cm]
	\bPhi 
	& \!\! = \!\! & 
	\bar{\bPhi}
	\; + \, 
	\bphi
	\; = \;
	\bar{\bPhi}
	\; + \, 
	\epsilon \bphi^{(1)}
	\; + \; 
	\epsilon^2 \bphi^{(2)}
	\; + \; 
	{\cal O} (\epsilon^3),
	\end{array}
	\end{equation}
and utilize the following relations between the components of $\bPsi$ and $\bPhi$,
	\begin{equation}
	\begin{array}{rcl}
	\Psi_1 
	& \!\! = \!\! & 
	\Phi_1,
	\\[0.1cm]
	\bPsi_2 
	& \!\! = \!\! & 
	\Phi_1 \bPhi_2,
	\\[0.1cm]
	\Psi_3 
	& \!\! = \!\! & 
	C_v \Phi_1 \Phi_3 \; + \; \dfrac{1}{2} \, \Phi_1 | \bPhi_2 |^2,
	\end{array}
	\end{equation}	
to obtain	
\begin{equation}
	\label{eq.CD}
	\begin{array}{rrcl}
	\!\!
	\mbox{$\cO(1)$:} 
	&
	\bar{\Phi}_1 
	& \!\! = \!\! & 
	\bar{\Psi}_1,
	~~
	\bar{\bPhi}_2
	\; = \;
	\dfrac{\bar{\bPsi}_2}{\bar{\Psi}_1},
	~~
	\bar{\bPhi}_3
	\; = \;
	\dfrac{1}{C_v \bar{\Psi}_1} 
	\!
	\left(
	\bar{\Psi}_3 \; - \; \dfrac{| \bar{\bPsi}_2 |^2}{2 \bar{\Psi}_1}
	\right),
	\\[0.35cm]
	\!\!
	\mbox{$\cO(\epsilon)$:} 
	&
	\left[
	\begin{array}{c}
	{\phi^{(1)}_{1}}  \\
	{\bphi^{(1)}_{2}}  \\
	\phi^{(1)}_{3}
	\end{array}
	\right] 
	& \!\! = \!\! &
	\underbrace{\dfrac{1}{\bar{\Phi}_{1}}
	\!
	\left[
	\begin{array}{ccc}
	\bar{\Phi}_{1} & 0 & 0  
	\\[0.1cm]
	-\bar{\bPhi}_{2} & I & 0 
	\\[0.1cm]
	\frac{1}{2 C_{v}} \, | \bar{\bPhi}_{2} |^2 - \bar{\Phi}_{3} & -\frac{1}{C_{v}} \, \bar{\bPhi}_{2}^T & \frac{1}{C_{v}}
	\end{array}
	\right]}_{\cC}	
	\left[
	\begin{array}{c}
	{\psi^{(1)}_{1}}  \\
	{\bpsi^{(1)}_{2}}  \\
	{\psi^{(1)}_{3}}
	\end{array}
	\right],
	\\[0.35cm]
	\!\!
	\mbox{$\cO(\epsilon^2)$:} 
	&
	\left[
	\begin{array}{c}
	{\phi^{(2)}_{1}}  \\
	{\bphi^{(2)}_{2}}  \\
	\phi^{(2)}_{3}
	\end{array}
	\right] 
	& \!\! = \!\! &
	\cC
	\left[
	\begin{array}{c}
	{\psi^{(2)}_{1}}  \\
	{\bpsi^{(2)}_{2}}  \\
	\psi^{(2)}_{3}
	\end{array}
	\right] 
	\; + \; 
	\underbrace{\frac{1}{\bar{\Phi}_{1}}
	\!
	\left[
	\begin{array}{cc}
	0 & 0  \\
	I & 0 \\
	0  & \frac{1}{2 C_{v}}
	\end{array}
	\right]}_{\cD}
	\left[
	\begin{array}{c}
	-{\phi^{(1)}_{1}} {\bphi^{(1)}_{2}}  \\
	-2 C_{v} {\phi^{(1)}_{1}} {\phi^{(1)}_{3}} - {\bar{\Phi}_{1} |{\bphi^{(1)}_{2}}|^{2}}
	\end{array}
	\right].
	\end{array}
	\end{equation}	
	}
		
	\vspace*{-2ex}
\section{Energy transport equation in streamline coordinates}
\label{app.terms}

The terms on the right-hand side of transport equation~(\refeq{eq:ketrans_symb}) for streamwise specific kinetic energy $\mathcal{E}_s \DefinedAs u_s^{\prime} u_s^{\prime}$ are determined by
\begin{align}
\begin{split}
\mbox{Production:}~\,\mathcal{P} & \, \DefinedAs \, -{u^\prime_{s} u^\prime_{n}} \partial_{n}  \bar{u}_{s} - {u^\prime_{s} u^\prime_{s}} \partial_{s} \bar{u}_{s} - {\rho^{\prime} u_{s}^{\prime}}  \frac{\bar{u}_{s}}{{\bar{\rho}}} \partial_{s} \bar{u}_{s}, \\
\mbox{Source:}~\,\mathcal{S} & \, \DefinedAs \, -\frac{u^\prime_{s}}{\bar{\rho}} {\partial_s} p^{\prime}, \\
\mbox{Viscous:}~\,\mathcal{V} 
	& \, \DefinedAs \, 
	\frac{\bar{\mu}}{\bar{\rho}}\left( 2u^\prime_{s} \partial_{s} u^\prime_{s}  
	\, + \, 
	u^\prime_{s} \partial_{n} 
	\!
	\left( \partial_{n} u^\prime_{s} 
	\, + \, 
	\partial_{s} u^\prime_{n} \right)  
	\, + \, 
	\partial_{z} u^\prime_{s}
	\! 
	\left( \partial_s w^\prime  \,+\, \partial_{z} u^\prime_s \right) \right),
	\\
\mbox{Curvature:}~\,\mathcal{K} 
	& \, \DefinedAs \, 
	-K_{c} u^\prime_{s} u^\prime_{n} 
	\, - \, 
	\frac{2\bar{\mu}}{\bar{\rho}\bar{u}^2_{s}}\left(K^2_{s} u^\prime_{s} u^\prime_{n} 
	\, + \, 
	K^2_{c} u^\prime_{s} u^\prime_{s}\right) ~+ 
	\\[0.1cm]
	& 
	\, \phantom{\DefinedAs} \,  
	~~
	\frac{1}{{\bar{\rho}}}\left(u^\prime_{s} \partial_{s} 
	\!	
	\left(\frac{2\bar{\mu}K_{c} u^\prime_{n}}{\bar{u}_{s}} \right) 
	\, - \, 
	u^\prime_{s} \partial_n
	\!
	\left( \frac{\bar{\mu} 
	(K_c u^\prime_s \,+\, K_s u^\prime_n)}{\bar{u}_{s}} \right) \right) 
	~ +
		\\[0.15cm]
	& 
	\, \phantom{\DefinedAs} \,  
	~~
	\frac{2\bar{\mu}}{\bar{u}_{s} \bar{\rho}} 
	\!
	\left( K_s u^\prime_{s} \partial_{s} u^\prime_{s}   
	\, + \, 
	K_c u^\prime_{s} 
	\! 
	\left({\partial_n} u^\prime_{s}  
	\,+\,  
	{\partial_s} u^\prime_{n} \right)
	\, - \, 
	K_s u^\prime_{s} {\partial_n} u^\prime_{n} \right).
	\end{split}
\end{align} 
Here, $K_{c}$ and $K_{s}$ denote contributions that arise from the curvature normal to the streamlines and from deceleration along the streamline direction, respectively, and $\bar{\mu}$ is the coefficient of viscosity associated with the laminar base flow. \tc{black}{The curvature terms $K_{c}$ and $K_{s}$ result from transformation into streamline coordinate system~\citep{yousefi_veron_2020} and they can be analytically expressed using the mean vorticity and the velocity gradients of the laminar 2D base flow~\citep{finnigan1983streamline}; see equation~\eqref{eq:KsKc}.}

	\vspace*{-2ex}
\section{Transport equation for the fluctuation shear stress}
\label{app.shear}

The transport equation that governs the evolution of time and spanwise averaged fluctuation shear stress ${R}_{sn}\DefinedAs \braket{u^\prime_{s}u^\prime_{n}}$ in $(s,n,z)$ coordinates is given by
\begin{align}
\bar{u}_{s} \frac{\partial {R}_{sn}}{\partial s} 
\; = \;
\mathcal{P}_{r} 
\; + \; 
\mathcal{S}_{r} 
\; + \; 
\mathcal{K}_{r}.
\end{align}
The viscous terms are neglected because they do not contribute to the transport of $R_{sn}$ and the terms on the right-hand-side are determined by
\begin{align}
\begin{split}
\mbox{Production:}~\,\mathcal{P}_{r} & \, \DefinedAs \,  -\, {R}_{s n} \partial_s \bar{u}_{s}  \,-\,   \braket{u^\prime_{n}u^\prime_{n}} \partial_n \bar{u}_{s} \,+\, \braket{\rho^{\prime} u^\prime_{s}} \frac{\partial_{n} \bar{p}}{\bar{\rho}^2} \,+\, \braket{\rho^{\prime} u^\prime_{n}} \frac{\partial_{s} \bar{p}}{\bar{\rho}^2}, \\
\mbox{Source:}~\,\mathcal{S}_{r} & \, \DefinedAs\, -\frac{1}{\bar{\rho}} \, 
({\braket{u^\prime_{n}{\partial_s} p^{\prime}}} \,+\,{\braket{u^\prime_{s}{\partial_n} p^{\prime}}} ), 
	\\[0.15cm]
\mbox{Curvature:}~\,\mathcal{K}_{r} & \, \DefinedAs\, \left( 2E_{s} \,-\, \braket{u^\prime_{n}u^\prime_{n}}\right) K_{c} \, - \, {R}_{s n} K_{s}.
\end{split}
\end{align}

	\vspace*{-2ex}
{\color{black}\section{Grid convergence for DNS}
\label{app.grid}
 \begin{figure}
     \centering
       \begin{tabular}{c}
       \includegraphics[width=0.55\linewidth, trim= 4 4 4 4, clip]{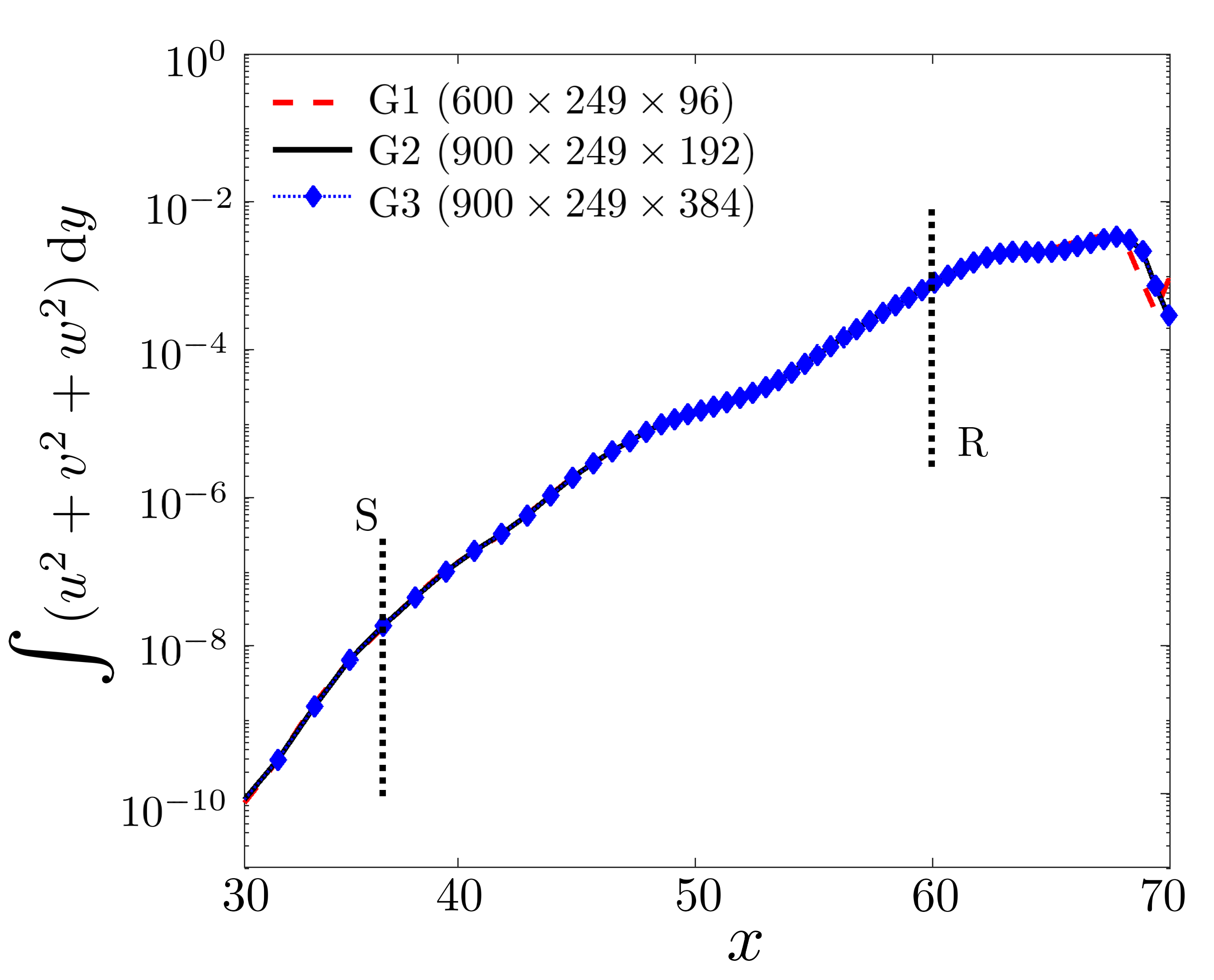}
     \end{tabular}
     \caption{Comparison of energy of streaks at $\lambda_{z} = 1.5$ for three different grid resolutions.} 
     \label{fig:gridcon}
 \end{figure}

Figure~\ref{fig:gridcon} plots the energy of streaks at $(\omega,\lambda_{z}) = (0,1.5)$ generated by the interaction of oblique waves with $(\omega,\lambda_{z}) = (\pm 0.4, 3.0)$ in a computational domain with the following number of grid points: (G1) $600 \times 249 \times 96$; (G2) $900 \times 249 \times 192$; and (G3) $900 \times 249 \times 384$. The disturbance amplitude is set to $a_\mathrm{ob} = 2.50 \times 10^{2} A_{0}$. As shown in figure~\ref{fig:gridcon}, the spatial evolution of energy of streaks is almost identical throughout the separation zone for all grids. All of our numerical computations in the paper are reported for (G3). In table~1, we also compare the resolution of (G3) with the discretization utilized in recent DNS studies of of supersonic and hypersonic transitional and turbulent flows.

\begin{table}
\centering
{
	\begin{tabular}{c c c c c c}
  			& Duan (1) & Mayer (2) & Pirozzoli (3) & Franko (4) & Our study\\ \hline
  Mach & $5.8$ & $3.0$ & $2.25$ & $6.0$ & $5.0$ \\
  $Re_{{\theta},\text{max.}}$ & $5775$& $1985$ & $4250$ &$2652$ & $1670$ \\
  $\Delta x^{+}_{\text{max.}}$ &$7.8$ & $3.3$ & $14.5$ &$4.4$ &  $4.2$ \\
  $(\Delta y^{+}_{wall})_{\text{max.}}$ &$0.3$ & $0.49$ & $1.0$ &$0.3$ &  $0.22$ \\
  $\Delta z^{+}_{\text{max.}}$ &$3.1$ & $1.4$ & $6.56$ &$2.97$ &  $2.5$ 
  	\end{tabular}
\caption{Summary of DNS computations of transition and turbulence reported in (1) \citet{duan_beekman_martin_2011}, (2) \citet{mayer2011direct}, (3) \citet{pirozzoli2011turbulence}, (4) \citet{franko2013breakdown}.}}
\label{tab:grid-comp}
\end{table}}

\vspace*{-2ex}
\section{\tc{black}{Distribution of wall temperature}}
\label{app.temp}

\begin{figure}
    \centering
    {
    \begin{tabular}{cc}
    {\begin{tabular}{c}
    \begin{tabular}{c}
    {
   \hspace*{-0.6cm}
    \subcaptionbox{}
    {\includegraphics[height= 2.35 cm, trim=5 5 5 5,clip]{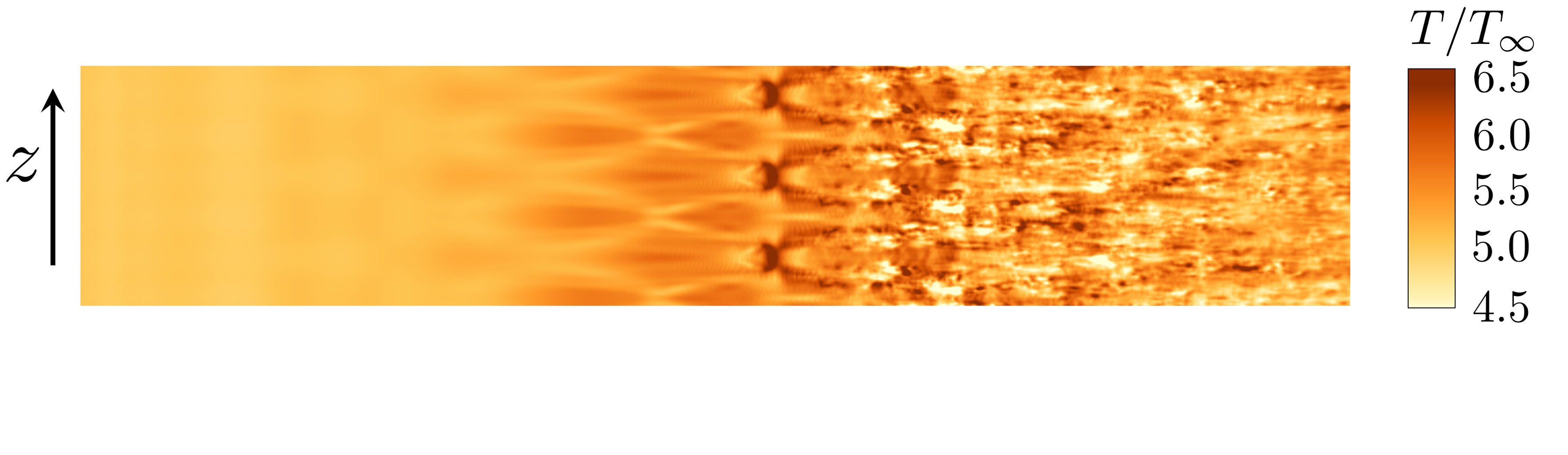}
           \label{fig.quad1}}}
    \end{tabular}
    \\
      \begin{tabular}{c}
    {
   \hspace*{-0.65cm}
    \subcaptionbox{}
    {\includegraphics[height= 2.35 cm, trim=5 5 5 5,clip]{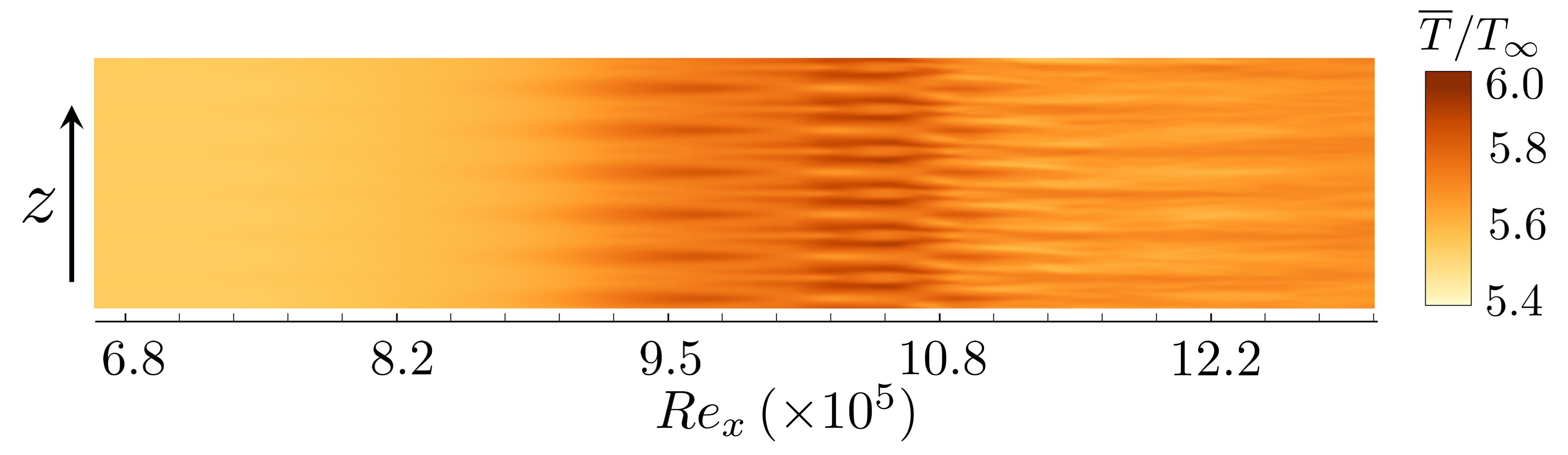}
           \label{fig.quad2}}}
    \end{tabular}
    \end{tabular}}
    &    
   {\begin{tabular}{c}
    {
   \hspace*{-0.75cm}
    \subcaptionbox{}
    {\includegraphics[height= 4.35 cm, trim=4 4 4 4,clip]{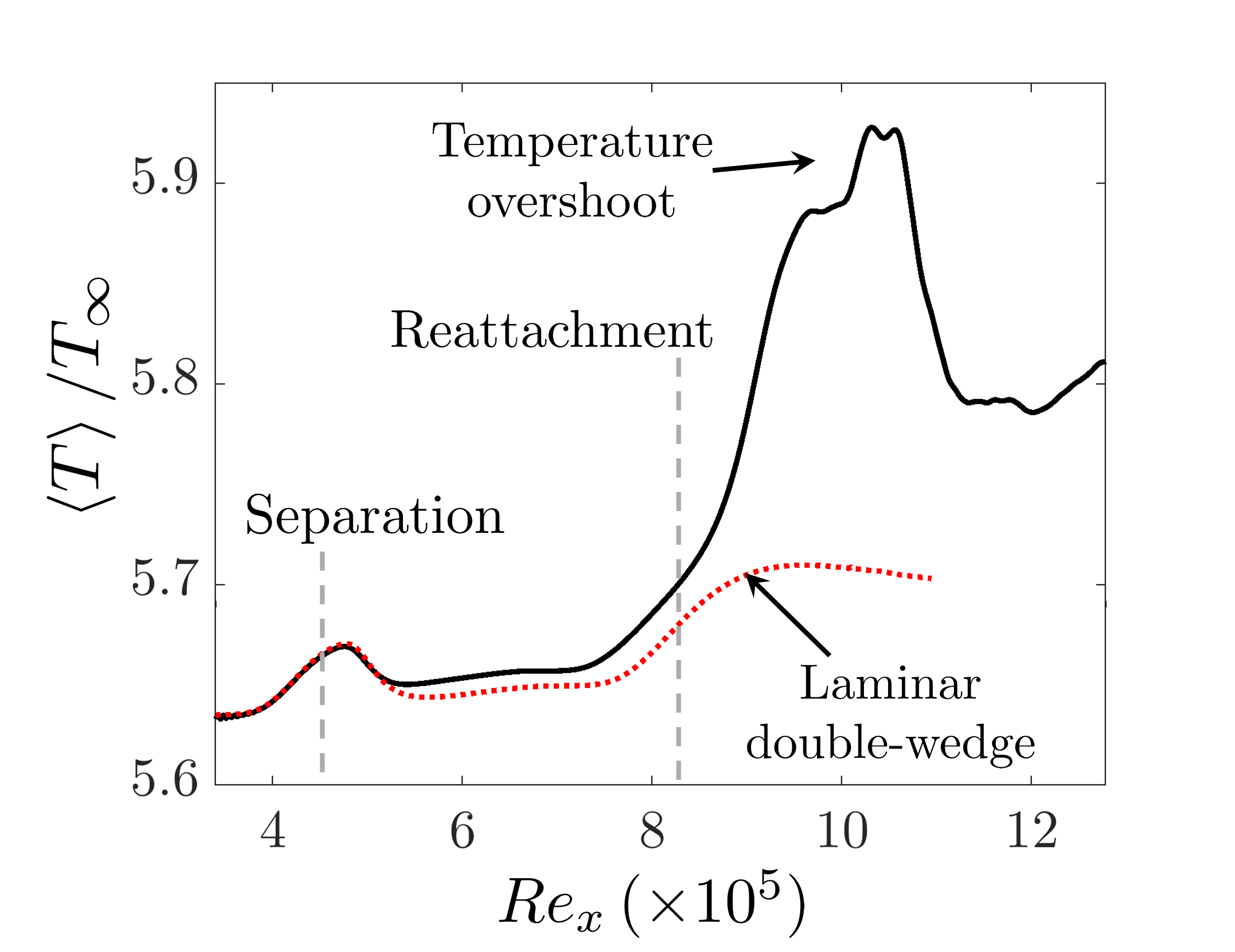}
           \label{fig.quad2}}}
    \end{tabular}}
    \end{tabular}
    }    
  \caption{Streamwise variations of (a) instantaneous; (b) time-averaged; and (c) time- and spanwise-averaged wall temperatures.}
    \label{fig:meanT}
    \end{figure}

{\color{black}In addition to the skin friction, the distribution of wall temperature provides insights into the thermal effects encountered in compressible boundary layer flows along the transition zone. In flows with adiabatic walls, there is no heat transfer to the wall and viscous dissipation near the wall converts kinetic into internal energy. This leads to high temperatures near the wall as well as in the associated thermal boundary layer and 3D patterns in the wall temperature are caused by temperature transport within the boundary layer by flow fluctuations.

	Figures~\ref{fig:meanT}(a) and~\ref{fig:meanT}(b) illustrate instantaneous and mean wall temperatures in the transition zone. In contrast to the instantaneous skin friction, where the streaks determine the spanwise modulation prior to flow transition, the instantaneous wall temperature contains strong imprints of unsteady oblique waves. The role of the streaks and higher spanwise harmonics becomes apparent when we examine the \mbox{time-averaged wall temperature.}

	In Figure~\ref{fig:meanT}(c), we illustrate the mean wall-temperature along the double-wedge. Even though temperature variations are not significant under present conditions, its analysis can be informative for different free-stream conditions. Comparison with the double wedge laminar solution shows that the wall temperature is higher in the 3D flow field immediately after the separation point and that it rises rapidly post-reattachment. In the transition zone, we observe an overshoot before reduction to its turbulent value.}

		\vspace*{-6ex}

\end{document}